\documentclass[twocolumn]{aastex62}
\hypersetup{linkcolor=red,citecolor=green,filecolor=cyan,urlcolor=magenta}

\def\ts     {\thinspace} 
\usepackage{amsmath}
\DeclareMathOperator{\erf}{erf}

\def\kms  {\ifmmode{{\rm \ts km\ts s}^{-1}}\else{\ts km\ts s$^{-1}$\ts}\fi}
\def\msol {\ifmmode{{\rm M}_{\odot}}\else{M$_{\odot}$\ts}\fi}
\def\lsun {\ifmmode{{\rm L}_{\odot}}\else{L$_{\odot}$\ts}\fi}
\def\cii  {\ifmmode{{\rm [C}{\rm \scriptstyle II}]}\else{[C\ts {\scriptsize II}]\ts}\fi}
\def\ci   {\ifmmode{{\rm C}{\rm \scriptstyle I}}\else{C\ts {\scriptsize I}\ts}\fi}
\def\m    {\ifmmode{\mu {\rm m}}\else{$\mu$m}\fi}
\def\hi   {\ifmmode{{\rm H}{\rm \scriptstyle I}}\else{H\ts {\scriptsize I}\ts}\fi}
\def\hii  {\ifmmode{{\rm H}{\rm \scriptstyle II}}\else{H\ts {\scriptsize II}\ts}\fi}
\def\nii  {\ifmmode{{\rm [N}{\rm \scriptstyle II}]}\else{[N\ts {\scriptsize II}]\ts}\fi}
\def\oiii {\ifmmode{{\rm [O}{\rm \scriptstyle III}]}\else{[O\ts {\scriptsize III}]\ts}\fi}
\def\hh   {\ifmmode{{\rm H}_2}\else{H$_2$\ts}\fi}
\def\nhh  {\ifmmode{N({\rm H}_2)}\else{$N$(H$_2$)\ts}\fi}
\def\microns {\ifmmode{\mu{\rm m}}\else{$\mu$m\ts}\fi}

\graphicspath{{./}{}}

\shorttitle{The ALMA Spectroscopic Survey in the HUDF: CO emission lines and 3 mm continuum sources}
\shortauthors{Gonz\'alez-L\'opez et al.}

\begin{document}

\title{The ALMA Spectroscopic Survey in the HUDF: CO emission lines and 3 mm continuum sources}

\correspondingauthor{Jorge Gonz\'alez-L\'opez}
\email{jorge.gonzalezl@mail.udp.cl}

\author[0000-0003-3926-1411]{Jorge Gonz\'alez-L\'opez}
\affil{N\'ucleo de Astronom\'ia de la Facultad de Ingenier\'ia y Ciencias, Universidad Diego Portales, Av. Ej\'ercito Libertador 441, Santiago, Chile}
\affil{Instituto de Astrof\'{\i}sica, Facultad de F\'{\i}sica, Pontificia Universidad Cat\'olica de Chile Av. Vicu\~na Mackenna 4860, 782-0436 Macul, Santiago, Chile}
\nocollaboration

\author[0000-0002-2662-8803]{Roberto Decarli}
\affil{INAF-Osservatorio di Astrofisica e Scienza dello Spazio, via Gobetti 93/3, I-40129, Bologna, Italy}

\author{Ricardo Pavesi}
\affil{Cornell University, 220 Space Sciences Building, Ithaca, NY 14853, USA}

\author[0000-0003-4793-7880]{Fabian Walter}
\affil{Max Planck Institute f\"ur Astronomie, K\"onigstuhl 17, 69117 Heidelberg, Germany}
\affil{National Radio Astronomy Observatory, Pete V. Domenici Array Science Center, P.O. Box O, Socorro, NM 87801, USA}

\author[0000-0002-6290-3198]{Manuel Aravena}
\affil{N\'ucleo de Astronom\'ia de la Facultad de Ingenier\'ia y Ciencias, Universidad Diego Portales, Av. Ej\'ercito Libertador 441, Santiago, Chile}

\author{Chris Carilli}
\affil{National Radio Astronomy Observatory, Pete V. Domenici Array Science Center, P.O. Box O, Socorro, NM 87801, USA}
\affil{Battcock Centre for Experimental Astrophysics, Cavendish Laboratory,
Cambridge CB3 0HE, UK}

\author[0000-0002-3952-8588]{Leindert Boogaard}
\affil{Leiden Observatory, Leiden University, PO Box 9513, NL-2300 RA Leiden, The Netherlands}

\author{Gerg\"{o} Popping}
\affil{Max Planck Institute f\"ur Astronomie, K\"onigstuhl 17, 69117 Heidelberg, Germany}

\author{Axel Weiss}
\affil{Max-Planck-Institut f\"ur Radioastronomie, Auf dem H\"ugel 69, 53121 Bonn, Germany}

\author{Roberto J. Assef}
\affil{N\'ucleo de Astronom\'ia de la Facultad de Ingenier\'ia y Ciencias, Universidad Diego Portales, Av. Ej\'ercito Libertador 441, Santiago, Chile}

\author[0000-0002-8686-8737]{Franz Erik Bauer}
\affil{Instituto de Astrof\'{\i}sica, Facultad de F\'{\i}sica, Pontificia Universidad Cat\'olica de Chile Av. Vicu\~na Mackenna 4860, 782-0436 Macul, Santiago, Chile}
\affil{Millennium Institute of Astrophysics (MAS), Nuncio Monse{\~{n}}or S{\'{o}}tero Sanz 100, Providencia, Santiago, Chile}
\affil{Space Science Institute, 4750 Walnut Street, Suite 205, Boulder, CO 80301, USA}

\author{Frank Bertoldi}
\affil{Argelander-Institut f\"ur Astronomie, Universit\"at Bonn, Auf dem H\"ugel 71, 53121 Bonn, Germany}

\author{Richard Bouwens}
\affil{Leiden Observatory, Leiden University, PO Box 9513, NL-2300 RA Leiden, The Netherlands}

\author{Thierry Contini}
\affil{Institut de Recherche en Astrophysique et Plan\'etologie (IRAP), 
Université de Toulouse, CNRS, UPS, 31400 Toulouse, France}

\author{Paulo C.~Cortes}
\affil{Joint ALMA Observatory - ESO, Av. Alonso de C\'ordova, 3104, Santiago, Chile}
\affil{National Radio Astronomy Observatory, 520 Edgemont Rd, Charlottesville, VA, 22903, USA} 

\author{Pierre Cox}
\affil{Institut d'astrophysique de Paris, Sorbonne Universit\'e, CNRS, UMR 7095, 98 bis bd Arago, 7014 Paris, France}

\author{Elisabete da Cunha}
\affil{Research School of Astronomy and Astrophysics, Australian National University, Canberra, ACT 2611, Australia}

\author{Emanuele Daddi}
\affil{Laboratoire AIM, CEA/DSM-CNRS-Universite Paris Diderot, Irfu/Service d'Astrophysique, CEA Saclay, Orme des Merisiers, 91191 Gif-sur-Yvette cedex, France}

\author{Tanio D\'iaz-Santos}
\affil{N\'ucleo de Astronom\'ia de la Facultad de Ingenier\'ia y Ciencias, Universidad Diego Portales, Av. Ej\'ercito Libertador 441, Santiago, Chile}

\author{Hanae Inami}
\affil{Univ. Lyon 1, ENS de Lyon, CNRS, Centre de Recherche Astrophysique de Lyon (CRAL) UMR5574, 69230 Saint-Genis-Laval, France}

\author{Jacqueline Hodge}
\affil{Leiden Observatory, Leiden University, PO Box 9513, NL-2300 RA Leiden, The Netherlands}

\author{Rob Ivison}
\affil{European Southern Observatory, Karl-Schwarzschild-Strasse 2, 85748, Garching, Germany}
\affil{Institute for Astronomy, University of Edinburgh, Royal Observatory, Blackford Hill, Edinburgh EH9 3HJ}

\author{Olivier Le F\`evre}
\affil{Aix Marseille Universit\'e, CNRS, LAM (Laboratoire d'Astrophysique de Marseille), UMR 7326, F-13388 Marseille, France}

\author{Benjamin Magnelli}
\affil{Argelander-Institut f\"ur Astronomie, Universit\"at Bonn, Auf dem H\"ugel 71, 53121 Bonn, Germany}

\author{Pascal Oesch}
\affil{Department of Astronomy, University of Geneva, Ch. des Maillettes 51, 1290 Versoix, Switzerland}

\author[0000-0001-9585-1462]{Dominik Riechers}
\affil{Cornell University, 220 Space Sciences Building, Ithaca, NY 14853, USA}
\affil{Max Planck Institute f\"ur Astronomie, K\"onigstuhl 17, 69117 Heidelberg, Germany}

\author{Hans--Walter Rix}
\affil{Max Planck Institute f\"ur Astronomie, K\"onigstuhl 17, 69117 Heidelberg, Germany}

\author{Ian Smail}
\affil{Centre for Extragalactic Astronomy, Department of Physics, Durham University, South Road, Durham, DH1 3LE, UK}

\author{A.M. Swinbank}
\affil{Centre for Extragalactic Astronomy, Department of Physics, Durham University, South Road, Durham, DH1 3LE, UK}

\author{Rachel S. Somerville}
\affil{Department of Physics and Astronomy, Rutgers, The State University of New Jersey, 136 Frelinghuysen Rd,
Piscataway, NJ 08854, USA}
\affil{Center for Computational Astrophysics, Flatiron Institute, 162 5th Ave, New York, NY 10010, USA}

\author{Bade Uzgil}
\affil{National Radio Astronomy Observatory, Pete V. Domenici Array Science Center, P.O. Box O, Socorro, NM 87801, USA}
\affil{Max Planck Institute f\"ur Astronomie, K\"onigstuhl 17, 69117 Heidelberg, Germany}

\author{Paul van der Werf}
\affil{Leiden Observatory, Leiden University, PO Box 9513, NL-2300 RA Leiden, The Netherlands}

\begin{abstract}

The ALMA SPECtroscopic Survey in the {\it Hubble} Ultra Deep Field is an ALMA large program that obtained a frequency scan in the 3\,mm band to detect emission lines from the molecular gas in distant galaxies. We here present our search strategy for emission lines and continuum sources in the HUDF. We compare several line search algorithms used in the literature, and critically account for the line-widths of the emission line candidates when assessing significance. We identify sixteen emission lines at high fidelity in our search. Comparing these sources to multi-wavelength data we find that all sources have optical/infrared counterparts. Our search also recovers candidates that have lower significance that can be used statistically to derive, e.g. the CO luminosity function. We apply the same detection algorithm to obtain a sample of six 3 mm continuum sources. All of these are also detected in the 1.2 mm continuum with optical/near-infrared counterparts. We use the continuum sources to compute 3 mm number counts in the sub-mJy regime, and find them to be higher by an order of magnitude than expected for synchrotron-dominated sources. However, the number counts are consistent with those derived at shorter wavelengths (0.85--1.3\,mm) once extrapolating to 3\,mm with a dust emissivity index of $\beta=1.5$, dust temperature of 35\,K and an average redshift of $z=2.5$. These results represent the best constraints to date on the faint end of the 3 mm number counts.

\end{abstract}


%%%%%%%%%%%%%%%%%%%%%%%%%%%%%%%%%%%%%%%%%%%%%%%%%%%%%%%%%%%%%%%%%%%%%%%%%%%%%%%%%%%%%%%%%%%%%%
%%%%%%%%%%%%%%%%%%%%%%%%%%%%%%%%%%%%%%%%%%%%%%%%%%%%%%%%%%%%%%%%%%%%%%%%%%%%%%%%%%%%%%%%%%%%%%
%%%%%%%%%%%%%%%%%%%%%%%%%%%%%%%%%%%%%%%%%%%%%%%%%%%%%%%%%%%%%%%%%%%%%%%%%%%%%%%%%%%%%%%%%%%%%%
\section{Introduction} \label{sec:intro}

One of the key goals in galaxy evolution studies is to obtain a detailed understanding of the origin of the cosmic star-formation history. We know that star--formation activity started during the so--called ``dark ages'' of the early universe and is, at least partially, responsible for the reionization of the Universe. Since the formation of the first stars, the cosmic star--formation density increased with cosmic time, peaking at $z\sim2$ and exponentially declining afterwards by a factor $\sim$8 to z\,=\,0. Several physical processes can shape the cosmic star-formation density, such as changes in star-formation efficiencies (e.g., galaxy merger rates and feedback) and changes in the available fuel for star-formation with time \citep[see review by][]{MadauDickinson2014}. This is a main motivation to map out the cosmic gas density  as a function of lookback time. As the rotational transitions of carbon monoxide (CO) are found to be reliable tracers of the molecular gas content in local galaxies (e.g. \citet{Bolatto2013}, these lines are also used in systems at high redshift (e.g. \citet{Carilli_Walter2013}). Historically, searches for molecular gas emission in high-redshift galaxies were restricted to single galaxies at a time. However, the increase in sensitivity and bandwidth of (sub-)millimeter facilities now enables studies of multiple sources at different redshifts in significant cosmic volumes.

In particular the spectral scan method (i.e. search for emission lines in a wide range of frequencies) has been demonstrated to be a unique tool to study the evolution of the molecular gas and dust emission in galaxies throughout cosmic time \citep{Walter2012,Walter2014,Decarli2014}. Molecular lines scan allow for the un-biased characterization of the molecular gas distribution within a volume--limited sample, since no pre--selection of galaxies is employed \citep{Carilli_Walter2013}. Under the same rationale, such observations have focused on legacy survey fields where deep and abundant ancillary data can be used to find and characterize the counterpart galaxy of any potential detection. The first fields targeted for molecular line scans were the {\it Hubble} Deep Field North \citep[HDF-N,][]{Williams1996,Decarli2014} and the {\it Hubble} Ultra-Deep Field \citep[UDF,][]{Beckwith2006,Walter2016}. 

\citet{Walter2016} presented the rationale and observational description of ASPECS: The ALMA SPECtroscopic Survey in the UDF (hereafter ASPECS-Pilot), which covered a $1\arcmin$ region within the UDF with full frequency scans over the bands 3 (84--115 GHz) and 6 (212--272 GHz) of ALMA. ASPECS-Pilot's spectroscopic coverage was designed to detect several CO rotational transitions in an almost continuous redshift window between $z=0-8$ and the ionized carbon \cii emission line at $z=6-8$. 
The main results of the ASPECS-Pilot are described in a series of papers, including detections of emission--line galaxies \citep{Aravena2016a,Aravena2016b,Decarli2016a,Decarli2016b}.

ASPECS-Pilot, and other line scans using ALMA and the Jansky Very Large Array (JVLA) \citep{Lentati2015,Matsuda2015,Kohno2016,GL2017c,Pavesi2018,Riechers2019}, have motivated further development of line search codes and algorithms to better understand the limitation and caveats of such type of surveys. The search and characterization of sources in 3D spectral line data cubes is also a topic of interest for \hi (neutral hydrogen) observations. Several codes and implementations have been developed to find and follow the complex structures of \hi and other lines in interferometric data \citep{Whiting2012,Serra2014,Loomis2018}. These codes focus mainly on the detectability of the spatially and spectrally extended lines, a different problem from the simpler detections expected in molecular line scans. In fact, the algorithms that more closely relate to the search problem have been developed for searches of emission lines in integral-field spectroscopy data cubes such as those obtained by MUSE \citep{HerenzWisotzki2017}. 

One of the consequences of observing line scans is that we also obtain a deep continuum image from collapsing the spectral axis, which allows for the detection of faint continuum sources over a large contiguous area. Such continuum observations have been used to constrain the number counts of sources over different ranges of flux density values as well as to obtain the properties of the faint population detected individually or by stacking analysis. In the ALMA era, such observations have focused mainly on the bands 6 and 7 observations ($850\mu m$ to 1.3 mm) since they offer the best combination of dust emission detectability and area coverage \citep{Lindner2011,Scott2012,Karim2013,Hatsukade2013,Ono2014,Simpson2015,Carniani2015,Oteo2016,Hatsukade2016,Aravena2016a,Fujimoto2016,MunozArancibia2017,Umehata2017,Geach2017,Dunlop2017,Franco2018}. 

New deep continuum observations at longer wavelengths ($>2$ mm) have been suggested to better constrain the population of dusty star-forming galaxies (DSFGs) at high redshift \citep{Bethermin2015,Casey2018a,Casey2018}. In fact, \citet{Aravena2016a} already presented the 3 mm deep continuum images obtained as part of the ASPECS-Pilot campaign, resulting in only one secure continuum detection. The 3 mm band offers a unique view compared to shorter wavelengths observations, since, at least in local galaxies, the bands sample potential emission from thermal dust, thermal bremsstrahlung (free-free emission), and non-thermal synchrotron emission \citep{Klein1988,CarlstromKronberg1991,Condon1992,YunCarilli2002}. Identifying and quantifying what emission dominates the DSFG population over different flux density ranges is crucial for understanding high redshift galaxies and their expected evolution \citep{Bethermin2011,Bethermin2012,Cai2013}.
Given the similarities of the search of continuum sources and emission lines, any advancement in the understanding of completeness and fidelity of line searches can be applied to source detection and number counts computation for both samples. 

Here we present the emission line and continuum detections in the band 3 (3 mm) data from the ASPECS Large Program (hereafter ASPECS-LP), an ALMA large program that expands the legacy of ASPECS-Pilot to over a $4.6\arcmin$ region within the UDF, i.e. five times larger than the ASPECS-Pilot coverage (Fig. \ref{fig:footprint}), using the same frequency setup and with similar sensitivity. A general presentation of the ASPECS-LP is given in a companion publication by \citet{Decarli2019}. ASPECS-LP offers the unique opportunity to test and compare line search algorithms by allowing the confirmation of line candidates with alternative methods. Targeting the HUDF, ASPECS-LP offers the opportunity to identify emission--line candidates using thousands of optical and near infrared (NIR) spectroscopic redshifts obtained in the field, therefore allowing for the independent confirmation of a greater number of emission--line candidates as well as securing lower significance detections \citep{Boogaard2019}. 

In this paper, we present the results from different algorithms and techniques to obtain reliable emission--line and continuum candidate samples in ASPECS-LP. Throughout this paper the quoted errors correspond to the inner $68\%$ confidence levels unless stated otherwise. We discuss the multi-wavelength properties of our most secure detections in the companion papers by \citet{Aravena2019} and \citet{Boogaard2019}. Implications for the CO luminosity function and the resulting cosmic evolution of the molecular gas density are discussed in \citet{Decarli2019} and \citet{Popping2019}.

%%%%%%%%%%%%%%%%%%%%%%%%%%%%%%%%%%%%%%%%%%%%%%%%%%%%%%%%%%%%%%%%%%%%%%%%%%%%%%%%%%%%%%%%%%%%%%
%%%%%%%%%%%%%%%%%%%%%%%%%%%%%%%%%%%%%%%%%%%%%%%%%%%%%%%%%%%%%%%%%%%%%%%%%%%%%%%%%%%%%%%%%%%%%%
%%%%%%%%%%%%%%%%%%%%%%%%%%%%%%%%%%%%%%%%%%%%%%%%%%%%%%%%%%%%%%%%%%%%%%%%%%%%%%%%%%%%%%%%%%%%%%
\section{Observations and data processing} \label{sec:observation}
\subsection{Survey design}

\begin{figure}
\epsscale{1.1}
\plotone{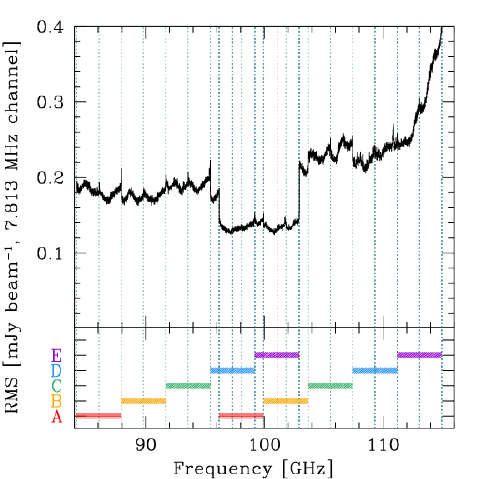}
\caption{Sensitivity across the observed frequencies for the ASPECS-LP band 3 data cube. The r.m.s. values are measured on 7.813 MHz width channels. The different spectral setups are plotted with different colors in the bottom panel. The spectral configuration lead to the central frequencies (96  to 103 GHz) being observed twice, resulting in a slightly lower r.m.s. values. \label{fig:RMS}}
\end{figure}

The data used in this work correspond to the band 3 observations from ASPECS-Pilot and ASPECS-LP. The details about the ASPECS-Pilot observations were presented in \citet{Walter2016} while the details about the ASPECS-LP are presented below and in \citet{Decarli2019}. 
The spectral setup of the ASPECS-LP is the same as the one used in the ASPECS-Pilot, with five tunings that cover most of the ALMA band 3. An overlap between the tunings meant that some channels in the range 96 to 103 GHz were observed twice, yielding lower noise levels in this frequency range (Fig. \ref{fig:RMS}). The higher r.m.s. values towards the higher frequencies are due to the lower atmospheric transmission at those frequencies. The ASPECS-LP observations are similar in sensitivity to the ASPECS-Pilot observations towards the low frequency range of band 3 while the sensitivity is slightly worse at the higher frequencies. We chose to work with both datasets separately because the combination injected artifacts in the data cubes while adding a modest increase in sensitivity in a small region of the map.

\begin{figure*}
\epsscale{1.1}
\plotone{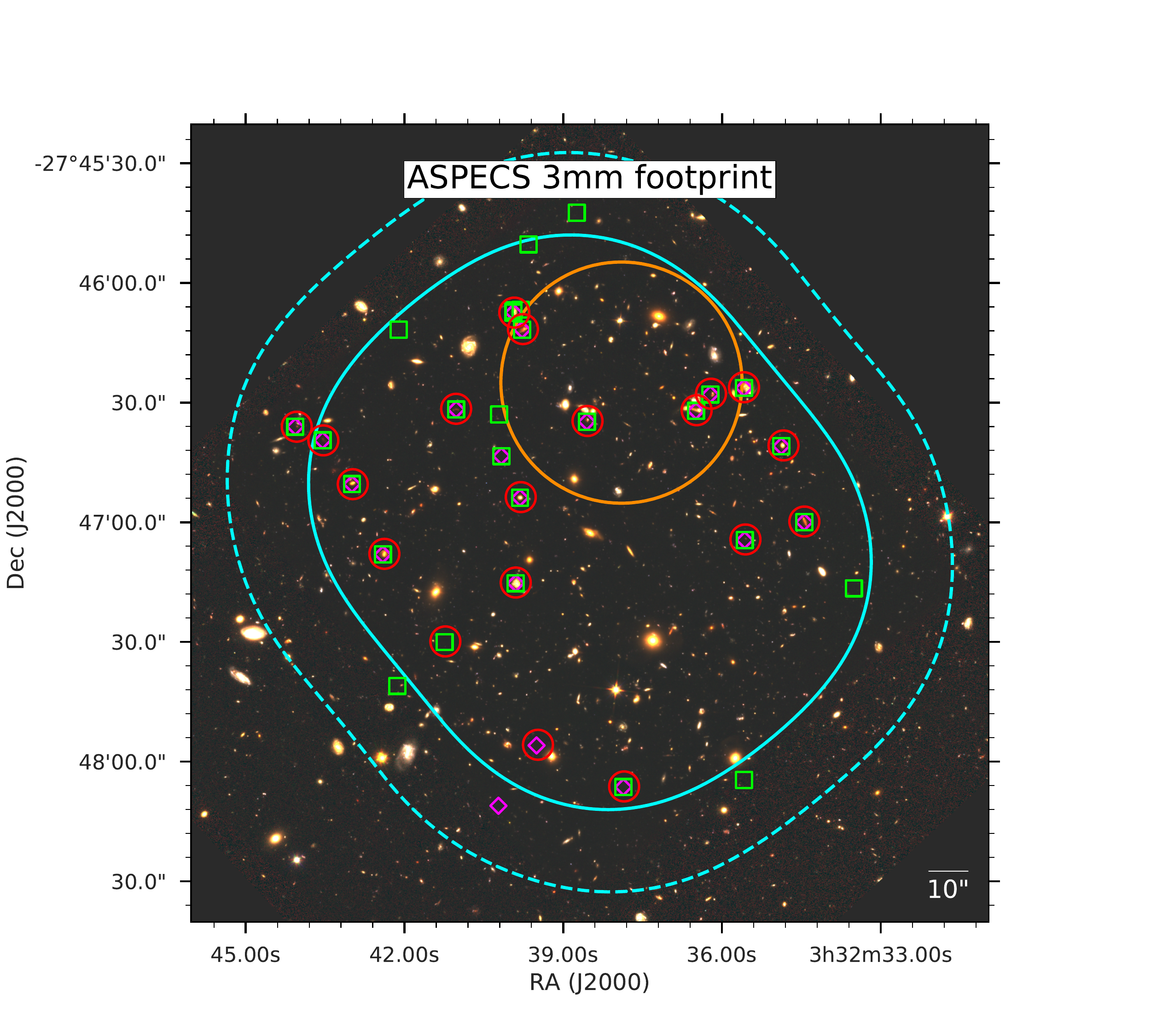}
\caption{Footprint of the band 3 ASPECS-LP observations. The orange solid line corresponds to size of the band 3 ASPECS-Pilot coverage primary beam. The solid cyan line encircles the area where the combined primary beam correction of the band 3 ASPECS-LP mosaic is $\geq$0.5. The dashed cyan line marks the region where the search for emission lines is done, where the combined primary beam correction is $\geq$0.2. 
The line candidates detected with ${\rm S/N}\geq6$ by LineSeeker are shown as red circles, the magenta diamonds correspond to MF3D detections and green squares to FindClump. The color image was created using a combination of {\it Hubble} space telescope images.\label{fig:footprint}}
\end{figure*}

The ASPECS-LP, which covers a five times larger area than ASPECS-Pilot (Fig. \ref{fig:footprint}), was mapped with 17 Nyquist-spaced pointings in band 3 cover an area of 4.6 arcmin$^2$ (within mosaic primary beam correction $\geq0.5$, see details in \citet{Decarli2019}). A small portion of the ASPECS-Pilot coverage is not covered by the ASPECS-LP. The ASPECS-LP observations in band 3 were made in antenna configuration C40-3. This antenna configuration should return a synthesized beam similar in size to the one that will be obtained for the ASPECS-LP band 6 observations. In both cases, the beam size was chosen to be as sensitive as possible to any extended emission.

\subsection{Data reduction, calibration and imaging}

The data was processed using both the {\sc CASA} ALMA calibration pipeline \citep[v. 4.7.0;][]{McMullin2007} and our own scripts \citep[see, e.g.,][]{Aravena2016a}, which follow a similar scheme to the ALMA manual calibration scripts.
Our independent inspection for data to be flagged allowed us to improve the depth of our scan in one of the frequency settings by up to 20\%. In all the other frequency settings, the final rms appears consistent with the one computed from the cube provided by the ALMA pipeline. As the cube created with our own procedures is at least as good (in terms of low noise) as the one from the pipeline, we will refer to the former in the remainder of the analysis.

We imaged the 3 mm cube with natural weighting using the task {\sc tclean}, resulting in synthesized beam sizes between $1.5\arcsec\times1.31\arcsec$ at the high frequencies and $2.05\arcsec\times1.68\arcsec$ at low frequencies. It is important to stress that the final cube does not have a common synthesized beam for all frequencies. Instead, a specific synthesized beam is obtained for each channel and that information is used in the simulation and analysis of the emission line search. 
The lack of very bright sources in our cubes allows us to perform our analysis on the `dirty' cube, thus preserving the intrinsic properties of the noise. 

The observations were taken using Frequency Division Mode (FDM) with a coverage per spectral window of 1875 MHz and original channel spacing of 0.488 MHz. Spectral averaging during the correlation and in the image processing result in a final channel resolution of 7.813 MHz, which corresponds to $\Delta v\approx23.5\kms$ at 99.5 GHz. This rebinning process (using `nearest' interpolation scheme) helps to mitigate any correlation among channels introduced by the Hanning weighting function applied to the original channels within the correlator. The Doppler tracking correction applied to the sky frequencies during the observations or during the off-line calibration should be smaller than the final channel resolutions. Because of this, we expect the spectral channels of the final cube to be fairly independent and no line broadening is expected as a product of the observations. The final spectral resolution is high enough to resolve in velocity emission lines with ${\rm FWHM}\approx40-50\kms$. Narrower lines can still be detected but their line width will not be well constrained. 

Finally, the continuum image was created by collapsing all the spectral windows and channels. The deepest region of the continuum image has an rms value of $3.8\ts{\rm \mu Jy\ts beam^{-1}}$ with a beam size of $2.08\arcsec\times1.71\arcsec$.

\begin{figure*}
\epsscale{1.2}
\plotone{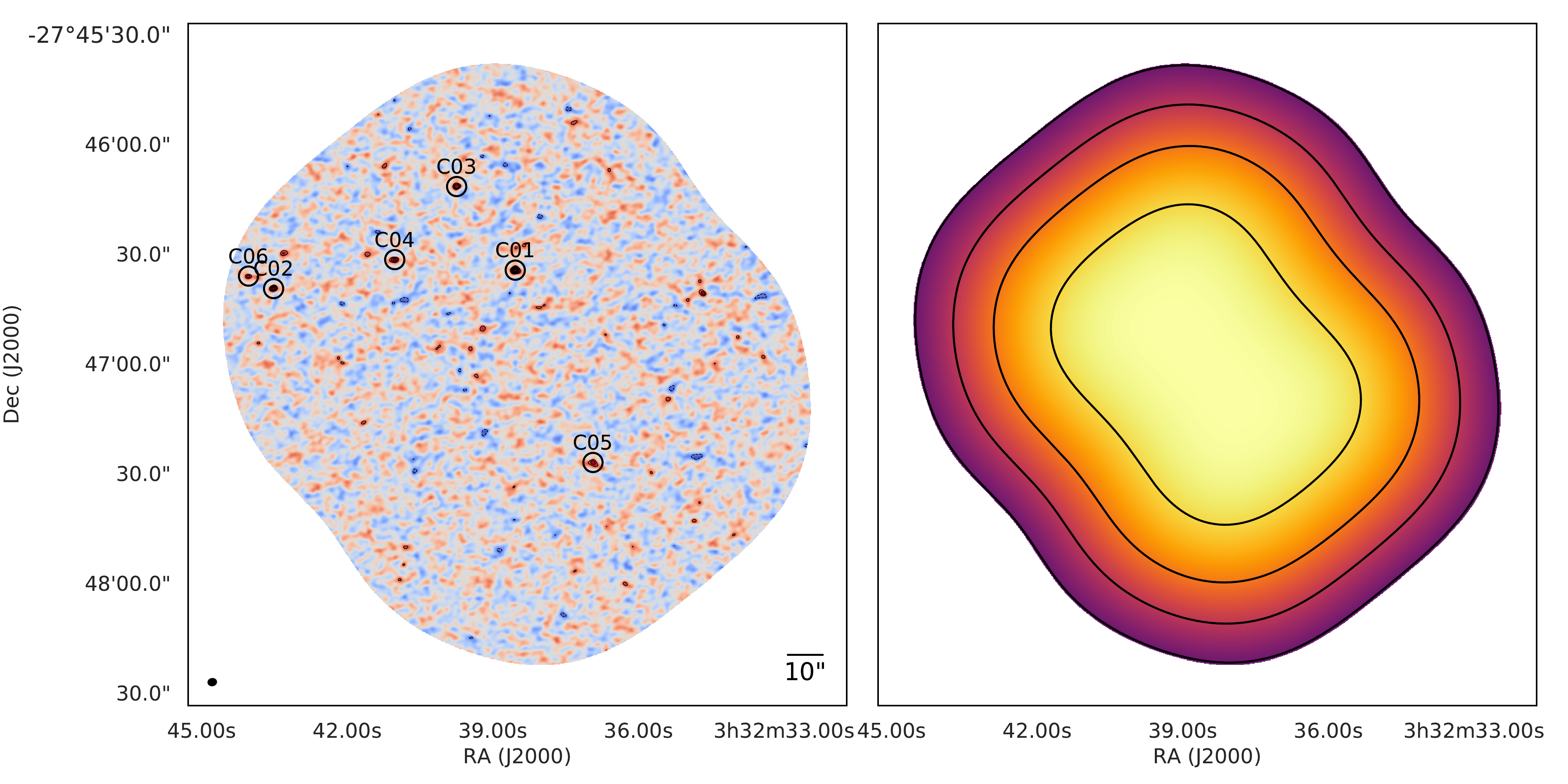}
\caption{The left panel shows the continuum image without mosaic primary beam correction (in color scale) obtained from the 3 mm ASPECS-LP observations. Black contours show the S/N levels starting with $\pm3\sigma$ with $\sigma=3.8{\rm \mu Jy \ts beam^{-1}}$. Black circles show the positions and IDs of the source candidates found to be significant ($\rm S/N\geq4.6$). The right panel shows the primary beam response of the continuum image mosaic. The contours show where the primary beam response is 0.3, 0.5, 0.7 and 0.9, respectively. \label{fig:ContinuumMapRMS}}
\end{figure*}

\begin{deluxetable}{cccc}
\tablecaption{Emission lines rest-frequency and corresponding redshift ranges for the band 3 line scan (84.176--114.928 GHz).\label{tab:LineRanges}}
\tablehead{
\colhead{Transition} & 
\colhead{$\nu_{0}$}& 
\colhead{$z_{\rm min}$}& 
\colhead{$z_{\rm max}$} \\
\colhead{} & 
\colhead{[GHz]}& 
\colhead{}& 
\colhead{}
} 
\colnumbers
\startdata
CO(1-0) & 115.271 & 0.0030 & 0.3694 \\
CO(2-1) & 230.538 & 1.0059 & 1.7387 \\
CO(3-2) & 345.796 & 2.0088 & 3.1080 \\
CO(4-3) & 461.041 & 3.0115 & 4.4771 \\
CO(5-4) & 576.268 & 4.0142 & 5.8460 \\
CO(6-5) & 691.473 & 5.0166 & 7.2146 \\
CO(7-6) & 806.652 & 6.0188 & 8.5829 \\
$[\ci]_{1-0}$ & 492.161 & 3.2823 & 4.8468 \\
$[\ci]_{2-1}$ & 809.342 & 6.0422 & 8.6148 \\
\enddata
\end{deluxetable}

%%%%%%%%%%%%%%%%%%%%%%%%%%%%%%%%%%%%%%%%%%%%%%%%%%%%%%%%%%%%%%%%%%%%%%%%%%%%%%%%%%%%%%%%%%%%%%
%%%%%%%%%%%%%%%%%%%%%%%%%%%%%%%%%%%%%%%%%%%%%%%%%%%%%%%%%%%%%%%%%%%%%%%%%%%%%%%%%%%%%%%%%%%%%%
%%%%%%%%%%%%%%%%%%%%%%%%%%%%%%%%%%%%%%%%%%%%%%%%%%%%%%%%%%%%%%%%%%%%%%%%%%%%%%%%%%%%%%%%%%%%%%

\section{Emission line Search} \label{sec:BlindSearch}

\subsection{Methods} \label{sec:method}

ASPECS-LP covers a frequency range where we expect to detect CO and/or other emission lines from many moderate to high redshift galaxies. Without any priori knowledge of positions and frequencies for the lines, we need to use some unbiased method to search for emission lines. We employ and compare three independent methods to search for emission lines in the data cubes, namely: LineSeeker, FindClump and MF3D. 
The three methods all rely on match filtering, wherein their combine different spectral channels and measuring the signal-to-noise ratio (S/N) in the resultant image, with the combination of channels is motivated by the shape and width of actual emission lines. The three methods implement different algorithms for the spectral channels combination and for how the high significance peaks are selected. Here we present a description of the three methods used.

\subsubsection{LineSeeker}
LineSeeker is the method used in the search for emission lines in the ALMA Frontier Fields survey \citep{GL2017c}. This method combines spectral channels using Gaussian kernels of different spectral widths. Each Gaussian kernel is controlled by the $\sigma_{{\rm GK}}$ parameter, which ranges from 0 up to 19. The Gaussian kernel generated with $\sigma_{{\rm GK}}=0$ is better suited for detecting single channel features while $\sigma_{{\rm GK}}=19$ is in the optimal range for detecting emission lines of ${\rm FWHM}\approx900-1200\kms$ based in the $\approx20\kms$ channel resolution of the ASPECS-LP 3 mm cube. 
The combination of the channels is done by convolving the cube in the spectral axis with the Gaussian kernel of the corresponding $\sigma_{{\rm GK}}$, with the search for high significance features is done on a channel by channel basis. The initial noise level per channel is estimated by taking the standard deviation of all the corresponding voxels. The noise level is then refined by repeating the calculation using only the voxels with absolute values lower than five times the initial noise estimate. This is done to mitigate the effects of bright emission lines artificially increasing the noise level in the corresponding channels. All the voxels above a given S/N ratio, calculated as the measured flux density in the collapsed channels divided by the refined noise value, are stored for each of the convolutions kernels. The final line candidates list is obtained by grouping the different voxels from the different channels using the Density-Based Spatial Clustering of Applications with Noise (DBSCAN) algorithm \citep{Ester1996} available in the {P}ython package Scikit-learn \citep{scikit-learn}. The S/N assigned to each emission line candidate is selected as the maximum value obtained from all the different convolutions. 

\subsubsection{FindClump}

FindClump is the method used in the molecular line scan of the HDF-N and the ASPECS-Pilot \citep{Decarli2014,Walter2016}. FindClump uses a top-hat convolution of the data cubes in the spectral axis. In each top-hat convolution, FindClump uses N number of channels on each side of the target channel to do the convolution, with N ranging from 1 up to 9. In the first convolution, FindClump combines the information from 3 consecutive channels while in the last convolution it uses the information of 19 consecutive channels. In the same manner as LineSeeker, FindClump searches for high significance features in channel by channel basis using SExtractor \citep{Bertin1996}. The sources found by SExtractor are grouped by selecting all the sources that fall within 2\arcsec and 0.1 GHz. 

\subsubsection{MF3D}

MF3D corresponds to the method used to search for emission lines in the JVLA CO luminosity density at high-z (COLDZ) survey \citep{Pavesi2018}. MF3D is similar to LineSeeker in the sense that it uses Gaussian kernels for the spectral axis convolutions with the caveat that it also implements convolutions in the spatial axis to look for spatially resolved emission lines. The spatial convolutions use 2D circular Gaussian kernels with FWHM of 0, 1, 2, 3 and 4 \arcsec as kernels. The extra dimensionality explored by MF3D means that it contains LineSeeker when the spatial kernel uses ${\rm FWHM}=0\arcsec$. 

\subsection{Comparison between methods}\label{Sec:ComparisonMethods}
\subsubsection{Lines detected from Simulations}

\begin{figure}
\epsscale{1.2}
\plotone{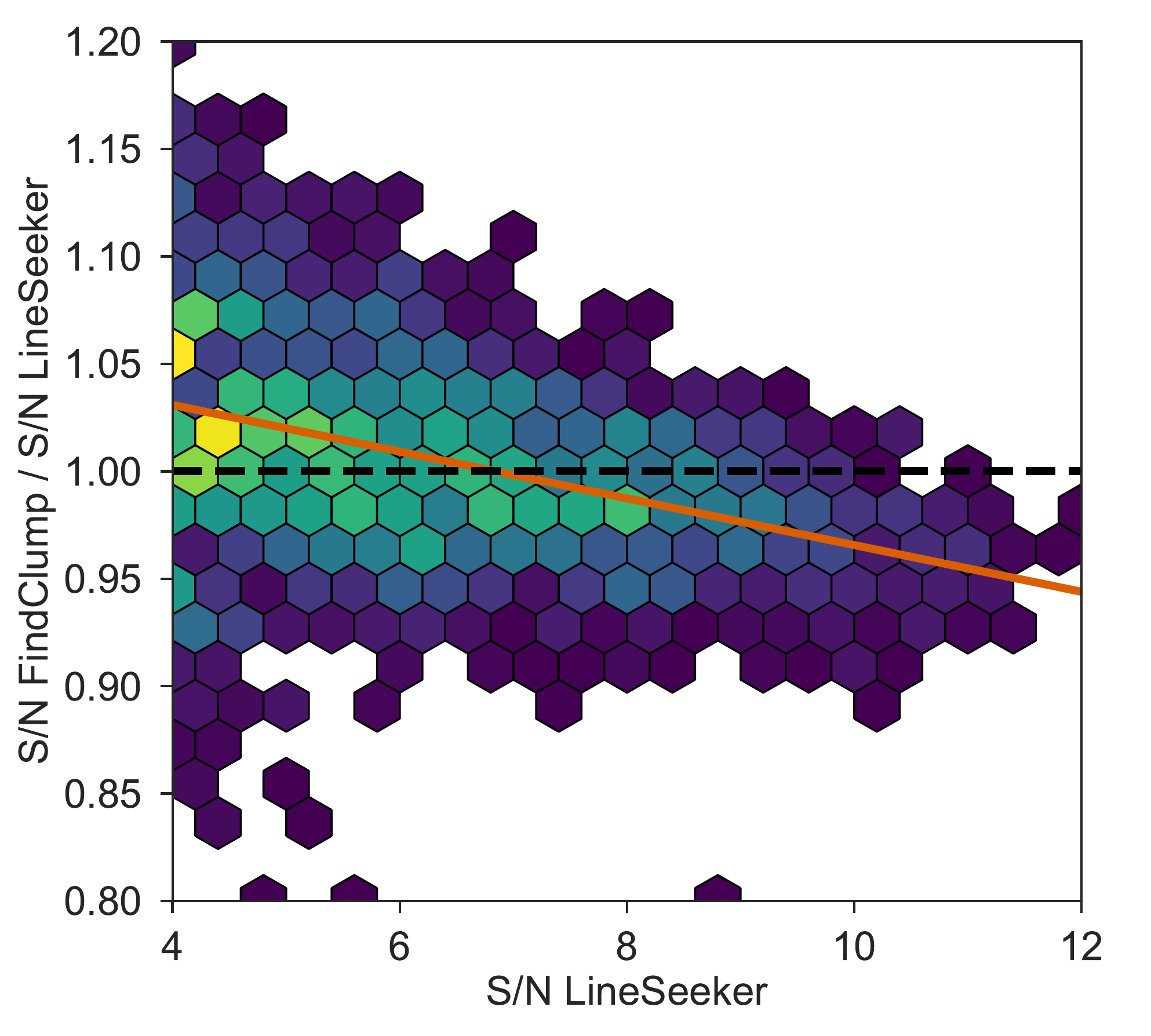}
\caption{Density map of the ratio between the S/N values obtained with FindClump and LineSeeker of an artificial/simulated data cube. The color of the cells follow a logarithm scale proportional to the number of points in each cell. The solid orange line corresponds to a linear fit to the points with a $7\%$ slope showing that the S/N values obtained with FindClump are slightly higher than the ones obtained with LineSeeker for the fainter sources. \label{fig:SnvsSN}}
\end{figure}

In this section, we use simulated data cubes to compare the results from the different methods. We chose to use simulated data cubes since they represent an ideal case of well behaved data. The data cubes were created using the {\sc CASA} task {\sc simobserve}  with a similar setup to the ASPECS-LP band 3 observations. The central frequency of the simulated cube was set to 100 GHz with 100 channels of width 7.813 MHz using the antenna configuration C40-3. To image the simulation, we used {\sc tclean}  with natural weighting and made images out to a primary beam correction of 0.2. 
The initial data cubes contain pure noise data without real emission. Simulated emission lines with Gaussian profiles are then injected to the cubes to be recovered by the searching methods. The injected lines were distributed homogeneously across the spatial and spectral axes. For simplicity, all the injected lines were simulated as point sources using the synthesized beam from the data cube. The emission line peaks were uniformly distributed between 1 and 3 times the median r.m.s. values across all channels ($\approx0.53\,\,{\rm mJy/beam}$) while the full width half maximum (FWHM) of the lines were uniformly distributed between 0 (technically the width of a single channel) and 500\kms. The number of lines injected per cube was limited to only ten per iteration to lower the chances of blending of two nearby emission lines. In each simulation iteration, the codes for LineSeeker and FindClump were run to obtain emission line candidates and their corresponding S/N ratios. The simulations were repeated until 5000 simulated lines were produced. 

Both methods performed similarly in the detection of the injected emission lines, recovering around $\approx99\%$ of the lines. The unrecovered lines correspond to the faint end of the injected lines, and mainly narrow lines (${\rm FWHM}\lesssim100\kms$) with low S/N ratios (${\rm S/N}\lesssim4$). The distribution of parameters for the lines not recovered by either method is very similar. For the recovered lines, the ratio between the S/N obtained with FindClump and  LineSeeker is of $1.00^{+0.06}_{-0.04}$. 
In Fig. \ref{fig:SnvsSN} we present the density map of the ratio between the S/N values obtained with FindClump and LineSeeker for all detected lines. All points lie close to unity within the scatter, showing an agreement between the results from FindClump and LineSeeker. We do note a second order trend whereby detections with lower S/N values tend to have slightly higher S/N values in FindClump than in LineSeeker. A linear fit to the full set of matched detections returns a slope of $7\%$ across the range of S/N values plotted. 
This effect can be explained by the different convolution kernels used by the different methods. We tested LineSeeker using a top-hat convolution kernel and obtained higher S/N values for less Gaussian-like emission lines, specially in the faint end. On the opposite case, a Gaussian convolution kernel will return higher S/N values for very bright simulated Gaussian emission lines, since the kernel manages to better recover the profile of the lines. 
One might infer from these results that a top-hat kernel is preferred over a Gaussian kernel. However, the same effect also boosts the S/N values of false lines, meaning that the higher S/N obtained with the top-hat convolution does not necessarily translate to a higher significance. 
While we notice the existence of the trend, this is well within the scatter of the distribution, therefore we conclude that the S/N values obtained from FindClump and LineSeeker are  equivalent. As we will see below, some of the bright emission lines show double-horn profiles typical of massive flat rotating disks \citep{Walter2008}. Fainter emission lines observed in less massive galaxies tend to be better described with single Gaussian functions, indication of gas dominated by turbulent motions instead of rotation. The faint emission lines we expect to detect could be associated to low mass high-redshift galaxies, for which a Gaussian profile should be a good choice for the kernel convolution. Furthermore, using Gaussian or tophat kernels for the convolution show very small differences when selecting by significance.
For detection purposes, using a more complex double-horn profile for the convolution is not efficient since more parameters need to be explored to sample all possible shapes. Gaussian and tophat functions offer a good description of the emission lines by only their peak and width.

MF3D was not included in the simulations because of the similarities between MF3D and LineSeeker and the point source nature of the injected emission lines. To confirm the equivalence between MF3D and LineSeeker for point source-like emission lines, we ran both codes in one of the simulated cubes, obtaining a ratio between the S/N values of $1.00^{+0.03}_{-0.04}$. The median ratio is consistent with unity while showing a slightly smaller scatter than the FindClump-LineSeeker comparison scenario. To conclude the three codes return effectively equivalent S/N values. 

\subsubsection{Lines detected from observations}

\begin{figure}
\epsscale{1.2}
\plotone{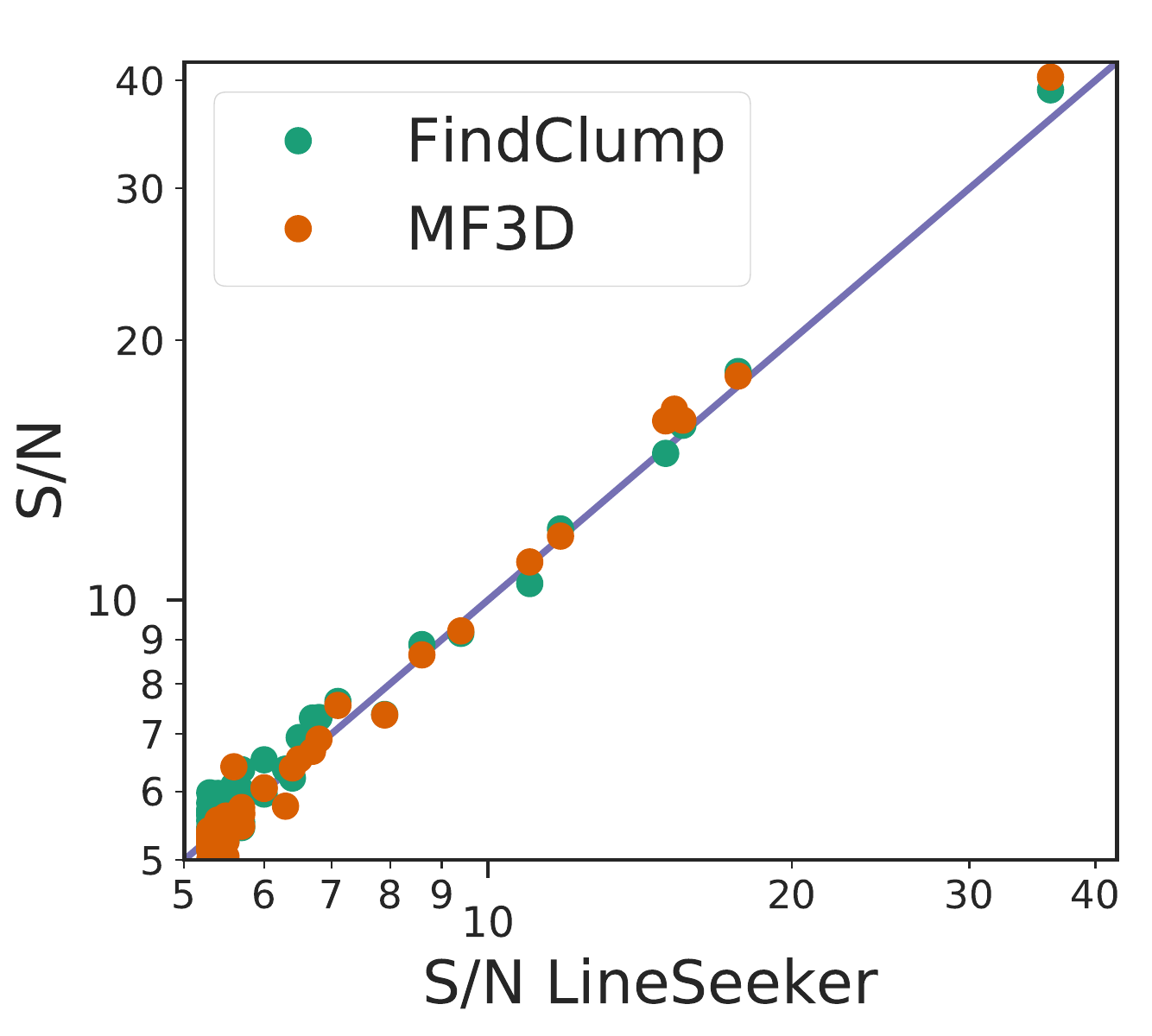}
\caption{Comparison of the S/N values obtained with LineSeeker, FindClump and MF3D for the ASPECS-LP emission line candidates. The green points show the candidates recovered by FindClump while the orange points by MF3D down to a ${\rm S/N}=5.3$ in LineSeeker. The three methods return similar S/N values for the bright end. At the low S/N end LineSeeker agrees better with MF3D than with FindClump, confirming the findings from the simulations. \label{fig:Comparison_SN}}
\end{figure}

\begin{deluxetable*}{ccccccccc}
\tablecaption{${\rm S/N}\geq6$ emission line candidates found by the three line search methods in ASPECS-LP band 3 cube. \label{tab:LP_LinesCandidates_3methods}}
\tablehead{
\colhead{ID} & 
\colhead{ID} & 
\colhead{ID} & 
\colhead{R.A.} & 
\colhead{Dec} & 
\colhead{Freq.} & 
\colhead{S/N} & 
\colhead{S/N} & 
\colhead{S/N}  \\
\colhead{LineSeeker} & 
\colhead{MF3D} & 
\colhead{FindClump} & 
\colhead{} & 
\colhead{} & 
\colhead{[GHz]} & 
\colhead{LineSeeker} & 
\colhead{MF3D} & 
\colhead{FindClump} 
}
\colnumbers
\startdata
LineSeeker.01  &  MF3D.01  &  FindClump.01  &  03:32:38.541  &  -27:46:34.620  &  97.58  &  37.7  &  40.4  &  39.0 \\
LineSeeker.02  &  MF3D.02  &  FindClump.02  &  03:32:42.379  &  -27:47:07.917  &  99.51  &  17.9  &  18.2  &  18.4 \\
LineSeeker.03  &  MF3D.04  &  FindClump.04  &  03:32:41.023  &  -27:46:31.559  &  100.135  &  15.8  &  16.2  &  15.9 \\
LineSeeker.04  &  MF3D.03  &  FindClump.03  &  03:32:34.444  &  -27:46:59.816  &  95.502  &  15.5  &  16.6  &  16.4 \\
LineSeeker.05  &  MF3D.05  &  FindClump.05  &  03:32:39.761  &  -27:46:11.580  &  90.4  &  15.0  &  16.1  &  14.8 \\
LineSeeker.06  &  MF3D.06  &  FindClump.06  &  03:32:39.897  &  -27:47:15.120  &  110.026  &  11.9  &  11.9  &  12.1 \\
LineSeeker.07  &  MF3D.07  &  FindClump.07  &  03:32:43.532  &  -27:46:39.474  &  93.548  &  10.9  &  11.1  &  10.4 \\
LineSeeker.08  &  MF3D.09  &  FindClump.10  &  03:32:35.584  &  -27:46:26.158  &  96.775  &  9.5  &  9.2  &  9.2 \\
LineSeeker.09  &  MF3D.08  &  FindClump.09  &  03:32:44.034  &  -27:46:36.053  &  93.517  &  9.3  &  9.3  &  9.6 \\
LineSeeker.10  &  MF3D.10  &  FindClump.11  &  03:32:42.976  &  -27:46:50.455  &  113.199  &  8.7  &  8.6  &  8.9 \\
LineSeeker.11  &  MF3D.12  &  FindClump.13  &  03:32:39.802  &  -27:46:53.700  &  109.972  &  7.9  &  7.4  &  7.4 \\
LineSeeker.12  &  MF3D.11  &  FindClump.12  &  03:32:36.208  &  -27:46:27.779  &  96.76  &  7.0  &  7.5  &  7.6 \\
LineSeeker.13  &  MF3D.13  &  FindClump.14  &  03:32:35.557  &  -27:47:04.318  &  100.213  &  6.8  &  6.9  &  7.3 \\
LineSeeker.14  &  MF3D.14  &  FindClump.15  &  03:32:34.838  &  -27:46:40.737  &  109.886  &  6.7  &  6.7  &  7.3 \\
LineSeeker.15  &  MF3D.15  &  FindClump.16  &  03:32:36.479  &  -27:46:31.919  &  109.964  &  6.5  &  6.5  &  6.9 \\
LineSeeker.16  &  MF3D.17  &  FindClump.21  &  03:32:39.924  &  -27:46:07.440  &  100.502  &  6.4  &  6.4  &  6.2 \\
LineSeeker.17  &  MF3D.24  &  FindClump.18  &  03:32:41.227  &  -27:47:29.878  &  85.094  &  6.3  &  5.8  &  6.4 \\
LineSeeker.18  &  MF3D.20  &  FindClump.30  &  03:32:39.477  &  -27:47:55.800  &  109.644  &  6.1  &  6.0  &  6.0 \\
LineSeeker.19  &  MF3D.19  &  FindClump.17  &  03:32:37.849  &  -27:48:06.240  &  111.066  &  6.1  &  6.1  &  6.5 \\
LineSeeker.28  &  MF3D.16  &  FindClump.23  &  03:32:40.17  &  -27:46:43.4  &  84.7741  &  5.7  &  6.4  &  6.1 \\
LineSeeker.2802  &  MF3D.18  &  \nodata  &  03:32:40.22  &  -27:48:11.1  &  86.4462  &  4.6  &  6.2  &  \nodata \\
LineSeeker.20  &  MF3D.21  &  FindClump.19  &  03:32:38.74  &  -27:45:42.4  &  109.6201  &  5.9  &  5.9  &  6.4 \\
LineSeeker.32  &  MF3D.29  &  FindClump.27  &  03:32:40.21  &  -27:46:33.0  &  86.8681  &  5.6  &  5.7  &  6.0 \\
LineSeeker.21  &  MF3D.30  &  FindClump.20  &  03:32:33.51  &  -27:47:16.6  &  110.4327  &  5.7  &  5.6  &  6.4 \\
LineSeeker.25  &  MF3D.32  &  FindClump.22  &  03:32:35.58  &  -27:48:04.6  &  100.6974  &  5.7  &  5.6  &  6.2 \\
LineSeeker.26  &  MF3D.34  &  FindClump.25  &  03:32:42.14  &  -27:47:41.0  &  107.4793  &  5.7  &  5.6  &  6.0 \\
LineSeeker.416  &  MF3D.185  &  FindClump.24  &  03:32:42.11  &  -27:46:11.8  &  99.0254  &  5.0  &  5.1  &  6.0 \\
LineSeeker.277  &  MF3D.187  &  FindClump.26  &  03:32:39.65  &  -27:45:50.3  &  105.9089  &  5.1  &  5.1  &  6.0 \\
\enddata
\tablecomments{
(1) Identification for emission line candidates found by LineSeeker. 
(2) Identification for emission line candidates found by MF3D.
(3) Identification for emission line candidates found by FindClump.
(4) Right ascension (J2000).
(5) Declination (J2000).
(6) Central frequency of the line.
(7) S/N value return by LineSeeker assuming an unresolved source.
(8) S/N value return by MF3D.
(9) S/N value return by FindClump assuming an unresolved source.
}
\end{deluxetable*}

Figure \ref{fig:Comparison_SN} shows the different S/N values obtained for the actual ASPECS-LP band 3 cube with the three different methods. The green points show the comparison between LineSeeker and FindClump while the orange points show the comparison between LineSeeker and MF3D. 
At the bright end of the emission--line candidate distribution, we see good agreement between the different methods. At the lower end, we see good agreement between LineSeeker and MF3D while FindClump shows the aforementioned increase in S/N ratio with respect to LineSeeker, similar to the simulation results.

In Table \ref{tab:LP_LinesCandidates_3methods} we present the ${\rm S/N}\geq6$ line candidates found independently by each of the three methods. The limit of ${\rm S/N}\geq6$ is motivated by the results obtained in ASPECS-Pilot \citep{Walter2016} where all the line candidates ${\rm S/N}\geq6$ are confirmed by their NIR counterpart redshift. All the line candidates found by LineSeeker with ${\rm S/N}\geq6$ show similar S/N values with the other two methods. LineSeeker appears to be the most conservative out of the three methods, finding 19 line candidates with ${\rm S/N}\geq6$ compared to 21 candidates with MF3D. FindClump finds 27 line candidates with the same S/N cut, an increase of $\approx35\%$ with respect to LineSeeker and MF3D, as expected and previously discussed. 
In Figure \ref{fig:footprint} we show the position of all the ${\rm S/N}\geq6$ line candidates identified in Table \ref{tab:LP_LinesCandidates_3methods}. Here we notice that most (6/8) of the FindClump candidates without ${\rm S/N}\geq6$ matches in the other two methods are found at the edges of the mosaic, where the PB correction value $\leq0.5$. This seems to indicate that something related to the position in the map is boosting the S/N value in the FindClump method. 
Finally, we comment on line candidate MF3D.18, which is the only one found by a single method with ${\rm S/N}\geq6$; it has considerably lower S/N in LineSeeker and is not detected by FindClump. Upon closer inspection, we see that the line candidate is spatially extended over multiple beams and therefore is only detected by the extra spatial filtering used by MF3D. Since the line is not detected by FindClump, and detected with a lower S/N value by LineSeeker, we leave it out of the list of selected line candidates. An independent confirmation will be done by using NIR counterpart and MUSE redshift in a posterior paper. The confirmation of this line would indicate the need to look for faint and extended emission lines in the future, although the lack of bright NIR counterpart makes it difficult to confirm at the moment. 

In conclusion, we find good agreement between the three methods in the bright end, with a closer agreement between LineSeeker and MF3D. The final sample of line candidates is created using the properties obtained with LineSeeker, since it is a fast code that can be run in simulated cubes.

%%%%%%%%%%%%%%%%%%%%%%%%%%%%%%%%%%%%%%%%%%%%%%%%%%%%%%%%%%%%%%%%%%%%%%%%%%%%%%%%%%%%%%%%%%%%%%
%%%%%%%%%%%%%%%%%%%%%%%%%%%%%%%%%%%%%%%%%%%%%%%%%%%%%%%%%%%%%%%%%%%%%%%%%%%%%%%%%%%%%%%%%%%%%%
%%%%%%%%%%%%%%%%%%%%%%%%%%%%%%%%%%%%%%%%%%%%%%%%%%%%%%%%%%%%%%%%%%%%%%%%%%%%%%%%%%%%%%%%%%%%%%

\startlongtable
\begin{deluxetable}{ccll}
\tablecaption{Completeness as function of emission line central frequency, integrated line flux and width.\label{tab:CompletenessFluxPeakFWHM}}
\tablehead{
\colhead{Freq.} & 
\colhead{Line Flux} & 
\colhead{FWHM} & 
\colhead{Completeness}\\
\colhead{[GHz]} & 
\colhead{[Jy km s$^{-1}$]} & 
\colhead{[km s$^{-1}$]} & 
\colhead{}} 
\colnumbers
\startdata
84.2--94.2 & 0.0--0.1 & 0--200 & $0.39_{-0.03}^{+0.03}$ \\
84.2--94.2 & 0.0--0.1 & 200--400 & $0.06_{-0.03}^{+0.04}$ \\
84.2--94.2 & 0.0--0.1 & 400--600 & $0.03_{-0.03}^{+0.05}$ \\
84.2--94.2 & 0.0--0.1 & 600--800 & $0.0_{-0.02}^{+0.04}$ \\
84.2--94.2 & 0.0--0.1 & 800--1000 & $0.06_{-0.05}^{+0.08}$ \\
84.2--94.2 & 0.1--0.2 & 0--200 & $1.0_{-0.01}^{+0.0}$ \\
84.2--94.2 & 0.1--0.2 & 200--400 & $0.91_{-0.04}^{+0.03}$ \\
84.2--94.2 & 0.1--0.2 & 400--600 & $0.69_{-0.08}^{+0.07}$ \\
84.2--94.2 & 0.1--0.2 & 600--800 & $0.44_{-0.09}^{+0.09}$ \\
84.2--94.2 & 0.1--0.2 & 800--1000 & $0.19_{-0.08}^{+0.09}$ \\
84.2--94.2 & 0.2--0.3 & 0--200 & $1.0_{-0.02}^{+0.01}$ \\
84.2--94.2 & 0.2--0.3 & 200--400 & $0.98_{-0.03}^{+0.02}$ \\
84.2--94.2 & 0.2--0.3 & 400--600 & $0.97_{-0.04}^{+0.03}$ \\
84.2--94.2 & 0.2--0.3 & 600--800 & $0.97_{-0.04}^{+0.03}$ \\
84.2--94.2 & 0.2--0.3 & 800--1000 & $0.96_{-0.06}^{+0.03}$ \\
84.2--94.2 & 0.3--0.4 & 0--200 & $1.0_{-0.07}^{+0.04}$ \\
84.2--94.2 & 0.3--0.4 & 200--400 & $0.99_{-0.02}^{+0.01}$ \\
84.2--94.2 & 0.3--0.4 & 400--600 & $1.0_{-0.03}^{+0.01}$ \\
84.2--94.2 & 0.3--0.4 & 600--800 & $1.0_{-0.05}^{+0.02}$ \\
84.2--94.2 & 0.3--0.4 & 800--1000 & $1.0_{-0.05}^{+0.02}$ \\
84.2--94.2 & 0.4--0.5 & 0--200 & $1.0_{-0.02}^{+0.01}$ \\
84.2--94.2 & 0.4--0.5 & 200--400 & $1.0_{-0.02}^{+0.01}$ \\
84.2--94.2 & 0.4--0.5 & 400--600 & $1.0_{-0.02}^{+0.01}$ \\
84.2--94.2 & 0.4--0.5 & 600--800 & $1.0_{-0.05}^{+0.02}$ \\
84.2--94.2 & 0.4--0.5 & 800--1000 & $1.0_{-0.05}^{+0.03}$ \\
94.2--104.2 & 0.0--0.1 & 0--200 & $0.45_{-0.03}^{+0.03}$ \\
94.2--104.2 & 0.0--0.1 & 200--400 & $0.23_{-0.05}^{+0.05}$ \\
94.2--104.2 & 0.0--0.1 & 400--600 & $0.13_{-0.05}^{+0.07}$ \\
94.2--104.2 & 0.0--0.1 & 600--800 & $0.0_{-0.02}^{+0.04}$ \\
94.2--104.2 & 0.0--0.1 & 800--1000 & $0.0_{-0.02}^{+0.05}$ \\
94.2--104.2 & 0.1--0.2 & 0--200 & $0.95_{-0.03}^{+0.02}$ \\
94.2--104.2 & 0.1--0.2 & 200--400 & $0.91_{-0.05}^{+0.04}$ \\
94.2--104.2 & 0.1--0.2 & 400--600 & $0.81_{-0.07}^{+0.06}$ \\
94.2--104.2 & 0.1--0.2 & 600--800 & $0.58_{-0.09}^{+0.08}$ \\
94.2--104.2 & 0.1--0.2 & 800--1000 & $0.36_{-0.09}^{+0.1}$ \\
94.2--104.2 & 0.2--0.3 & 0--200 & $1.0_{-0.03}^{+0.01}$ \\
94.2--104.2 & 0.2--0.3 & 200--400 & $0.98_{-0.03}^{+0.02}$ \\
94.2--104.2 & 0.2--0.3 & 400--600 & $0.97_{-0.04}^{+0.02}$ \\
94.2--104.2 & 0.2--0.3 & 600--800 & $0.95_{-0.07}^{+0.04}$ \\
94.2--104.2 & 0.2--0.3 & 800--1000 & $1.0_{-0.05}^{+0.02}$ \\
94.2--104.2 & 0.3--0.4 & 0--200 & $1.0_{-0.05}^{+0.02}$ \\
94.2--104.2 & 0.3--0.4 & 200--400 & $0.99_{-0.02}^{+0.01}$ \\
94.2--104.2 & 0.3--0.4 & 400--600 & $0.97_{-0.05}^{+0.03}$ \\
94.2--104.2 & 0.3--0.4 & 600--800 & $1.0_{-0.03}^{+0.02}$ \\
94.2--104.2 & 0.3--0.4 & 800--1000 & $1.0_{-0.06}^{+0.03}$ \\
94.2--104.2 & 0.4--0.5 & 0--200 & $1.0_{-0.31}^{+0.21}$ \\
94.2--104.2 & 0.4--0.5 & 200--400 & $1.0_{-0.02}^{+0.01}$ \\
94.2--104.2 & 0.4--0.5 & 400--600 & $1.0_{-0.02}^{+0.01}$ \\
94.2--104.2 & 0.4--0.5 & 600--800 & $0.97_{-0.05}^{+0.03}$ \\
94.2--104.2 & 0.4--0.5 & 800--1000 & $1.0_{-0.08}^{+0.04}$ \\
104.2--114.2 & 0.0--0.1 & 0--200 & $0.2_{-0.03}^{+0.03}$ \\
104.2--114.2 & 0.0--0.1 & 200--400 & $0.0_{-0.01}^{+0.02}$ \\
104.2--114.2 & 0.0--0.1 & 400--600 & $0.0_{-0.01}^{+0.03}$ \\
104.2--114.2 & 0.0--0.1 & 600--800 & $0.0_{-0.02}^{+0.04}$ \\
104.2--114.2 & 0.0--0.1 & 800--1000 & $0.0_{-0.03}^{+0.05}$ \\
104.2--114.2 & 0.1--0.2 & 0--200 & $0.89_{-0.03}^{+0.03}$ \\
104.2--114.2 & 0.1--0.2 & 200--400 & $0.47_{-0.06}^{+0.06}$ \\
104.2--114.2 & 0.1--0.2 & 400--600 & $0.17_{-0.05}^{+0.06}$ \\
104.2--114.2 & 0.1--0.2 & 600--800 & $0.06_{-0.04}^{+0.06}$ \\
104.2--114.2 & 0.1--0.2 & 800--1000 & $0.06_{-0.05}^{+0.08}$ \\
104.2--114.2 & 0.2--0.3 & 0--200 & $0.98_{-0.03}^{+0.02}$ \\
104.2--114.2 & 0.2--0.3 & 200--400 & $0.96_{-0.03}^{+0.02}$ \\
104.2--114.2 & 0.2--0.3 & 400--600 & $0.85_{-0.06}^{+0.05}$ \\
104.2--114.2 & 0.2--0.3 & 600--800 & $0.74_{-0.08}^{+0.07}$ \\
104.2--114.2 & 0.2--0.3 & 800--1000 & $0.25_{-0.1}^{+0.11}$ \\
104.2--114.2 & 0.3--0.4 & 0--200 & $1.0_{-0.06}^{+0.03}$ \\
104.2--114.2 & 0.3--0.4 & 200--400 & $1.0_{-0.02}^{+0.01}$ \\
104.2--114.2 & 0.3--0.4 & 400--600 & $0.96_{-0.06}^{+0.04}$ \\
104.2--114.2 & 0.3--0.4 & 600--800 & $1.0_{-0.04}^{+0.02}$ \\
104.2--114.2 & 0.3--0.4 & 800--1000 & $1.0_{-0.06}^{+0.03}$ \\
104.2--114.2 & 0.4--0.5 & 0--200 & $1.0_{-0.31}^{+0.21}$ \\
104.2--114.2 & 0.4--0.5 & 200--400 & $1.0_{-0.02}^{+0.01}$ \\
104.2--114.2 & 0.4--0.5 & 400--600 & $1.0_{-0.03}^{+0.01}$ \\
104.2--114.2 & 0.4--0.5 & 600--800 & $1.0_{-0.04}^{+0.02}$ \\
104.2--114.2 & 0.4--0.5 & 800--1000 & $1.0_{-0.07}^{+0.04}$ \\
\enddata
\tablecomments{
(1) Range of line central frequency.
(2) Range of emission line flux. 
(3) Range of line widths as given by FWHM.
(4) Completeness level for emission lines within the given ranges. 
}
\end{deluxetable}
% \clearpage

\subsection{Fidelity} \label{sec:Fidelity}

As emission lines of galaxies found in field spectral surveys are in many cases faint, it is crucial to assess how reliable each line identification is. The reliability of an emission line candidate is determined by the probability $P$ that such a line is due to noise alone and therefore not real. We define the quantity ${\rm Fidelity}=1-P$ that contains this information. ${\rm Fidelity}=0$ indicates a line candidate consisting with being produced by noise while ${\rm Fidelity}=1$ indicates a secure detection. The probability $P$ will depend on the significance of the line candidate and it is the main factor to determine Fidelity.

The significance of an emission line candidate is a difficult property to estimate because of the hidden nature of the noise distribution of the data cubes. Previous works have linked the significance of a detection to its S/N ratio while using the negative data or simulated cubes as noise references \citep{Walter2016,GL2017c}. We argue here that the usage of the S/N value as the only indicator for the significance only works when the number of independent elements is constant across all the search, which is not the case for the search of emission lines with different widths (or resolved sizes) in data cubes. 

In the scenario of a data cube with $N$ independent elements per channel and $M$ channels, the total number of independent elements for a search done across the whole cube will be of $N\times M$. If we decide to do the search in a convolved cube (along the velocity/frequency axis), as it is the case for the methods described above, the number of independent elements will be lower since the $M$ channels are no longer independent. The extreme case that exemplifies the latter is when a data cube is collapsed to form a continuum image where the number of independent elements will be only $N\times 1$. For any spectral convolution of the data cube, the number of independent elements will be in between $N$ and $N\times M$. The exact value will be determined by the nature of the convolution kernel and the amount of channels combined. 

It should be noted that the difference in the number of independent elements in a cube and in the corresponding continuum image explains the reason why the significance of emission lines and continuum sources detected on same observations with the same S/N values are not the same. This is the reason of why we can explore lower S/N candidates when searching for sources in the continuum regime. 

In the case of the ASPECS-LP band 3 data cube, the number of independent elements per channel is defined by the size of the synthesized beam and the size of the mosaic map. A good estimate of the number of independent elements is twice the number of beams contained in the map \citep{Condon1997,Condon1998,Dunlop2017}. 
The number of independent elements for a spectral convolution will depend on the width of the kernel used. This will have the effect that the significance for a line emission candidate will depend on its width. Broader emission line candidates will have higher significance than narrower lines with the same S/N, since the number of independent elements for the search of broader emission lines is lower than for the more narrow ones.

We estimate the significance of the emission line candidates by taking into account the width of the line (and the corresponding convolution kernel for which the S/N ratio is the highest) as well as their S/N. 

We use the negative lines as reference to estimate the fidelity, assuming noise around 0, typical of interferometric data. The usage of negative sources is based on the fact that the negative lines are expected to be produced only by noise and should give a good representation of what one would have detected if no real detection were present in the cube. 
The fidelity is estimated as follows 

\begin{equation}
{\rm Fidelity}=1 - \frac{N_{\rm Neg}}{N_{\rm Pos}},
\end{equation}

with $N_{\rm Neg}$ and $N_{\rm Pos}$ being the number of negative and positive emission line candidates detected with a given S/N value in a particular kernel convolution. The positive and negative lines are searched over the total area of the cube, which is of 7 arcmin$^2$ within mosaic primary beam correction $\geq0.2$.
To avoid the effects of low number statistics in the tails of the distribution, we fit a function of the form $N(1 - \erf(\rm SN / \sqrt{2}\sigma))$ to the S/N histogram of negative lines, with $\erf$ being the error function and $N$ and $\sigma$ free parameters. We do this to estimate the shape of the underlying negative rate distribution. We select as reliable all the emission lines that have ${\rm Fidelity}\geq0.9$. 

\subsection{Completeness} \label{sec:Completeness}

\begin{figure}
\epsscale{1.2}
\plotone{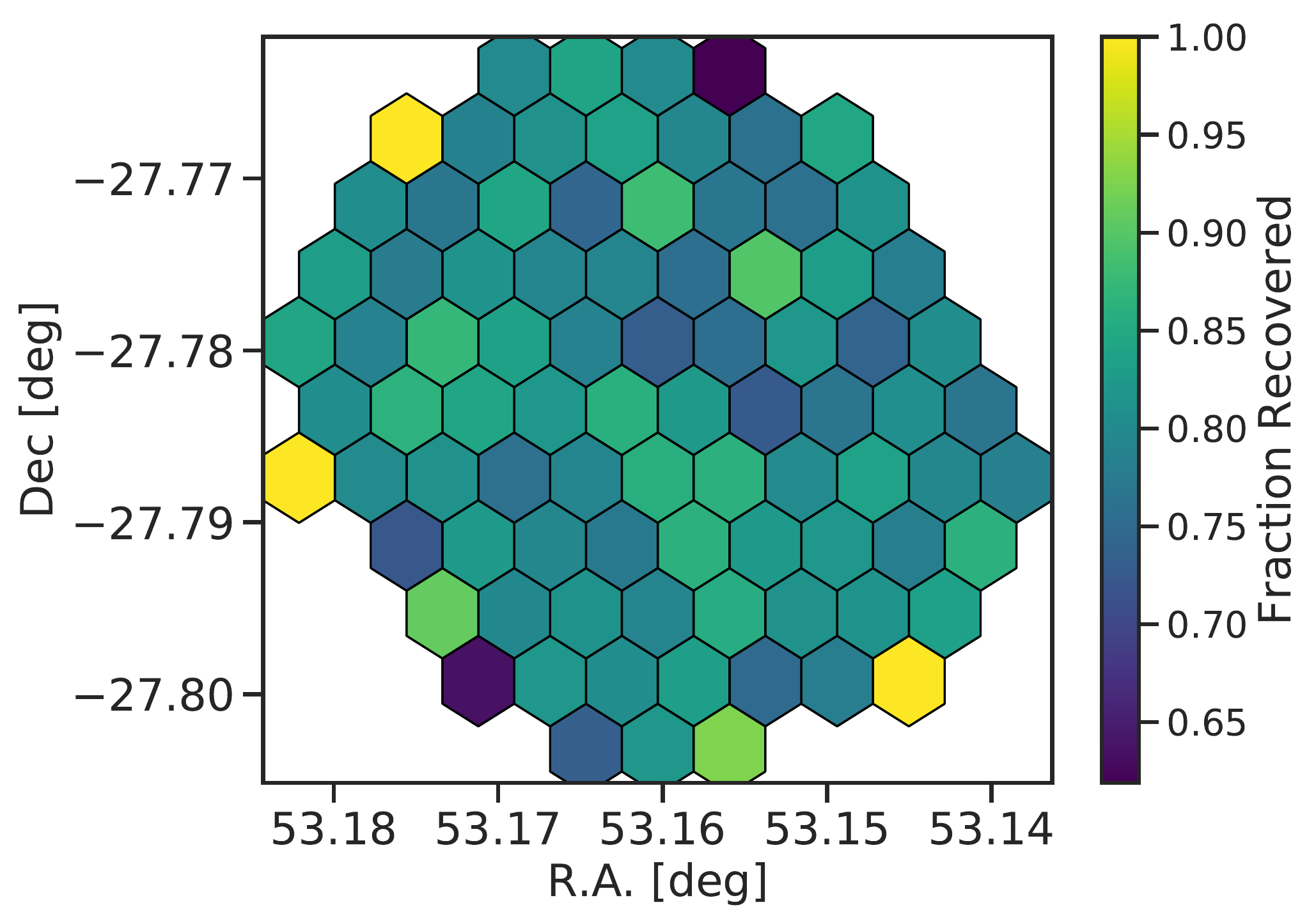}
\epsscale{1.2}
\plotone{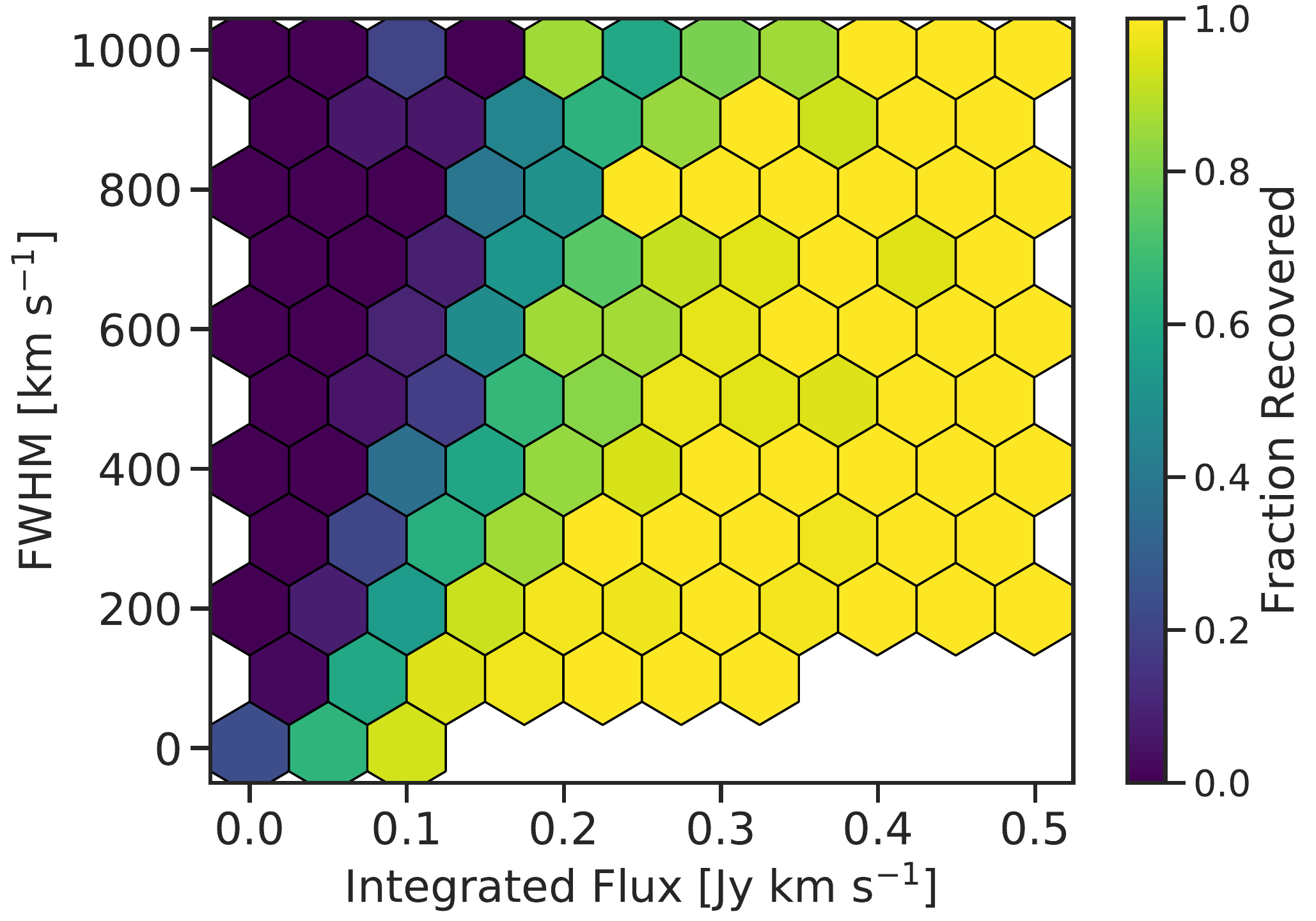}
\epsscale{1.2}
\plotone{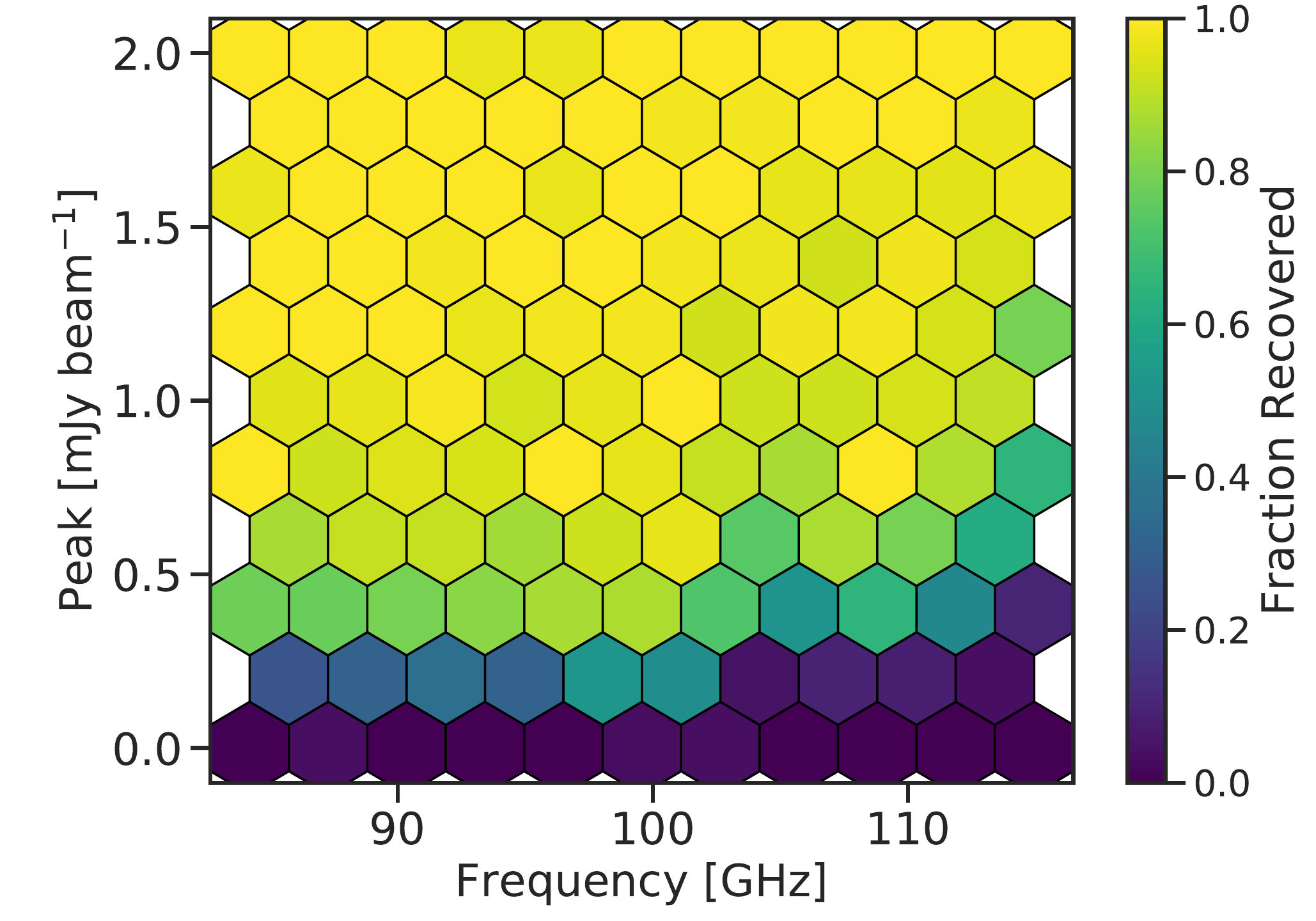}
\caption{Completeness of the emission lines marginalized to different properties. The color in each map represent the fraction of emission lines recovered in the corresponding cell. The top panel shows the recovery fraction for different positions in the ASPECS-LP mosaic. The central panel shows the recovery fraction for different values of integrated flux and FWHM (over the full range of frequencies). The bottom panel shows the recovery fraction for different values of line peak and central frequency (over the full range of FWHMs). \label{fig:completeness}}
\end{figure}

Determining completeness of the sample of CO line emitting galaxies identified in the ASPECS-LP cube is crucial in deriving the CO luminosity function. We need to estimate the possibility of detecting an emission line with a given set of properties within the real data cube.
The completeness is estimated by injecting simulated emission lines with Gaussian profiles and point source spatial profile to the real cube and checking if they are recovered. The line peaks range between 0 and 2 mJy beam$^{-1}$, the FWHM are in between 0 and $\sim1000\kms$ and the central frequencies between 84.2 and 114.2 GHz. The recovery of the lines was tested using LineSeeker. In each iteration, we inject 50 simulated lines to limit the chances of having two nearby emission lines blended or confused as one. We repeated this process until 5000 emission lines were injected. 
We say that an injected line is recovered when it has ${\rm Fidelity}\geq0.9$ in the output from LineSeeker.  As shown below, some of the detected emission line are not well described by a Gaussian function (e.g. ASPECS-LP.3mm.05 and ASPECS-LP.3mm.06 in Fig. \ref{fig:LP_spectra_postamp2}). Such lines are in the bright end of the detections and are well identified when using a Gaussian kernel. On the other hand, the faint end of the detected lines are reasonable well described by Gaussian functions (within the errors), which supports the assumption of Gaussian lines for the completeness calculation. 

In Fig. \ref{fig:completeness}, we present the completeness values marginalized to different pairs of emission line properties (while taking into account the full range of the remaining parameters). We can see that the position of the line within the map does not play an important role in the probability that the line is recovered. The reason for this is that the search for emission lines is done in the cubes where the noise distribution is flat across the spatial axes, before correcting by the mosaic sensitivity correction. As expected, the integrated flux of the emission line plays a very important role in our ability to recover it. In the middle panel of Fig. \ref{fig:completeness}, we see how the recovery fraction depends on the integrated flux as well as in the FWHM. For the same integrated line flux, narrower lines are easier to recover, as shown for the higher recovery fraction in the bottom of the panel, where the peaks of the lines are higher.
Finally, in the bottom panel of Fig. \ref{fig:completeness}, we show the recovery fraction as a function of the peak of the line and the central frequency. There is a clear dependence between the recovery fraction and the central frequency of the emission lines. That dependency can be explained by the lower sensitivity of the higher frequencies (Fig. \ref{fig:RMS}). In table \ref{tab:CompletenessFluxPeakFWHM} we present the completeness values calculated within different ranges of central frequencies, line fluxes and line widths.

\subsection{Search of continuum sources}

Our search for sources in a continuum image is very similar to the search of emission lines presented above. A continuum source is equivalent to an emission line with width equal to one channel, such that the methods used for emission lines already described above can also be used for search of continuum sources. In this manner, we obtain the fidelity of continuum source candidates.
We estimate completeness values for the candidates found in the continuum image. The procedure is the same as for the emission line search, 10,000 point sources of flux densities between 0 and 10 $\rm \mu Jy \ts beam^{-1}$ are injected into the continuum image and checked to determine if they are recovered with $\rm S/N\geq4.6$. This is done in the continuum image without the mosaic primary beam correction. The completeness values are presented in Table \ref{tab:CompletenessContinuum}. 
The primary beam corrected completeness values were obtained using the following steps. First we replace each pixel in the continuum image with the intrinsic flux density we want to test (here the pixels are just a discretization of the observed area). The next step is to correct by the mosaic primary beam response, which converts the intrinsic flux density to a distribution of observed flux density pixels. We then create a completeness map by assigning to each observed flux density pixel a completeness values interpolated from Table \ref{tab:CompletenessContinuum}. Finally, the average of the completeness map is the completeness value for the intrinsic flux density. The error associated with the primary beam corrected completeness values are obtained by propagating the errors associated to each bin in Table \ref{tab:CompletenessContinuum} and are $\ll 0.01$.

\begin{deluxetable}{cc}
\tablecaption{Completeness for the continuum image. \label{tab:CompletenessContinuum}}
\tablehead{
\colhead{Flux density range [$\rm \mu Jy \ts beam^{-1}$]} & 
\colhead{Completeness}} 
\startdata
6--9 &  0.02\\
9--12 &  0.08\\
12--15 &  0.23\\
15--18 &  0.53\\
18--21 &  0.77\\
21--24 &  0.93\\
24--27 &  0.98\\
27--30 &  1.00
\enddata
\tablenotetext{}{The average error for the completeness values is  $<\pm0.01$.}
\end{deluxetable}

%%%%%%%%%%%%%%%%%%%%%%%%%%%%%%%%%%%%%%%%%%%%%%%%%%%%%%%%%%%%%%%%%%%%%%%%%%%%%%%%%%%%%%%%%%%%%%
%%%%%%%%%%%%%%%%%%%%%%%%%%%%%%%%%%%%%%%%%%%%%%%%%%%%%%%%%%%%%%%%%%%%%%%%%%%%%%%%%%%%%%%%%%%%%%
%%%%%%%%%%%%%%%%%%%%%%%%%%%%%%%%%%%%%%%%%%%%%%%%%%%%%%%%%%%%%%%%%%%%%%%%%%%%%%%%%%%%%%%%%%%%%%
\section{Results}\label{sec:Results}

\subsection{Detected emission lines} \label{sec:ResultsBlind}

\begin{figure*}
\epsscale{0.6}
\plotone{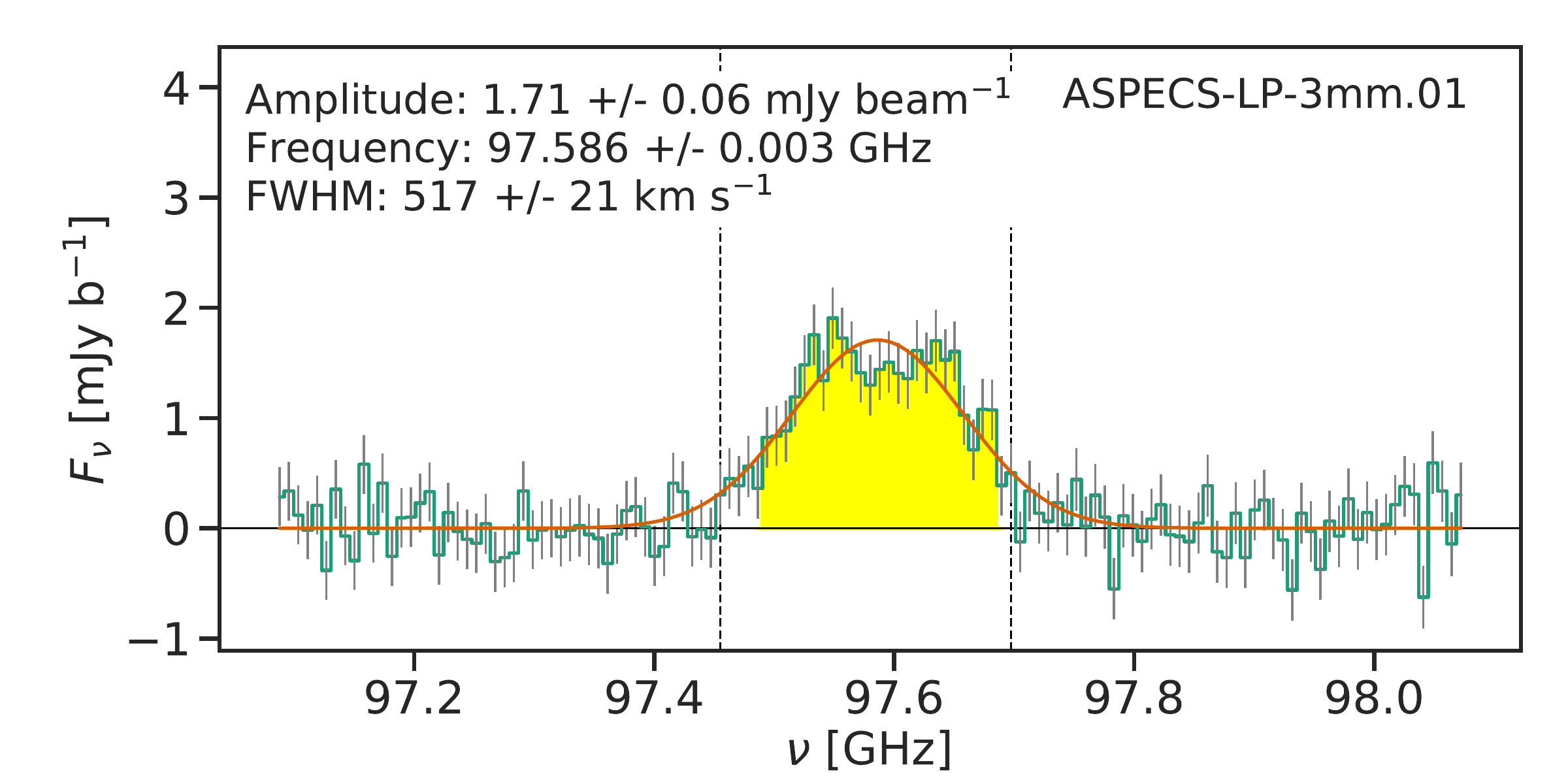}
\epsscale{0.37}
\plotone{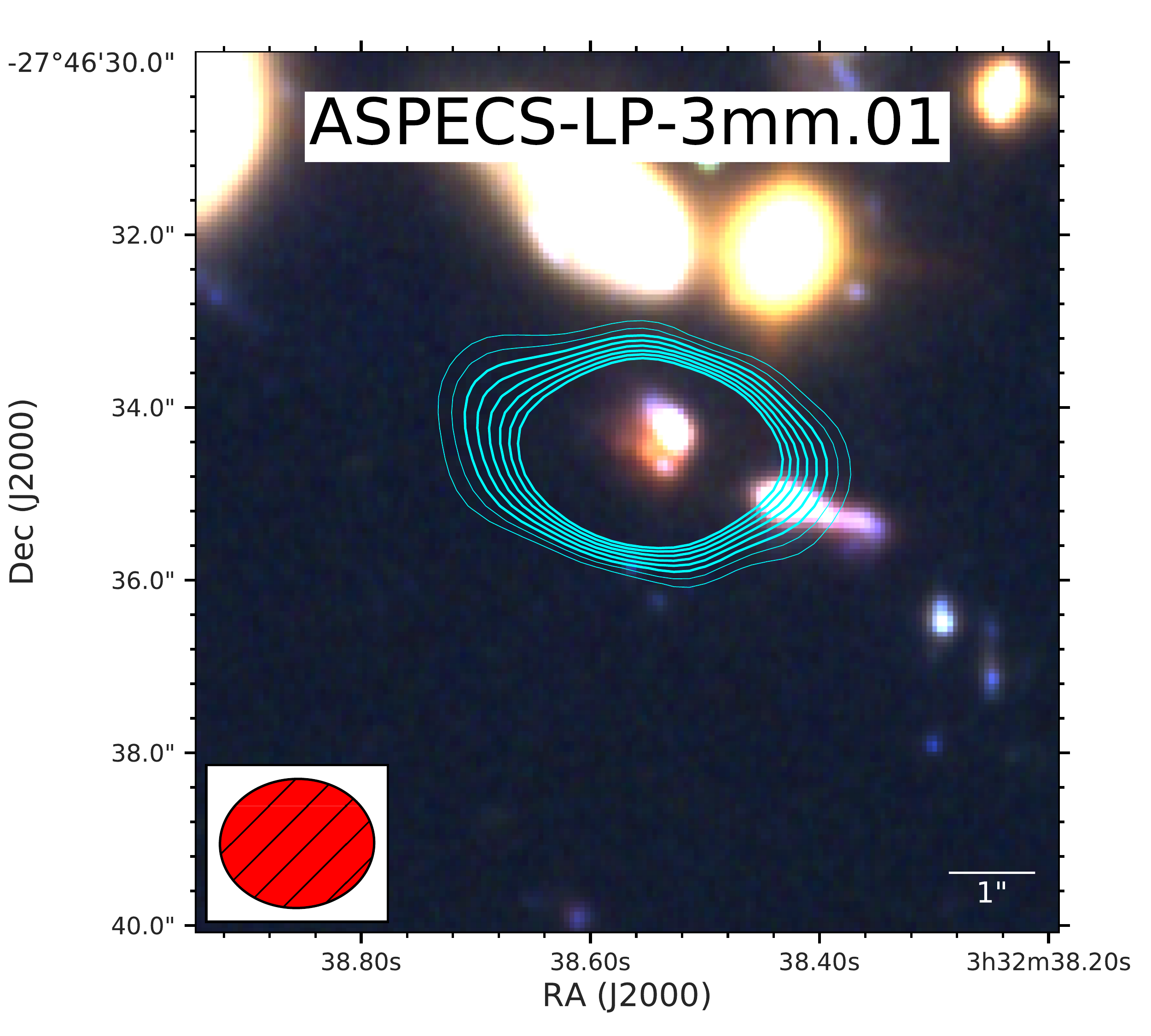}

\epsscale{0.6}
\plotone{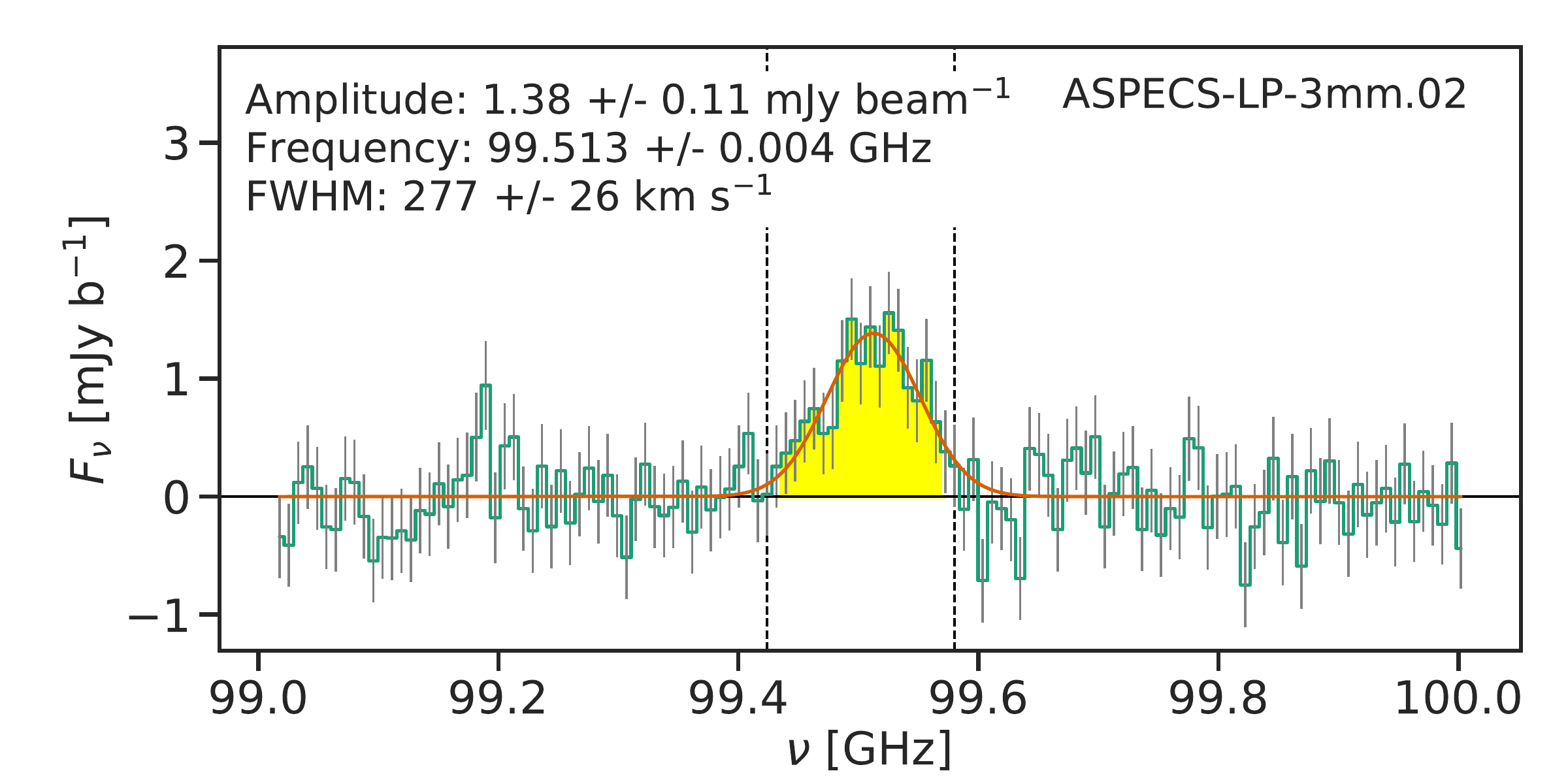}
\epsscale{0.37}
\plotone{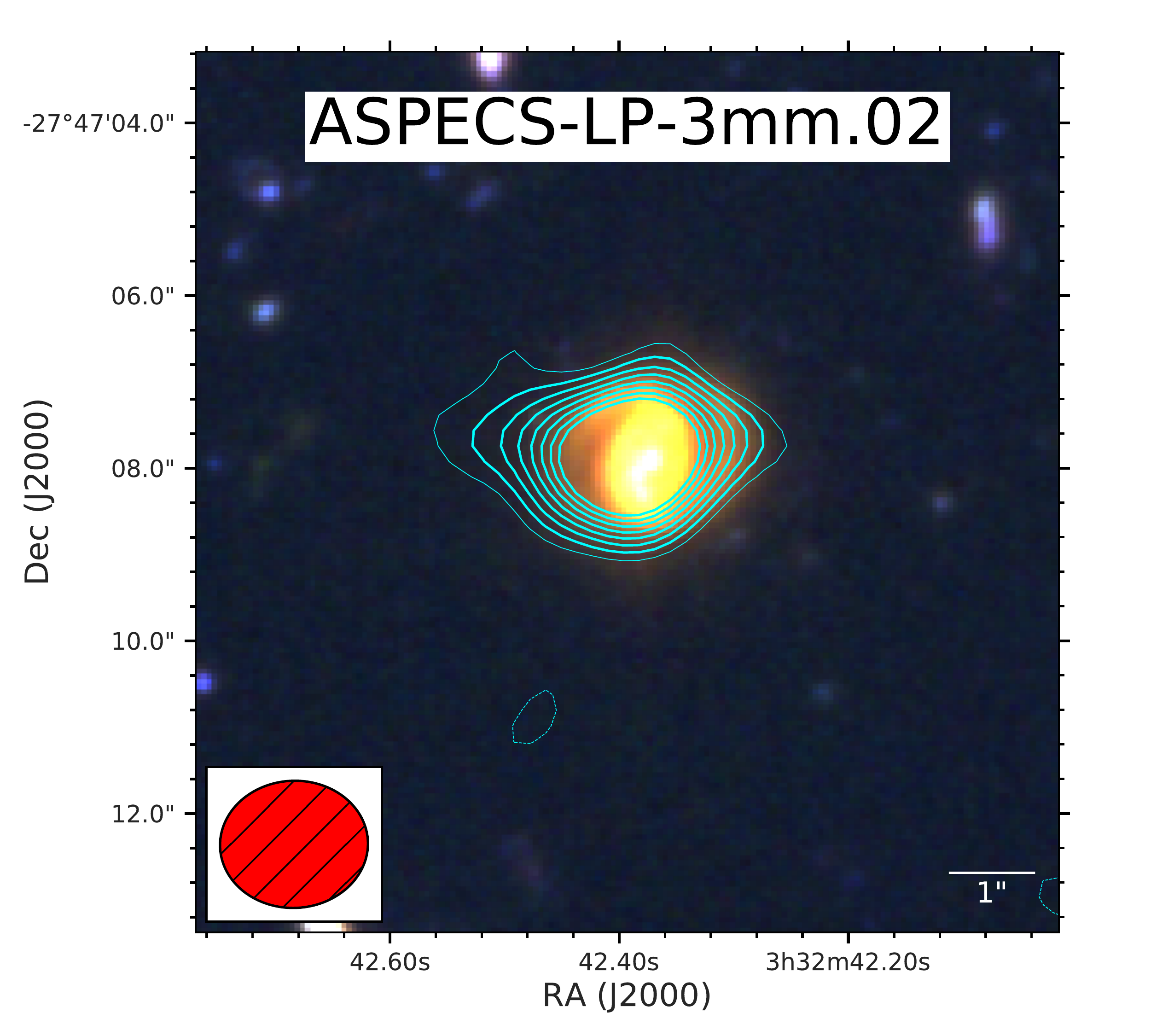}

\epsscale{0.6}
\plotone{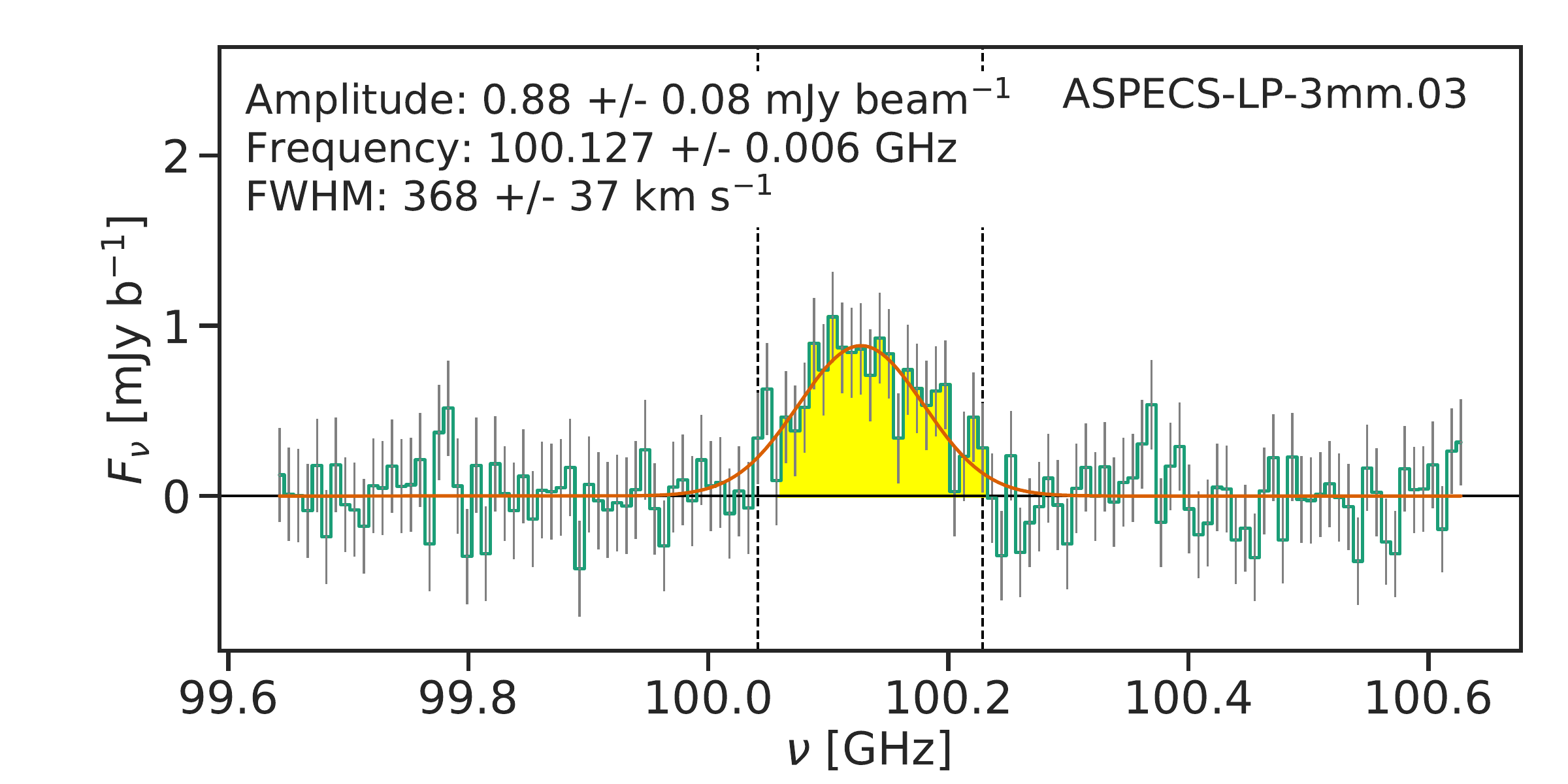}
\epsscale{0.37}
\plotone{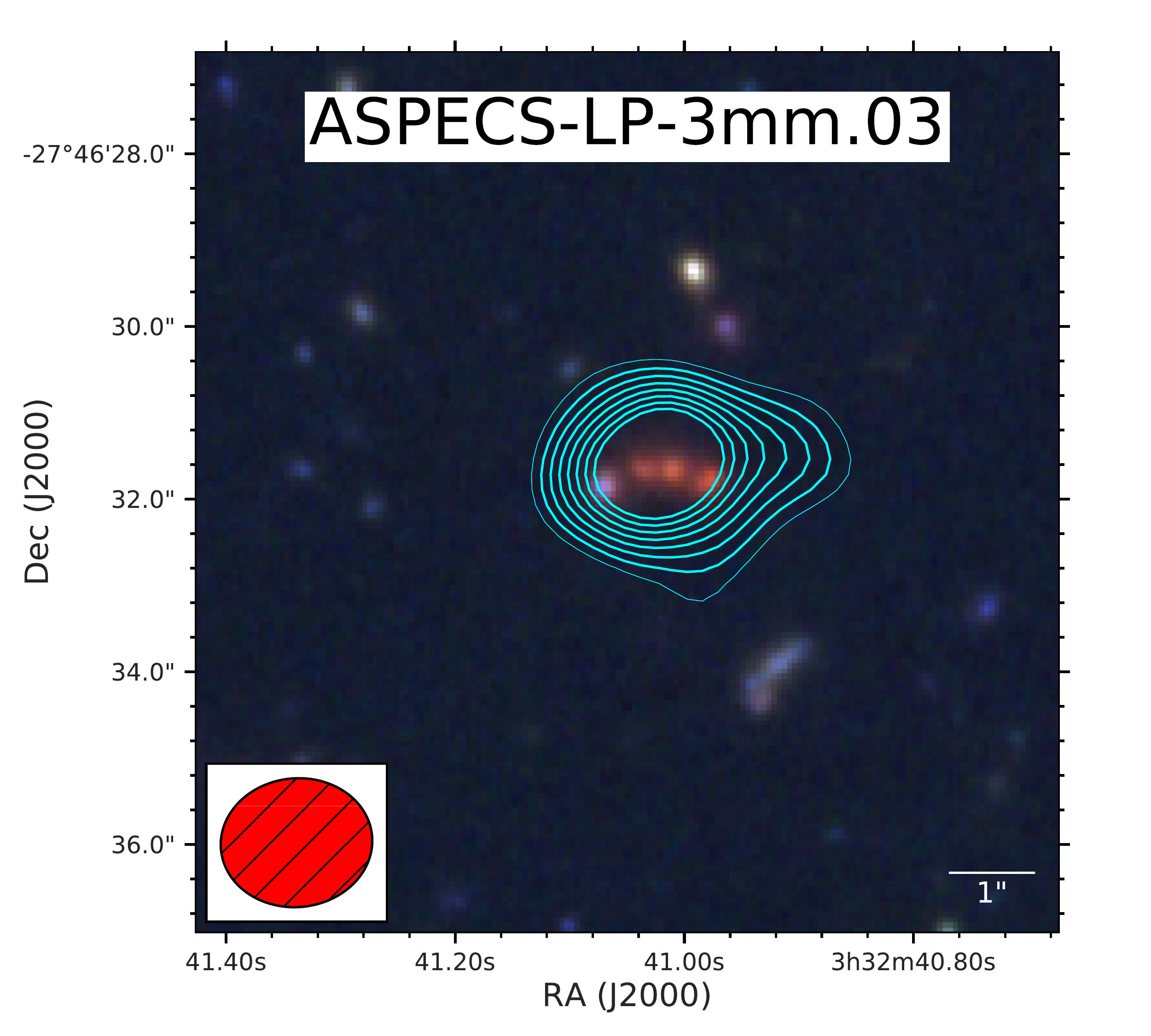}

\epsscale{0.6}
\plotone{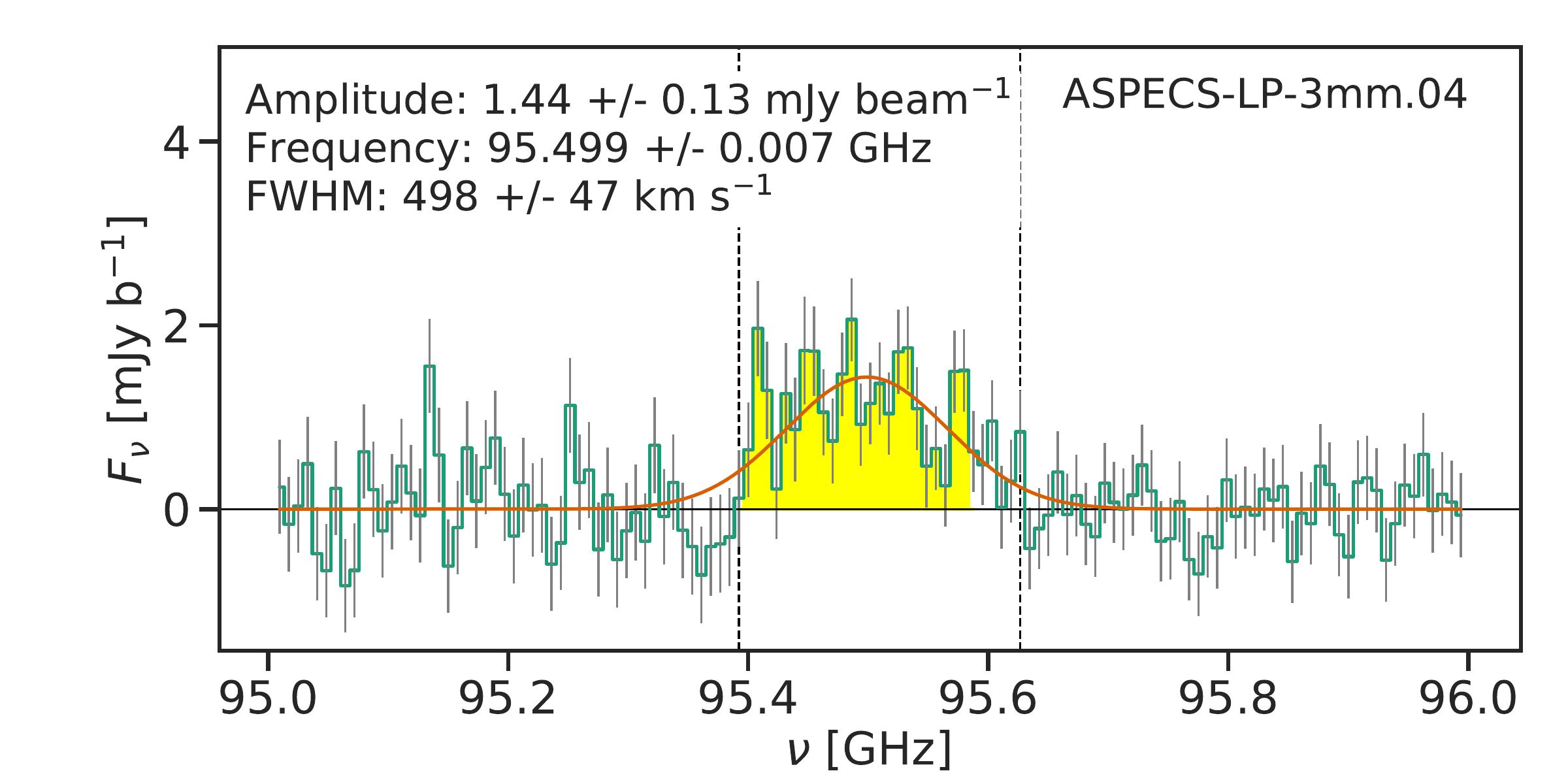}
\epsscale{0.37}
\plotone{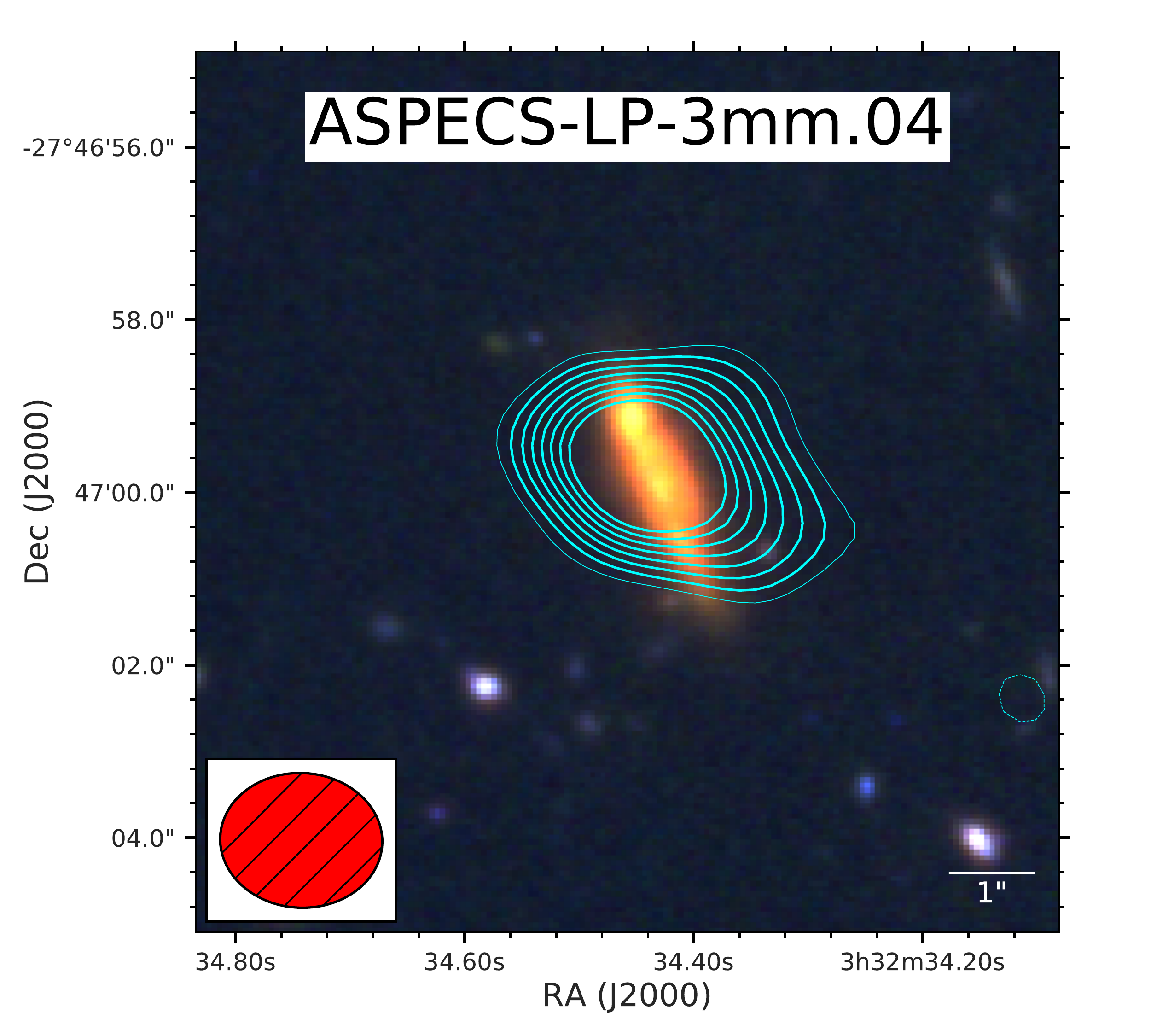}

\caption{Extracted spectra and color postage stamp of the secure emission lines detected in the ASPECS-LP band 3 cube. The vertical lines show the channels where the integrated line flux was measured. The contour levels go from $\pm3\sigma$ up to $10\sigma$ in steps of $1\sigma$. The color scale is the same as in Fig. \ref{fig:footprint} and the synthesized beam is shown in the bottom left corner.  \label{fig:LP_spectra_postamp1}}
\end{figure*}

\begin{figure*}
\epsscale{1.1}
\epsscale{0.6}
\plotone{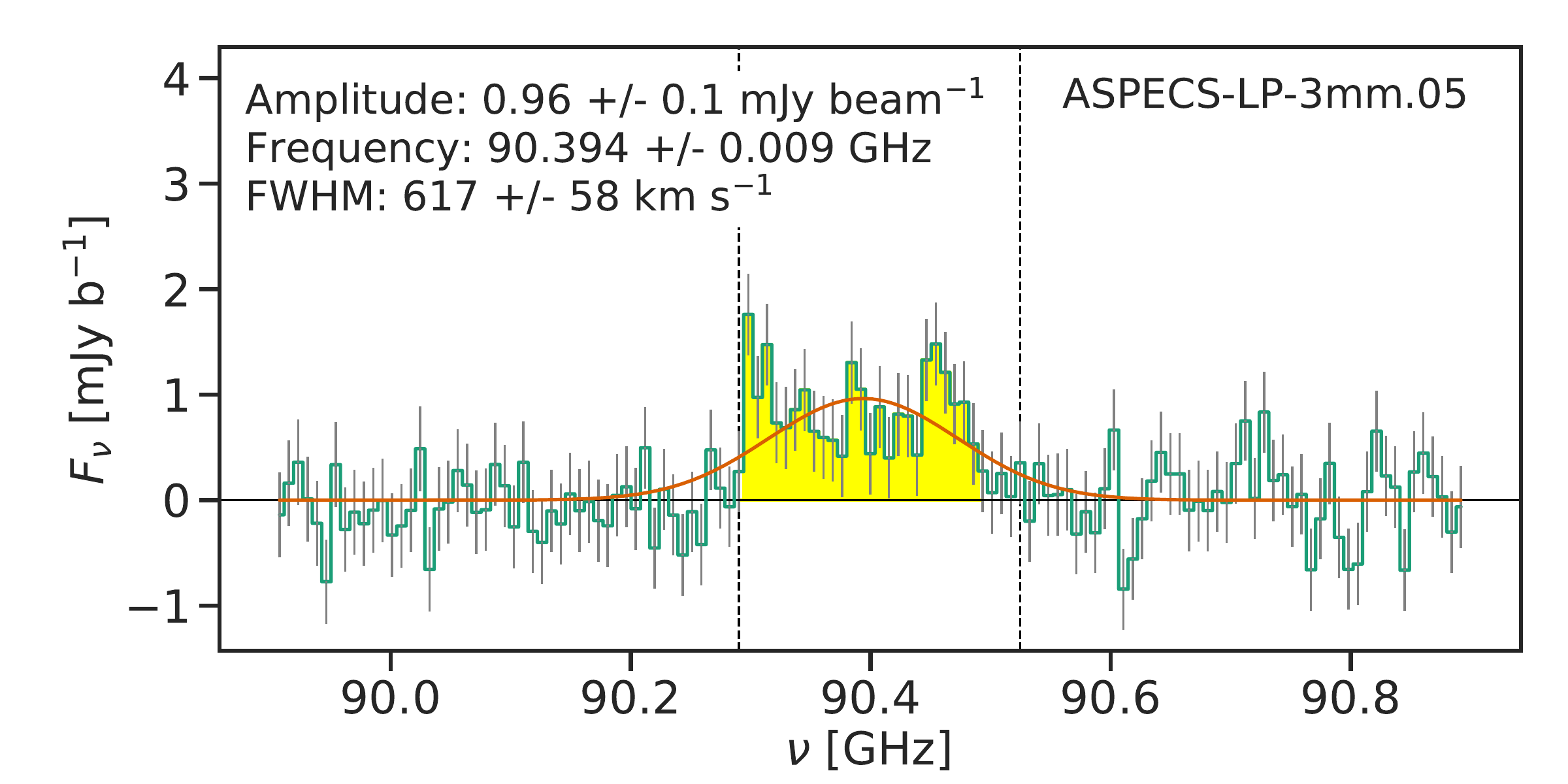}
\epsscale{0.37}
\plotone{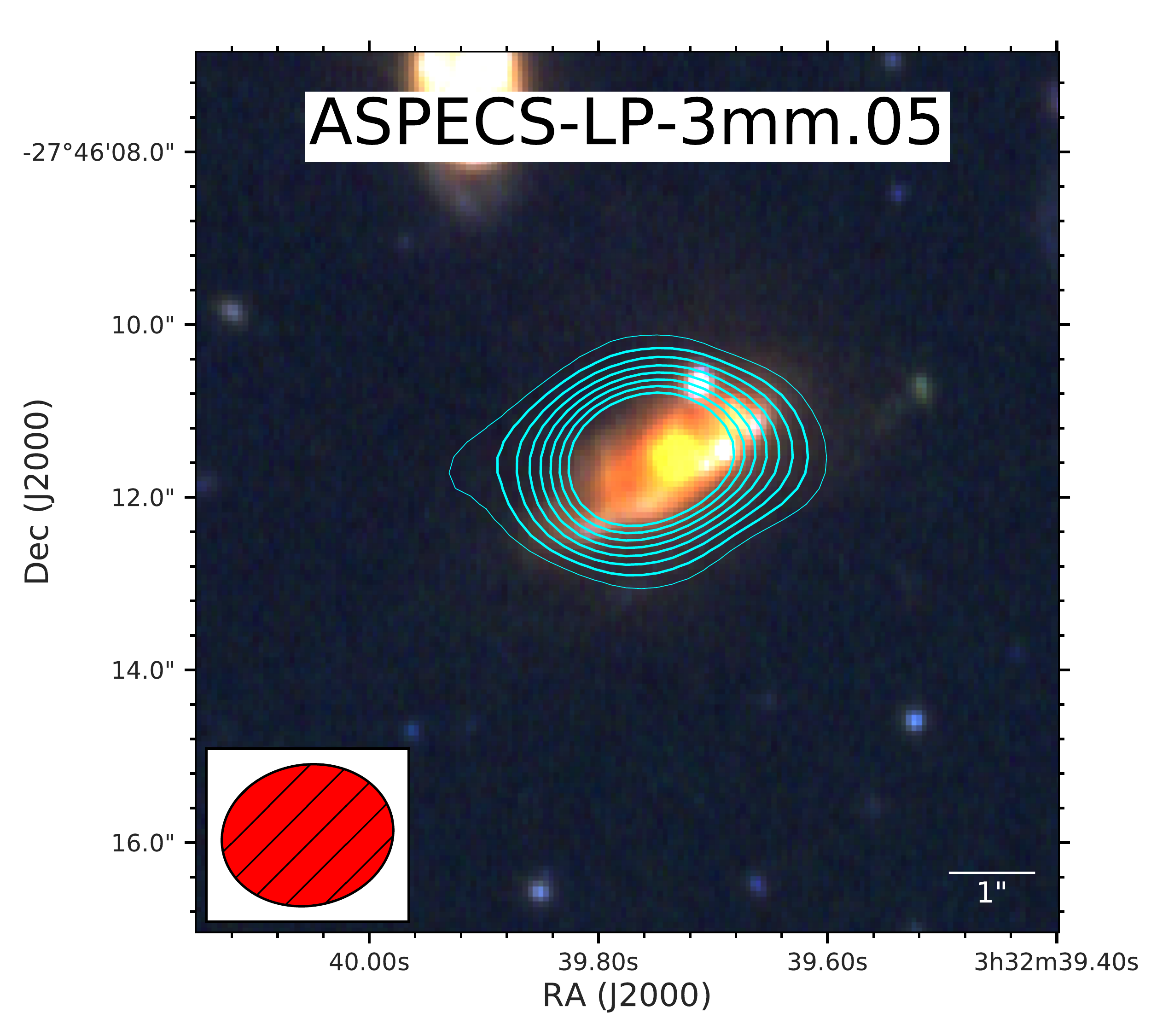}

\epsscale{0.6}
\plotone{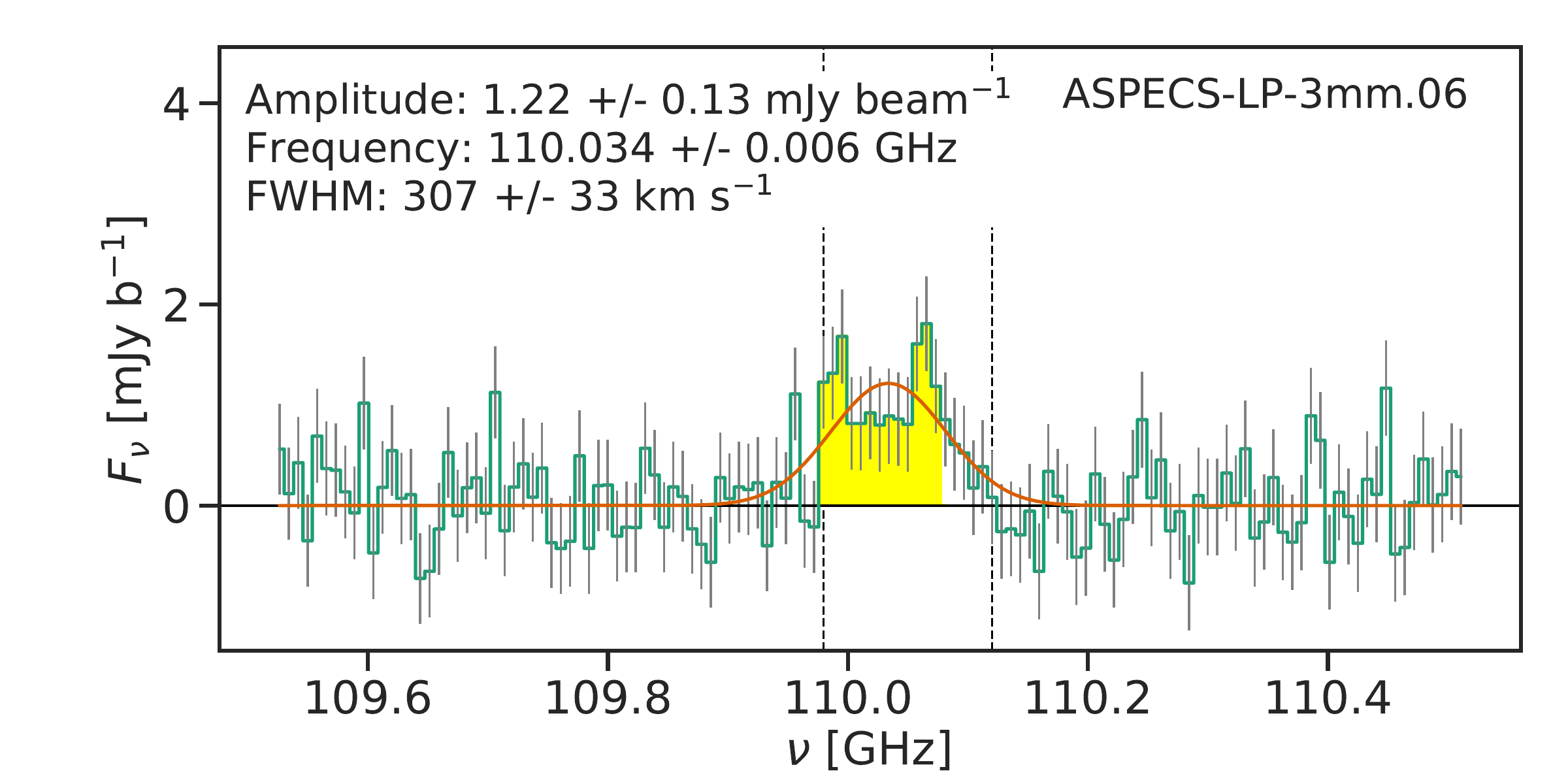}
\epsscale{0.37}
\plotone{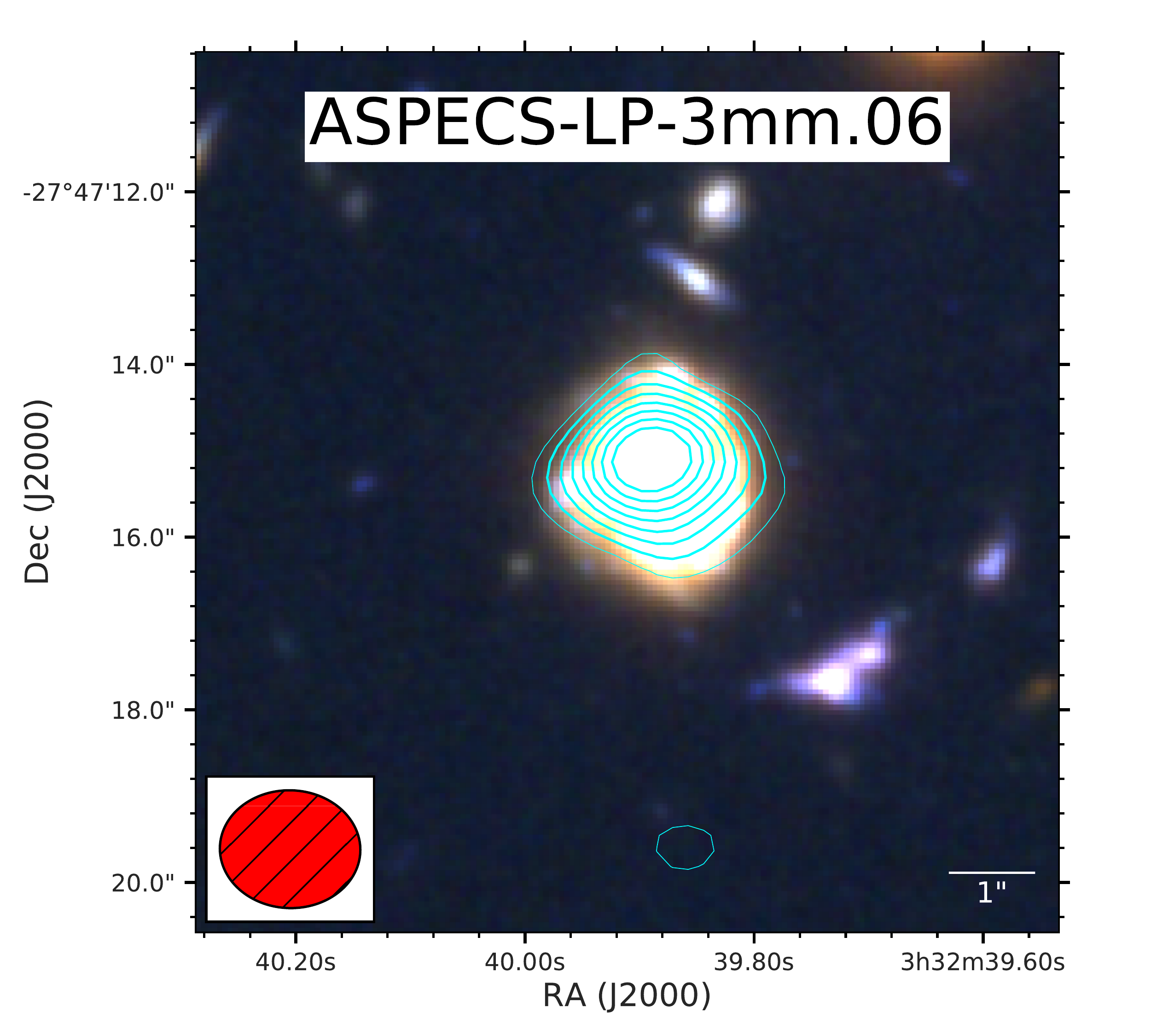}

\epsscale{0.6}
\plotone{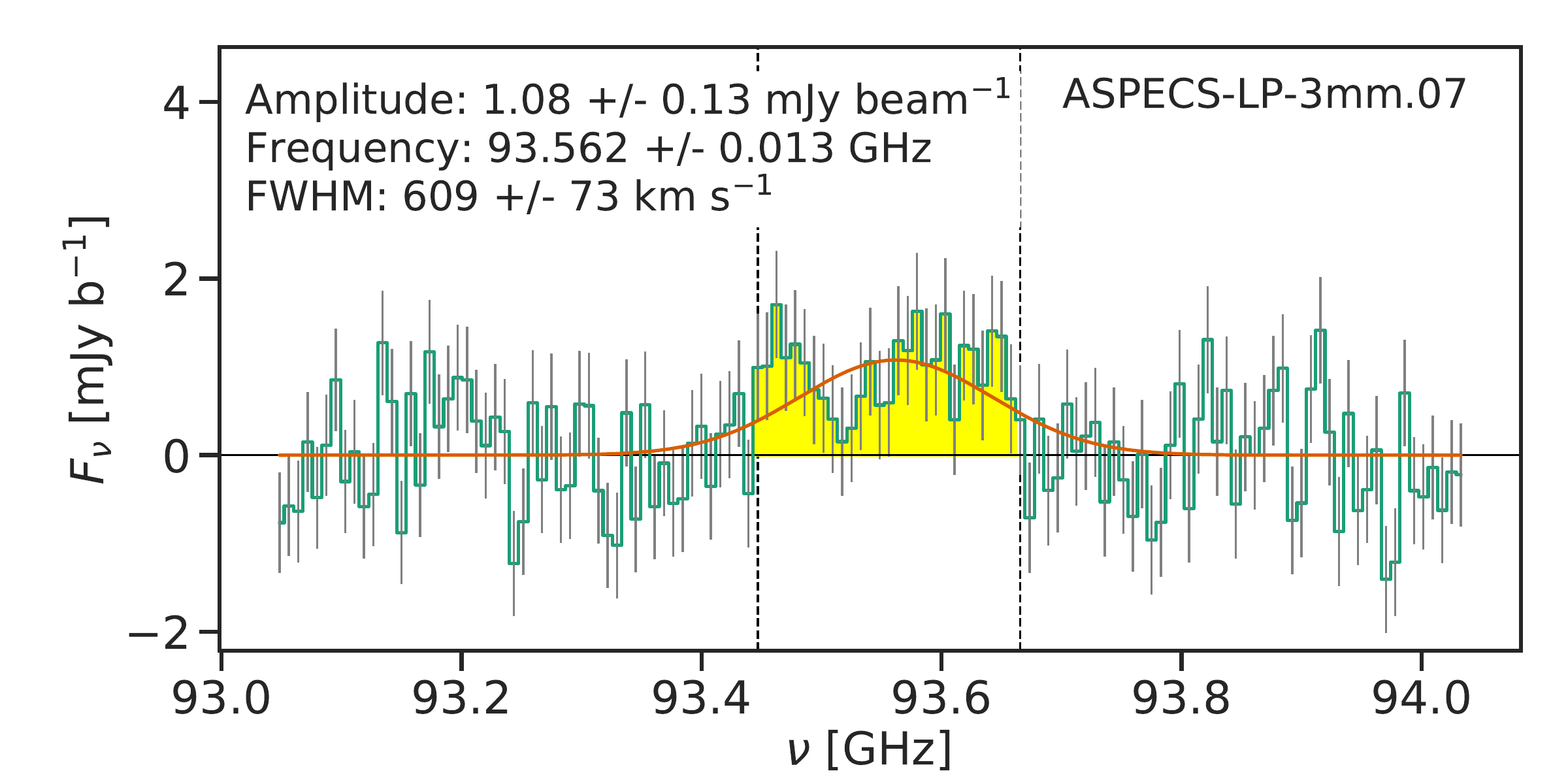}
\epsscale{0.37}
\plotone{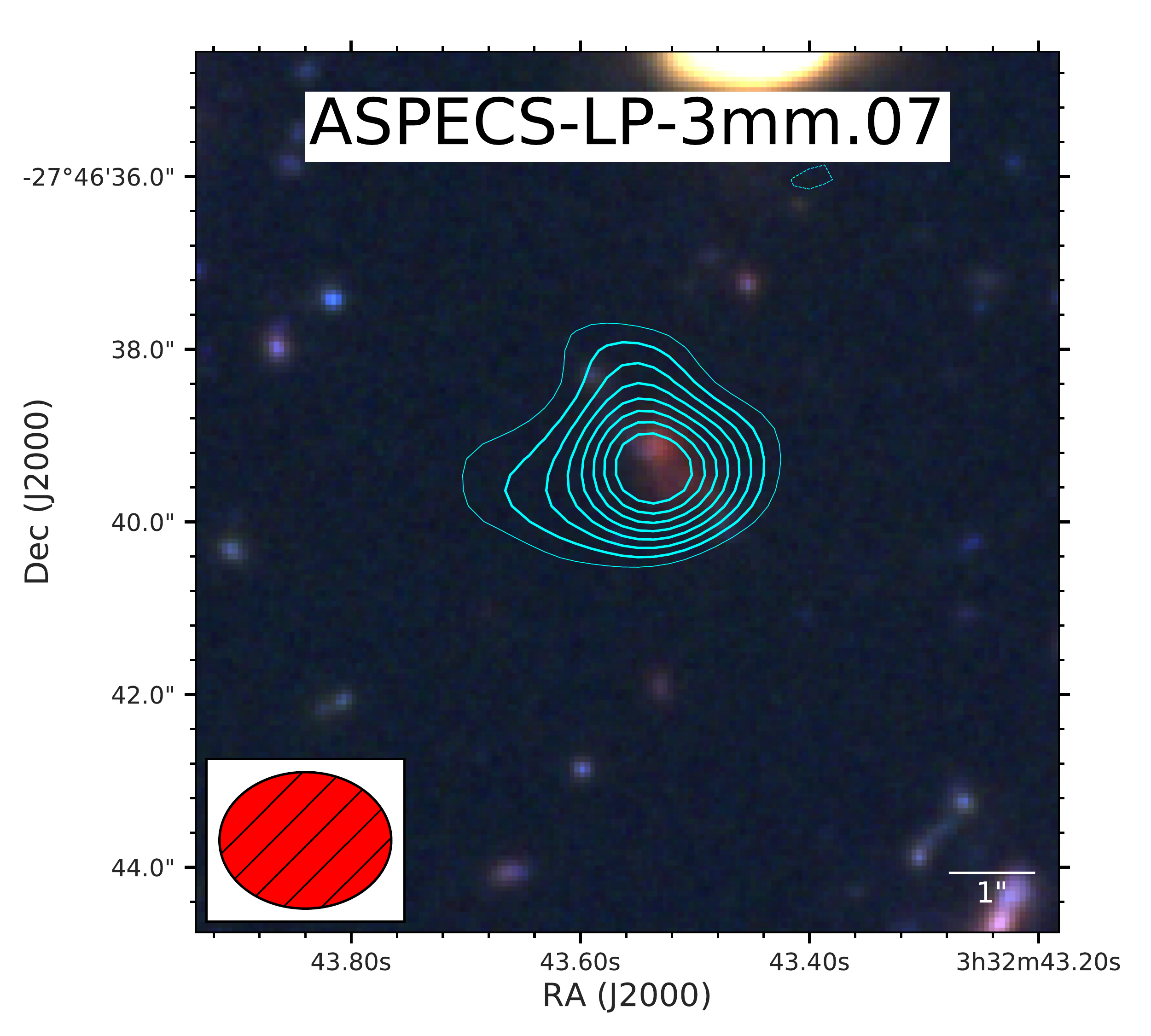}

\epsscale{0.6}
\plotone{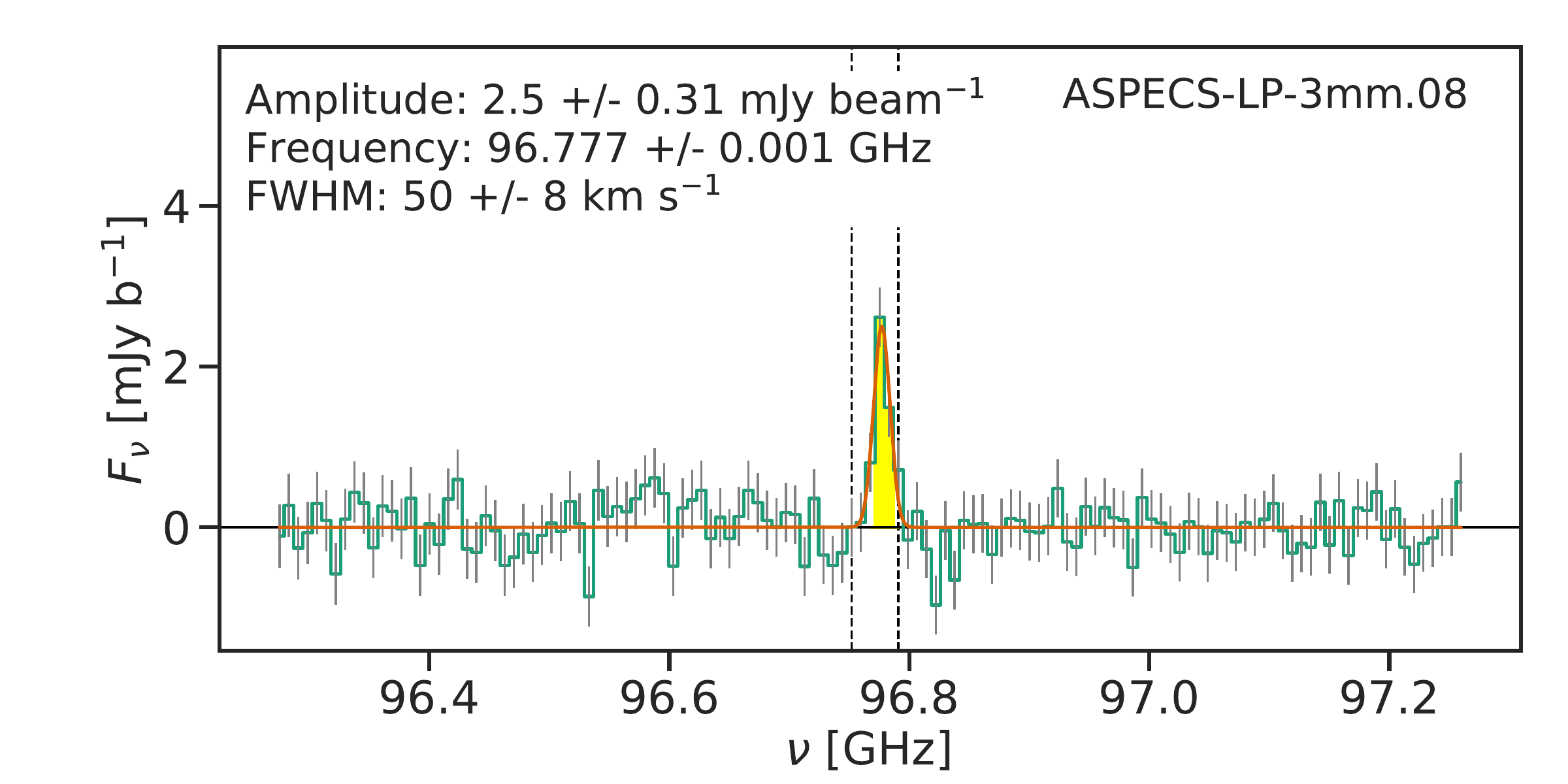}
\epsscale{0.37}
\plotone{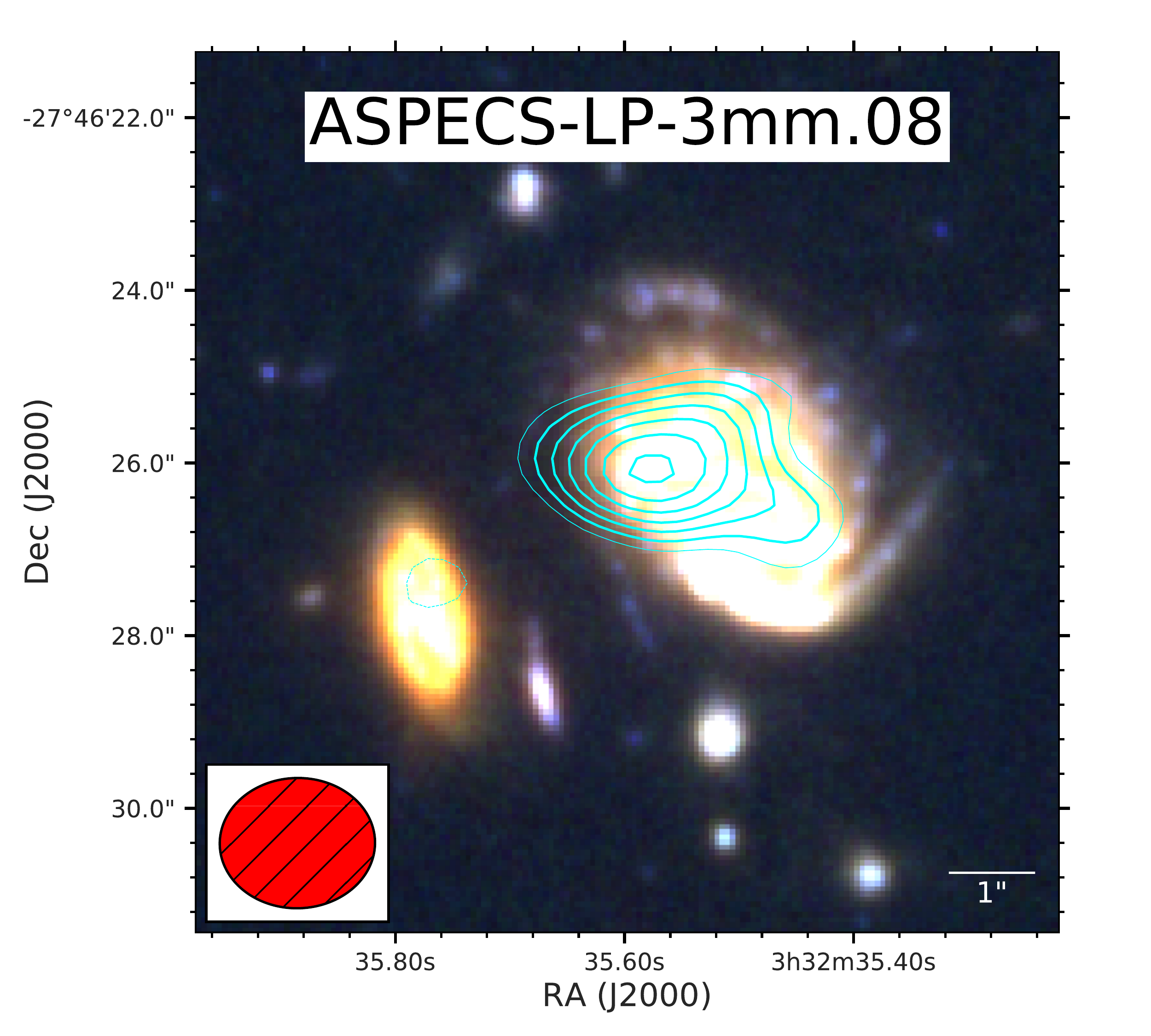}

\caption{Continuation from Fig. \ref{fig:LP_spectra_postamp1}.\label{fig:LP_spectra_postamp2}}
\end{figure*}

\begin{figure*}
\epsscale{1.1}
\epsscale{0.6}
\plotone{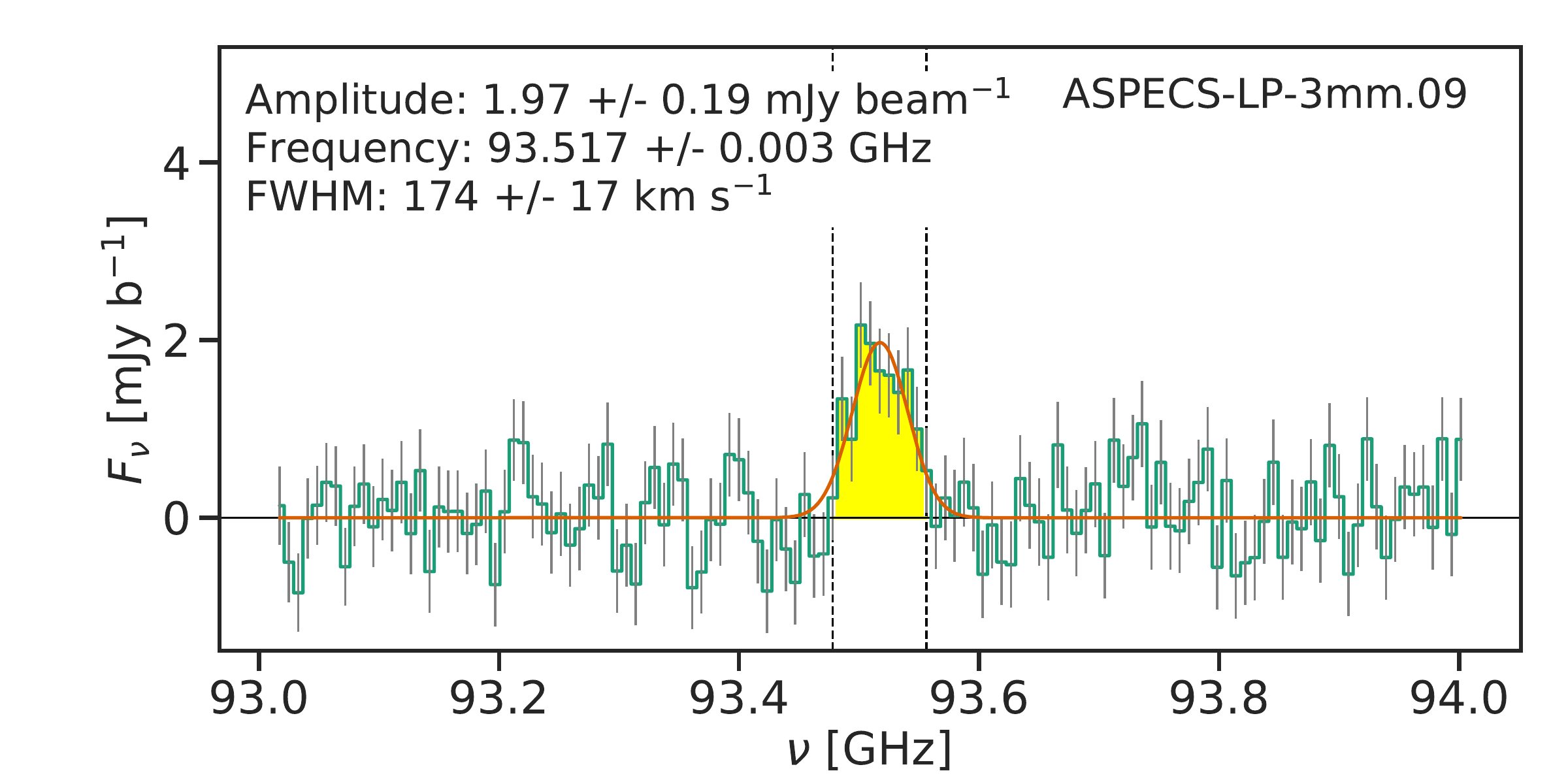}
\epsscale{0.37}
\plotone{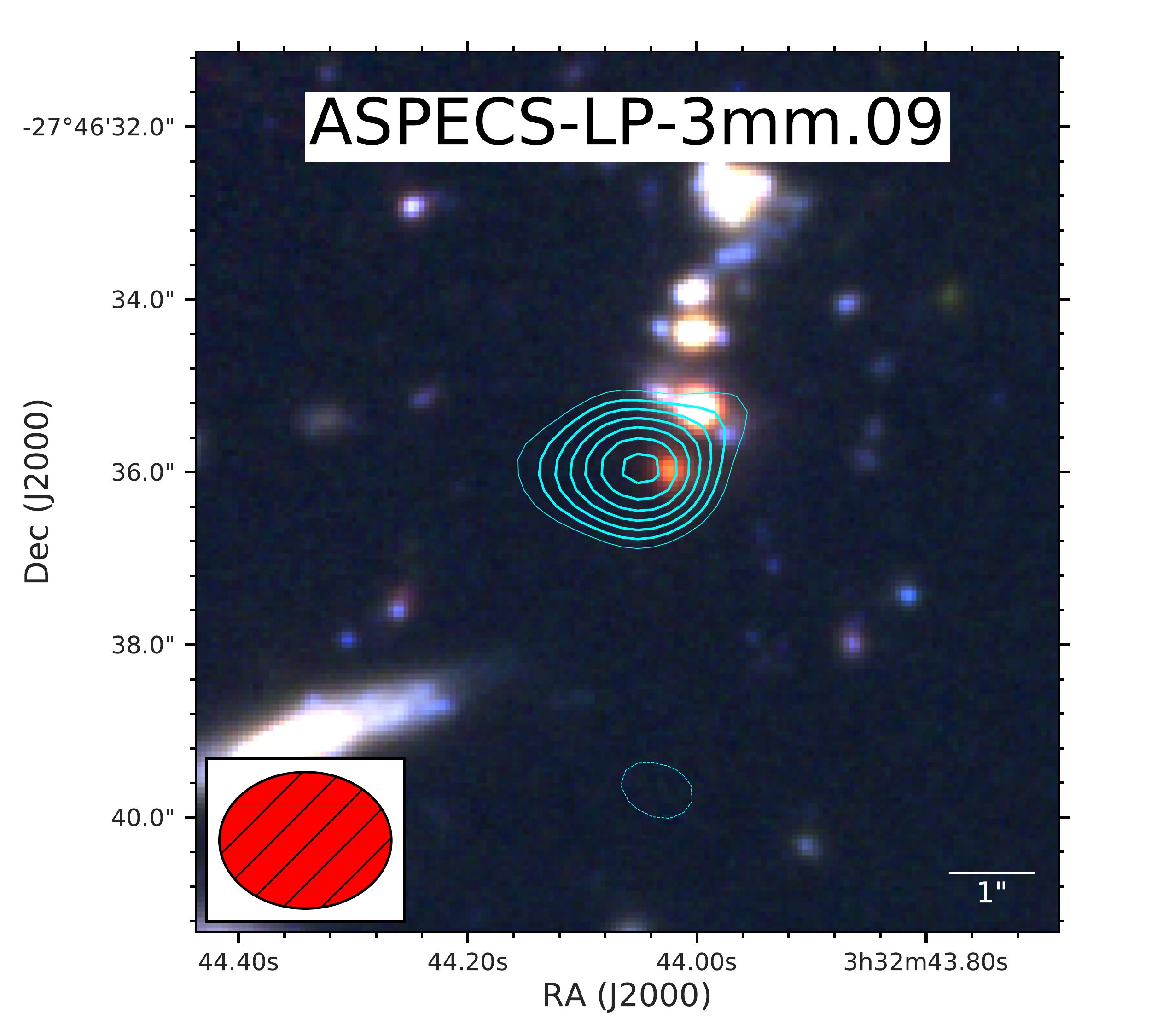}

\epsscale{0.6}
\plotone{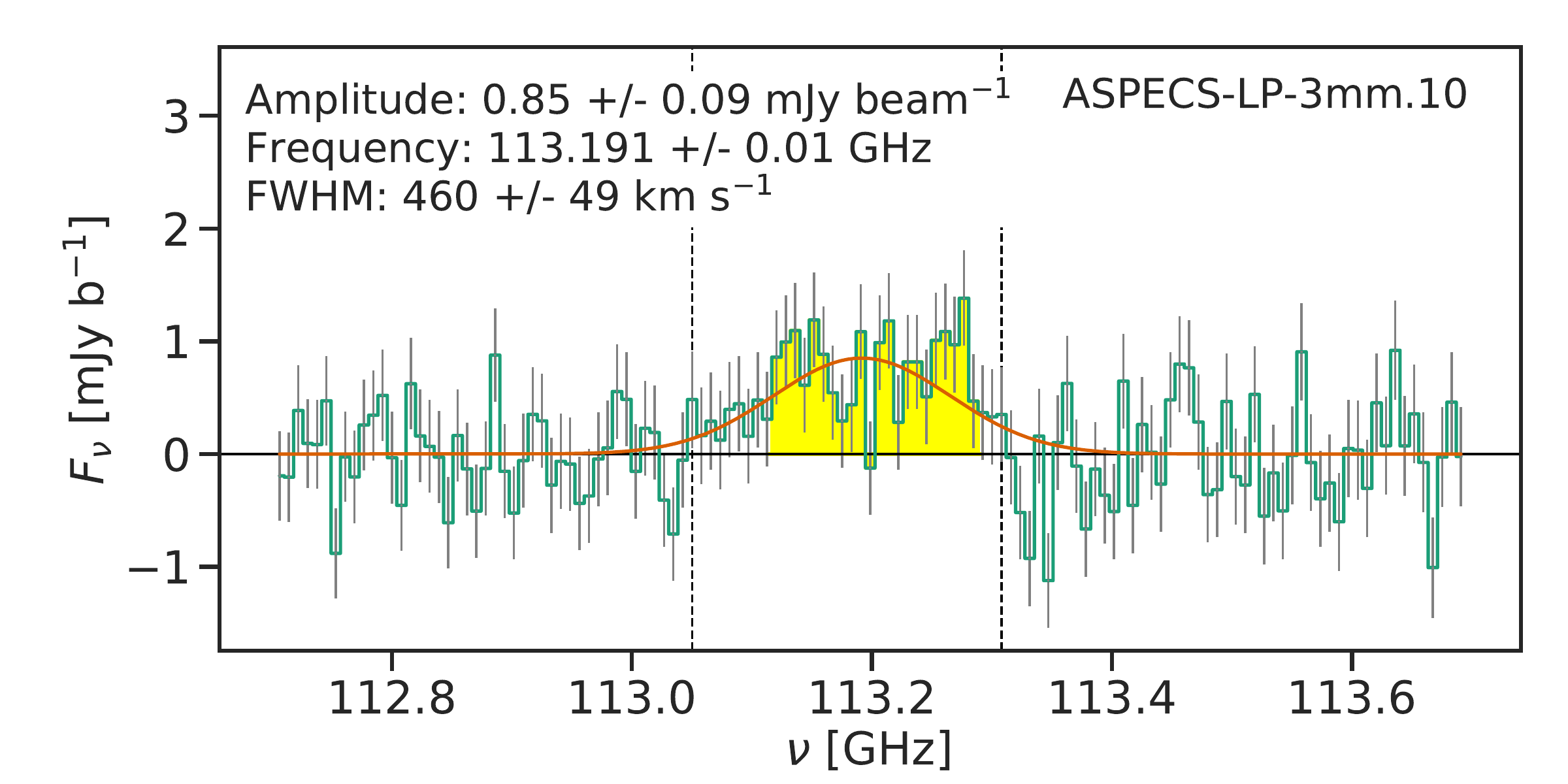}
\epsscale{0.37}
\plotone{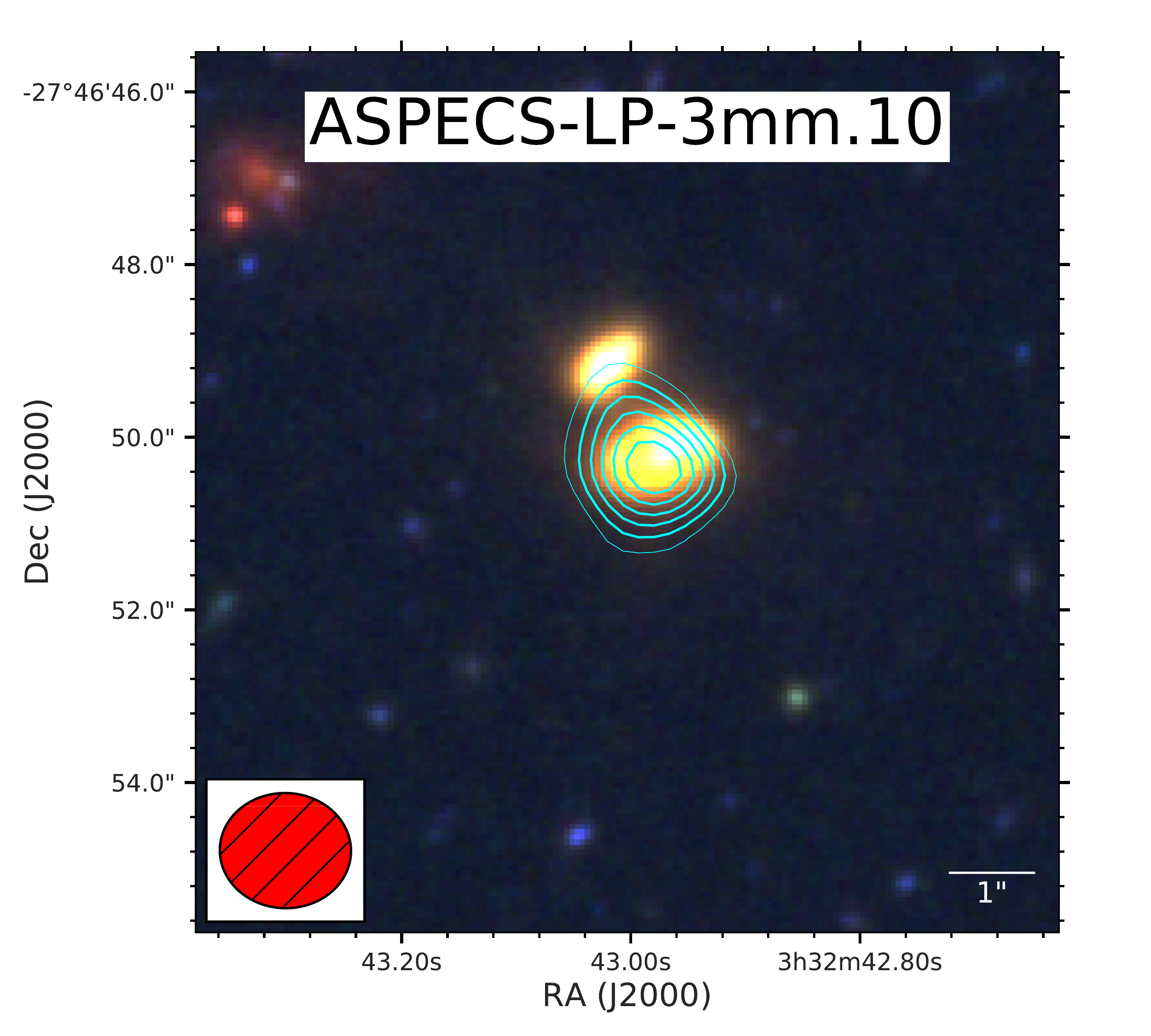}

\epsscale{0.6}
\plotone{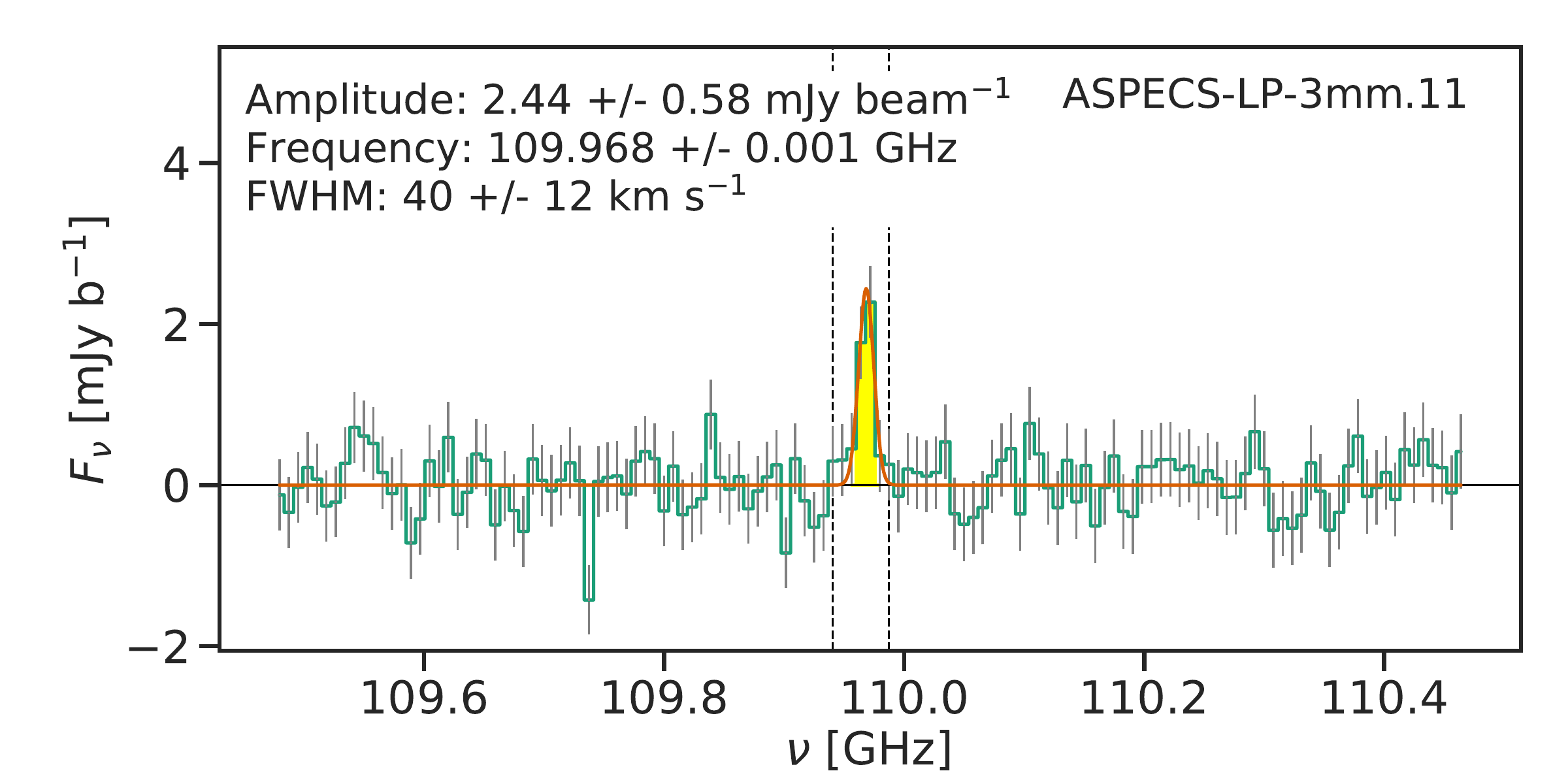}
\epsscale{0.37}
\plotone{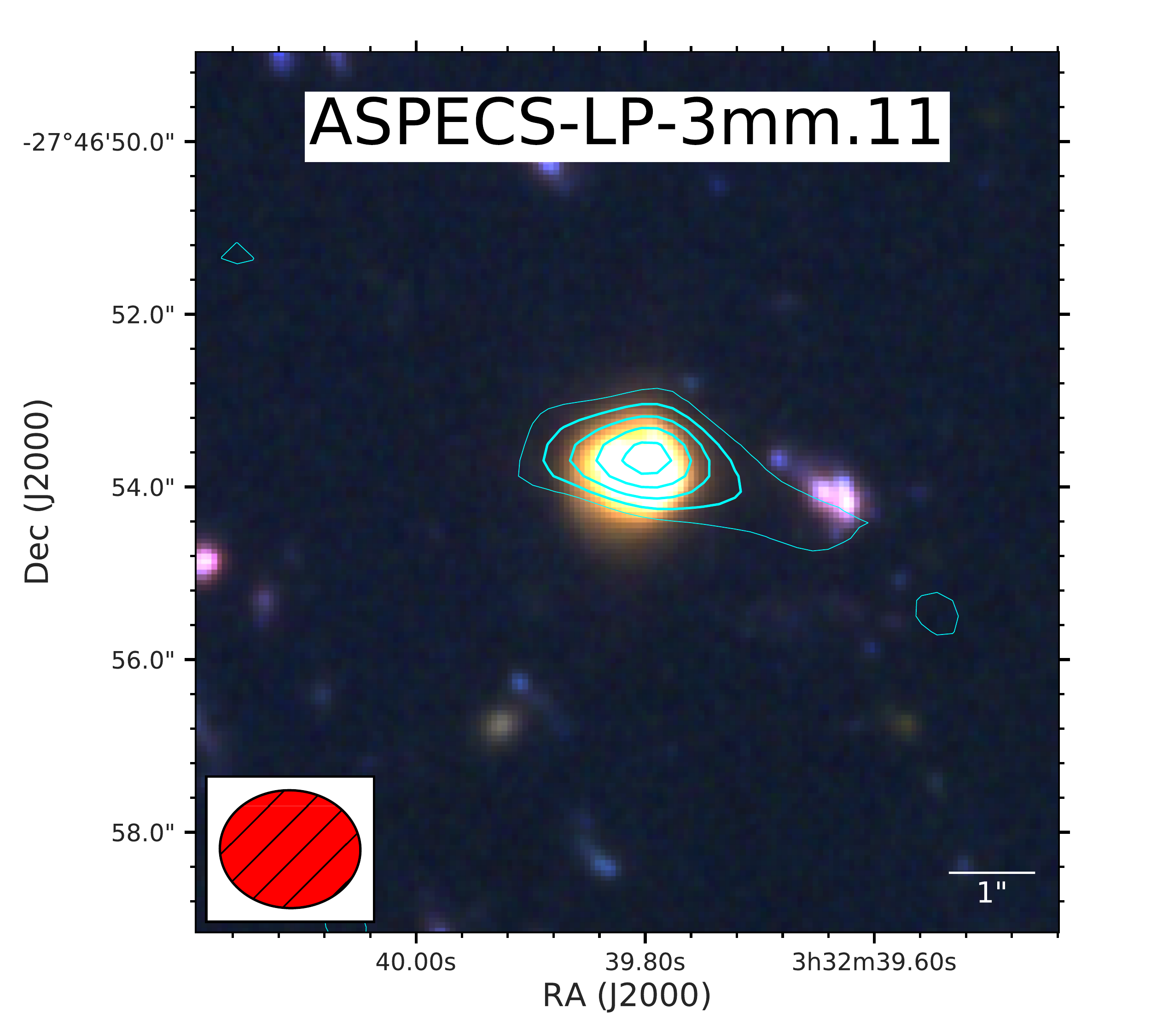}

\epsscale{0.6}
\plotone{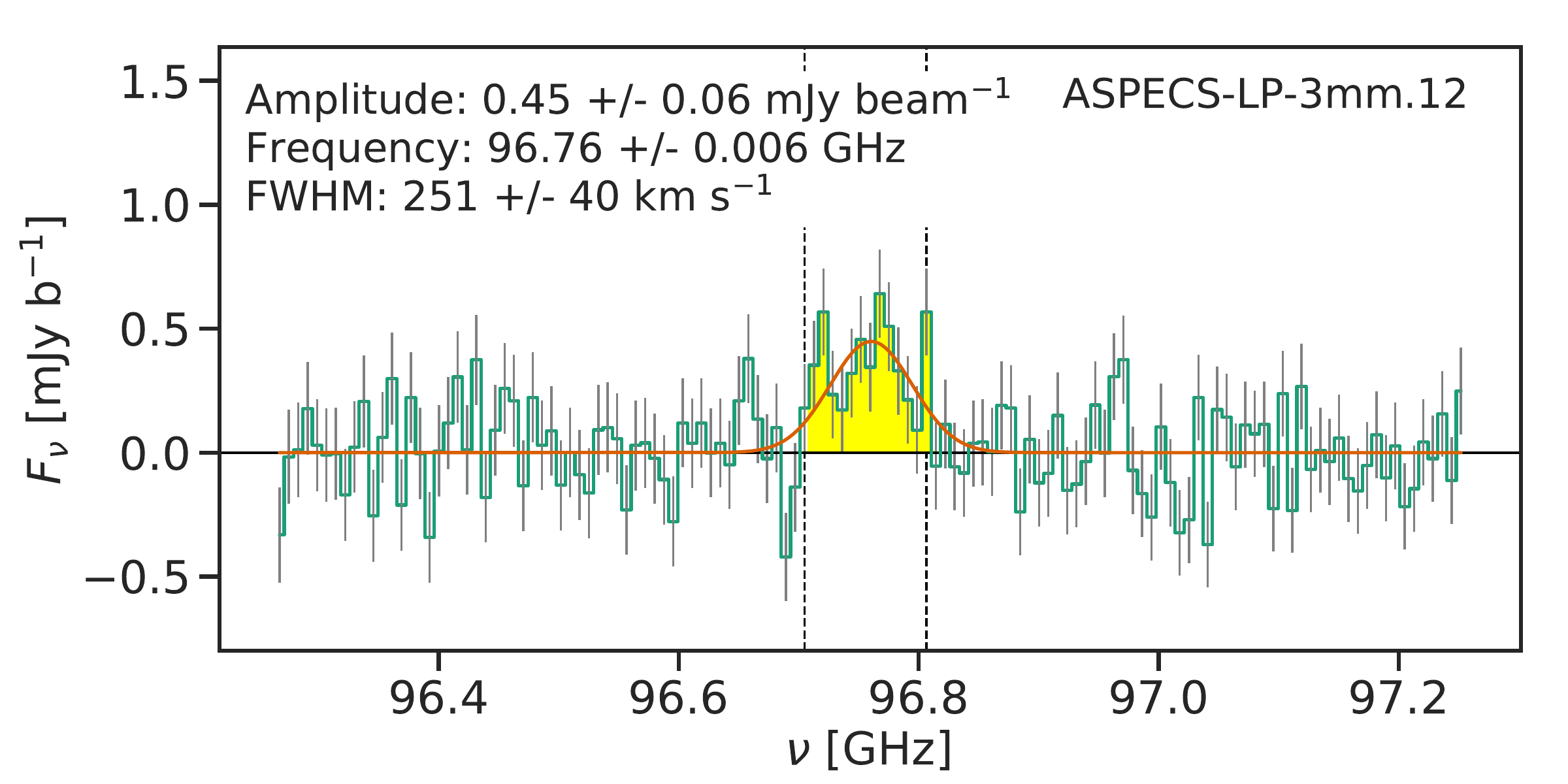}
\epsscale{0.37}
\plotone{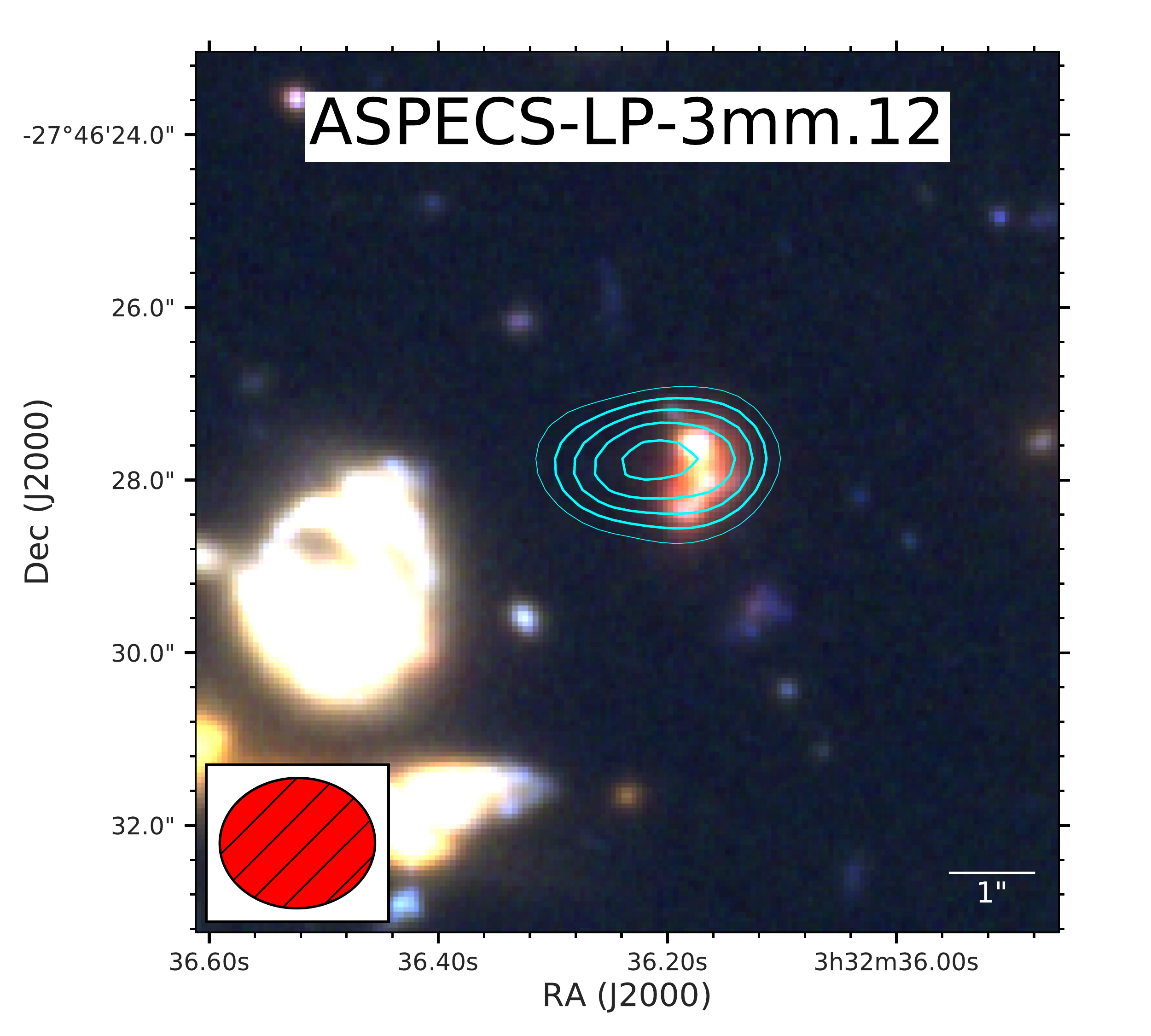}
\caption{Continuation from Fig. \ref{fig:LP_spectra_postamp1}.\label{fig:LP_spectra_postamp3}}
\end{figure*}

\begin{figure*}
\epsscale{1.1}
\epsscale{0.6}
\plotone{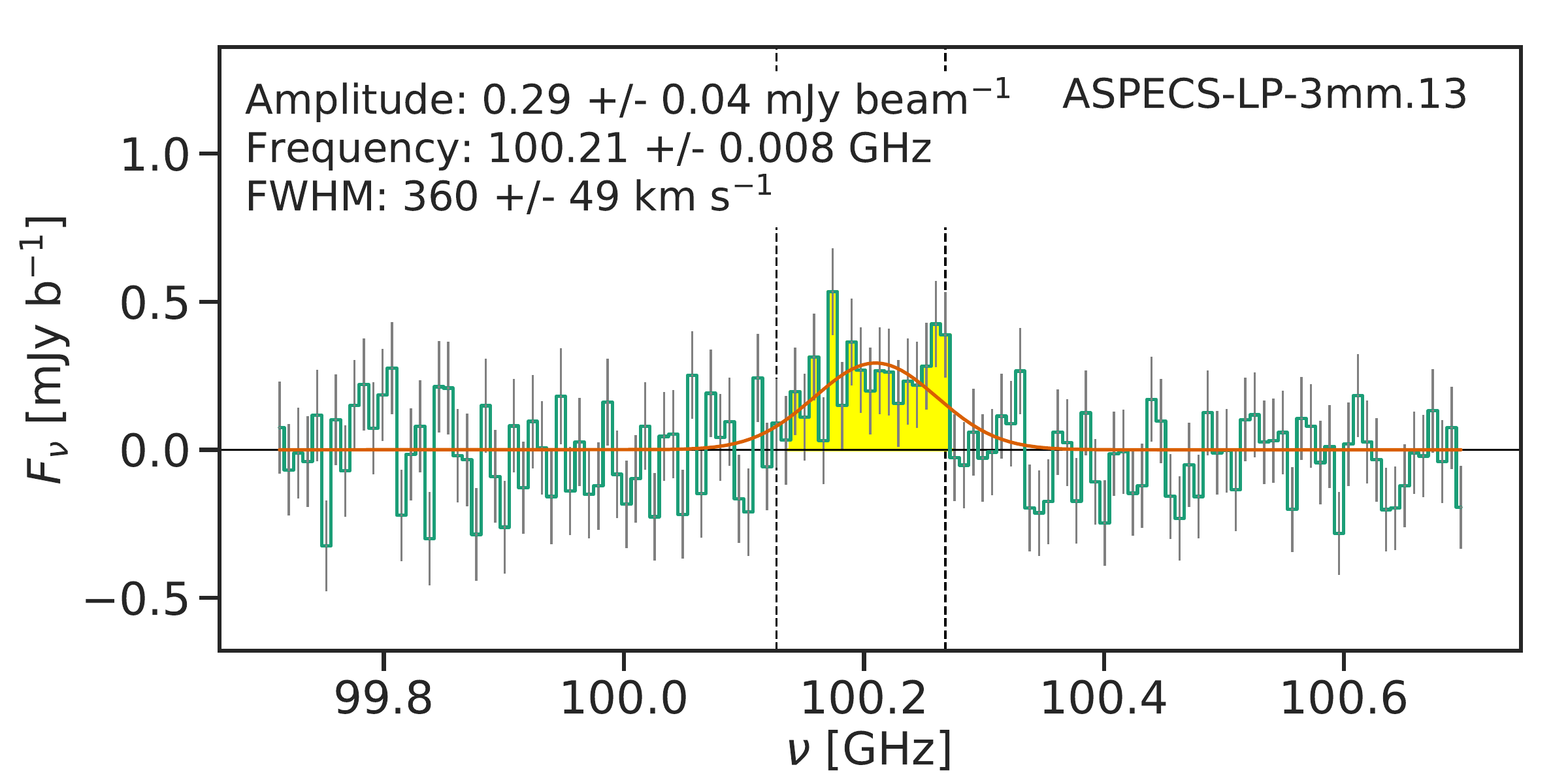}
\epsscale{0.37}
\plotone{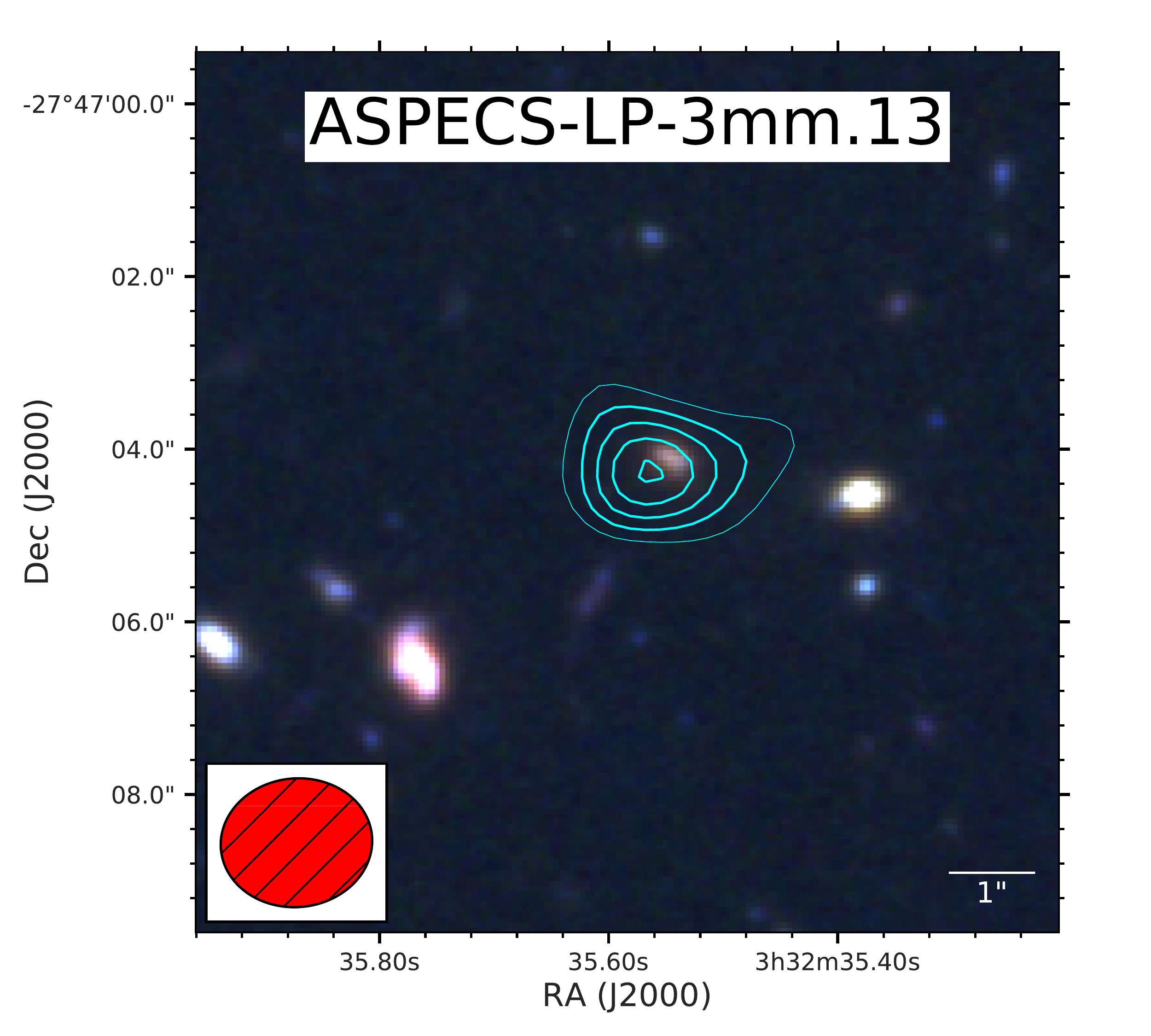}

\epsscale{0.6}
\plotone{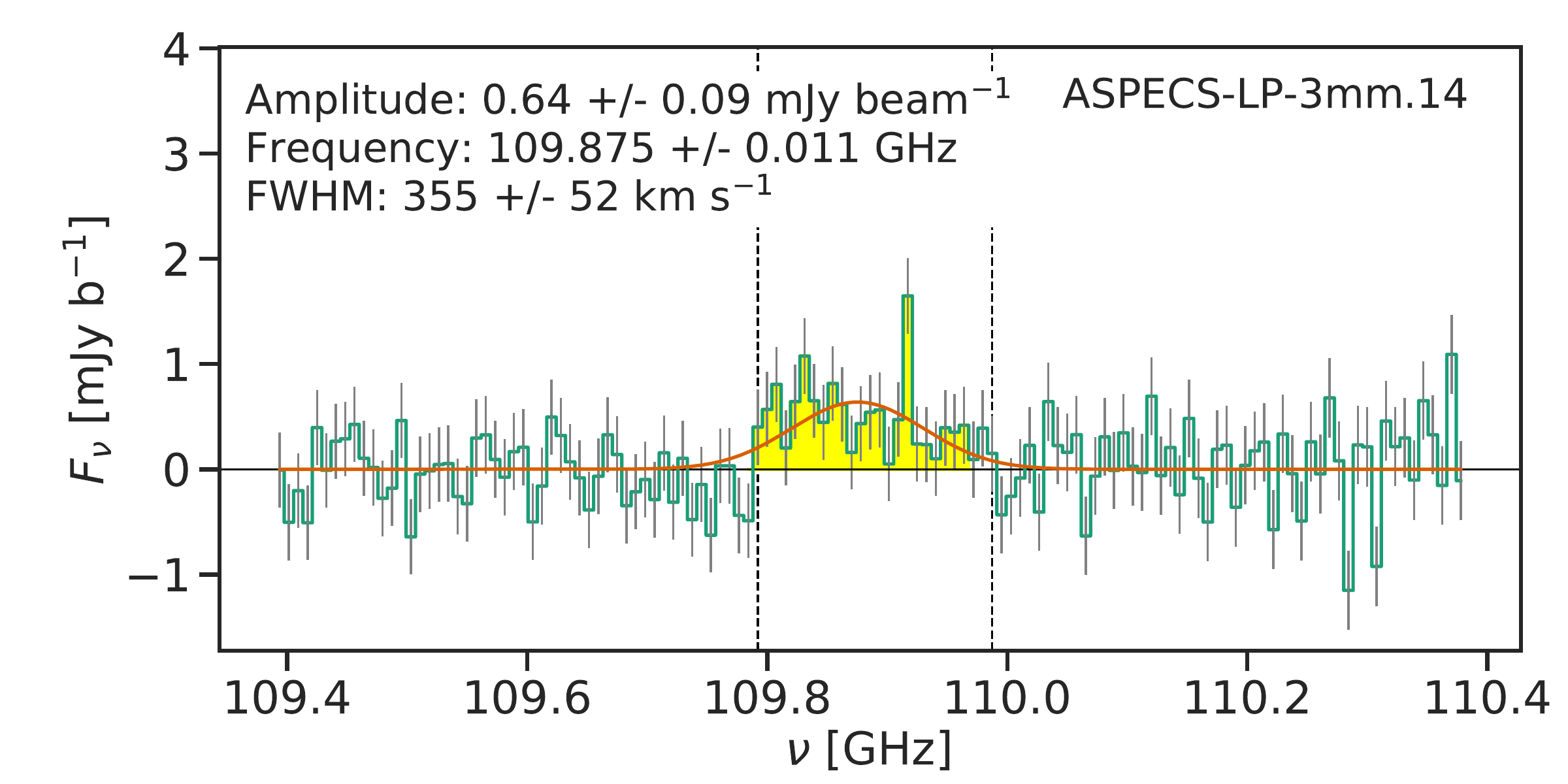}
\epsscale{0.37}
\plotone{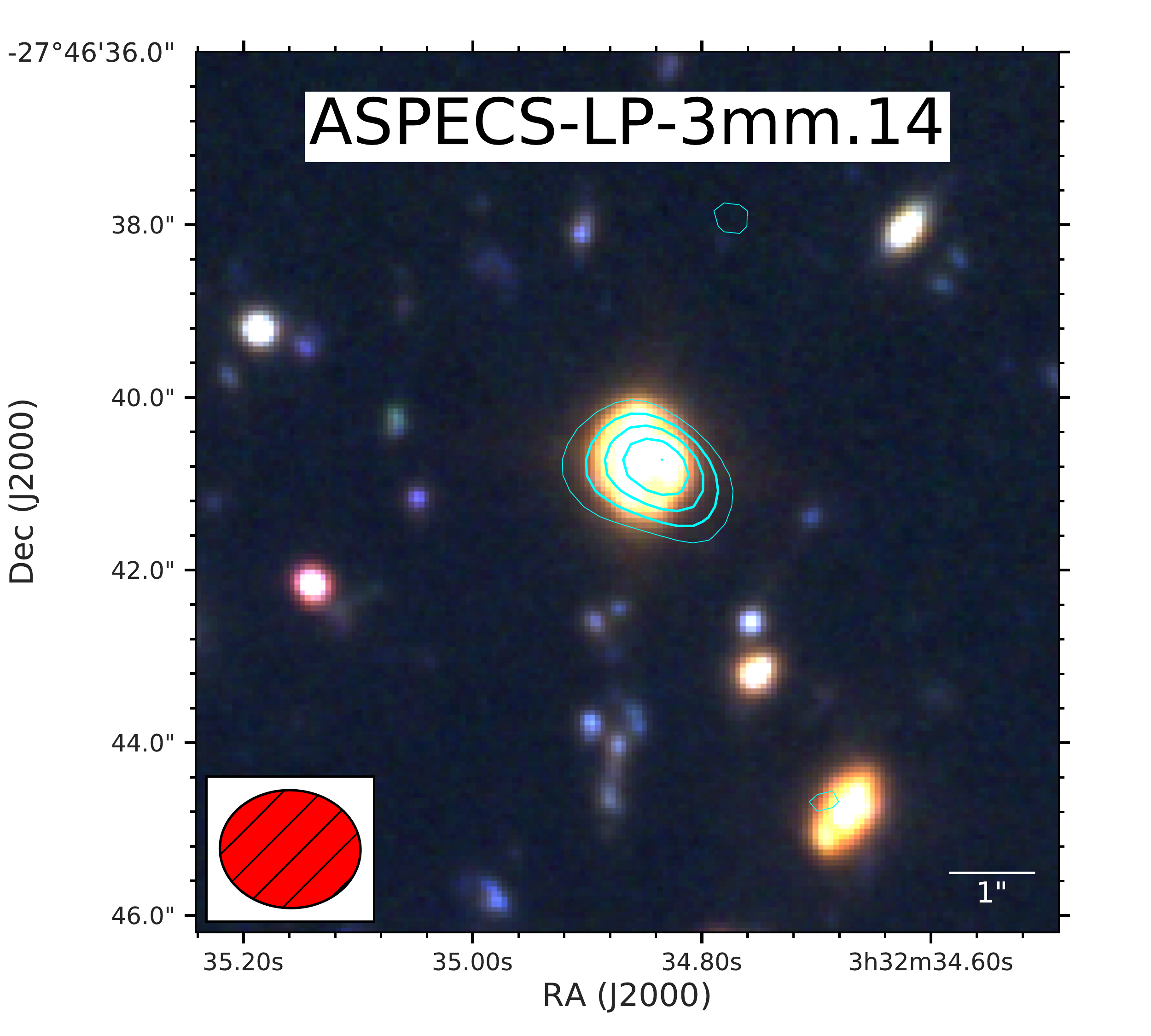}

\epsscale{0.6}
\plotone{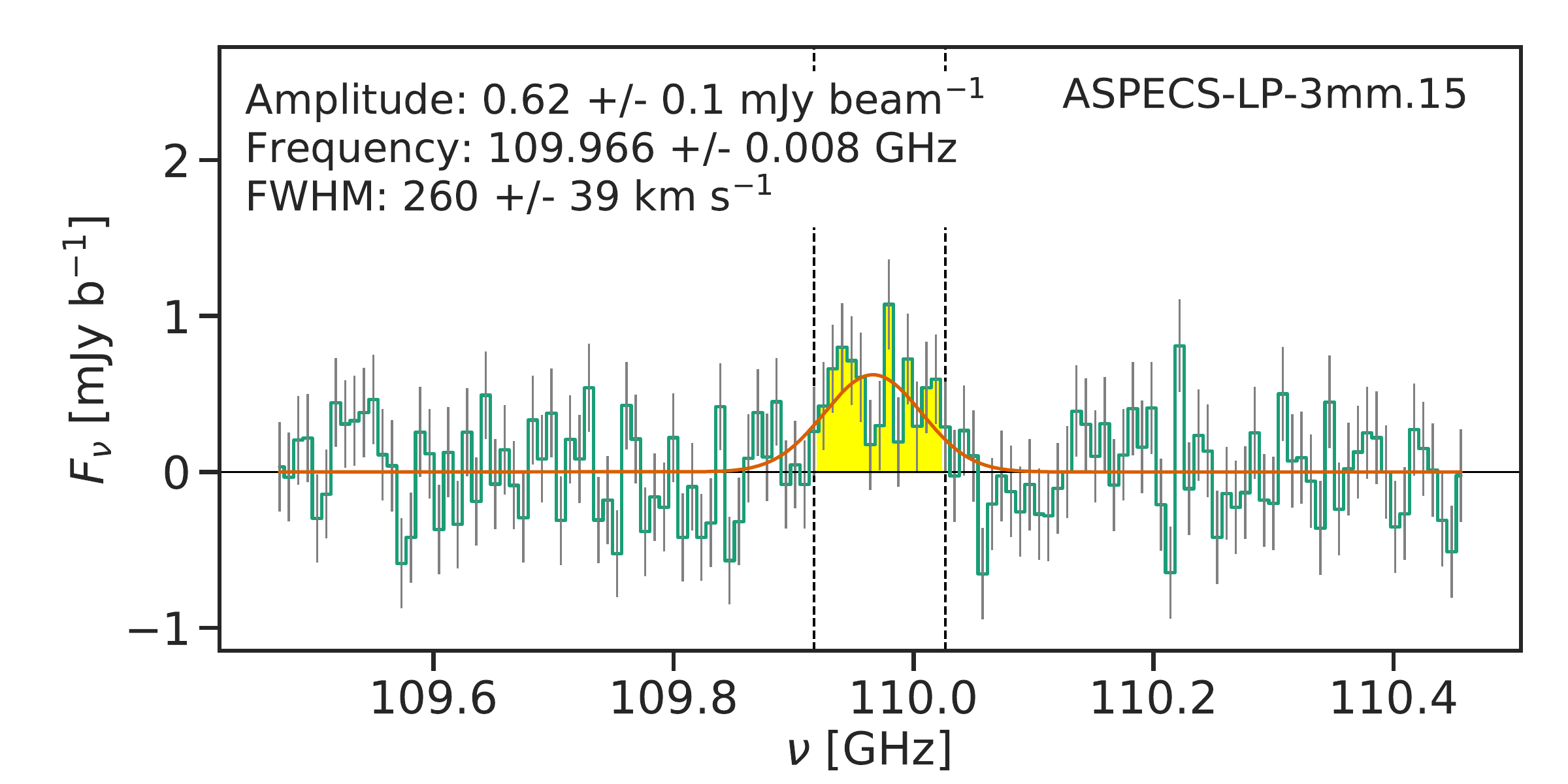}
\epsscale{0.37}
\plotone{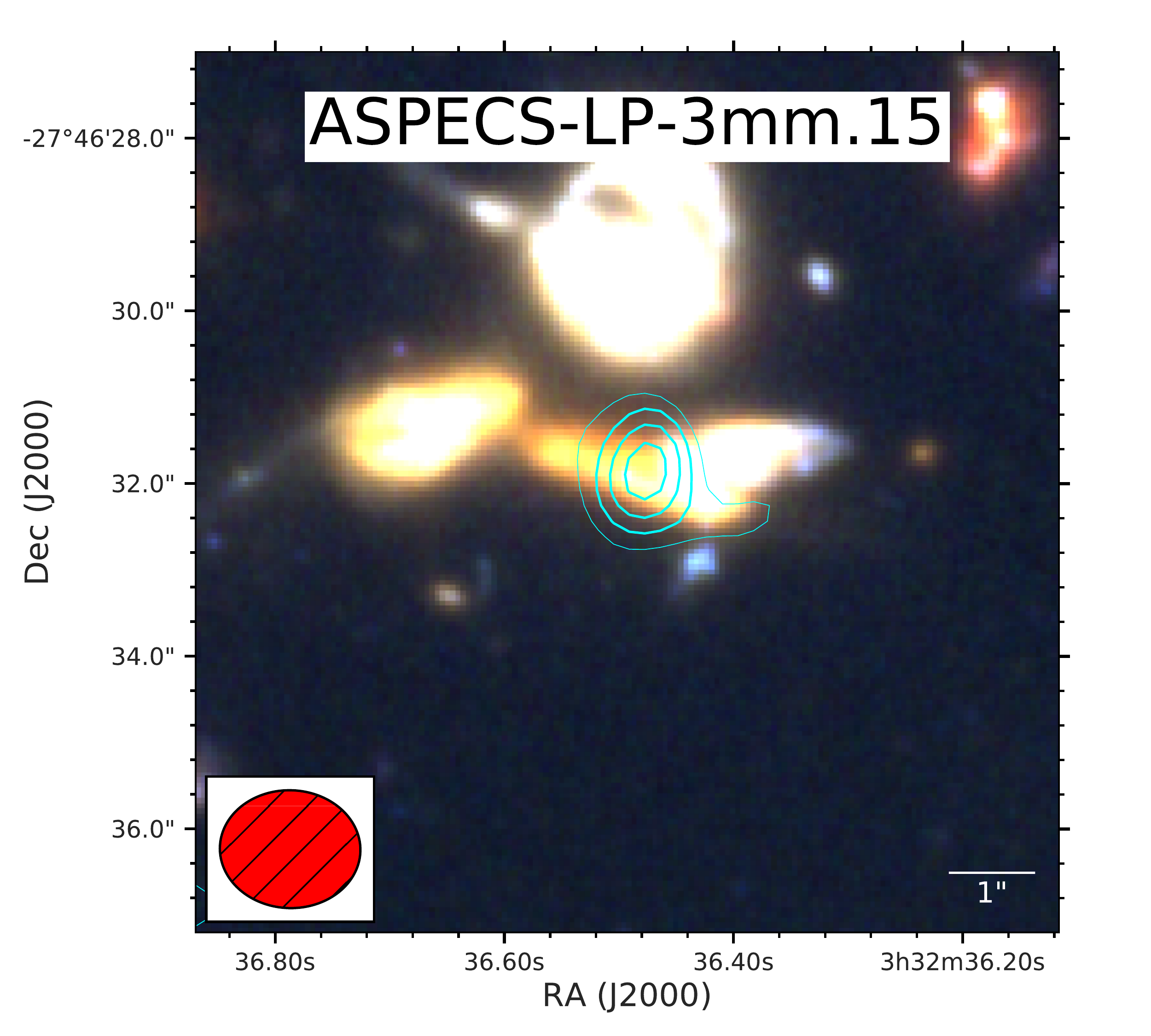}

\epsscale{0.6}
\plotone{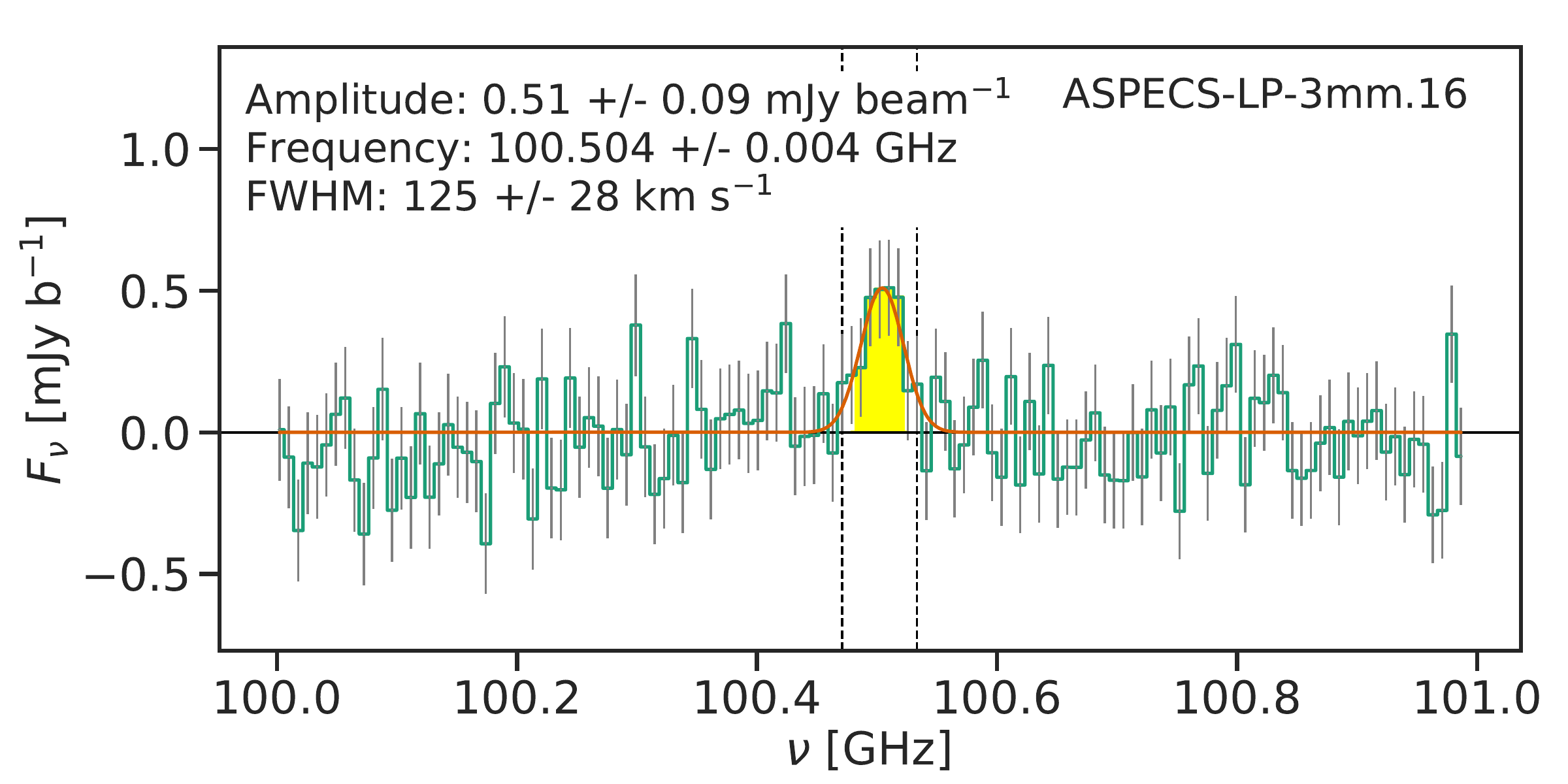}
\epsscale{0.37}
\plotone{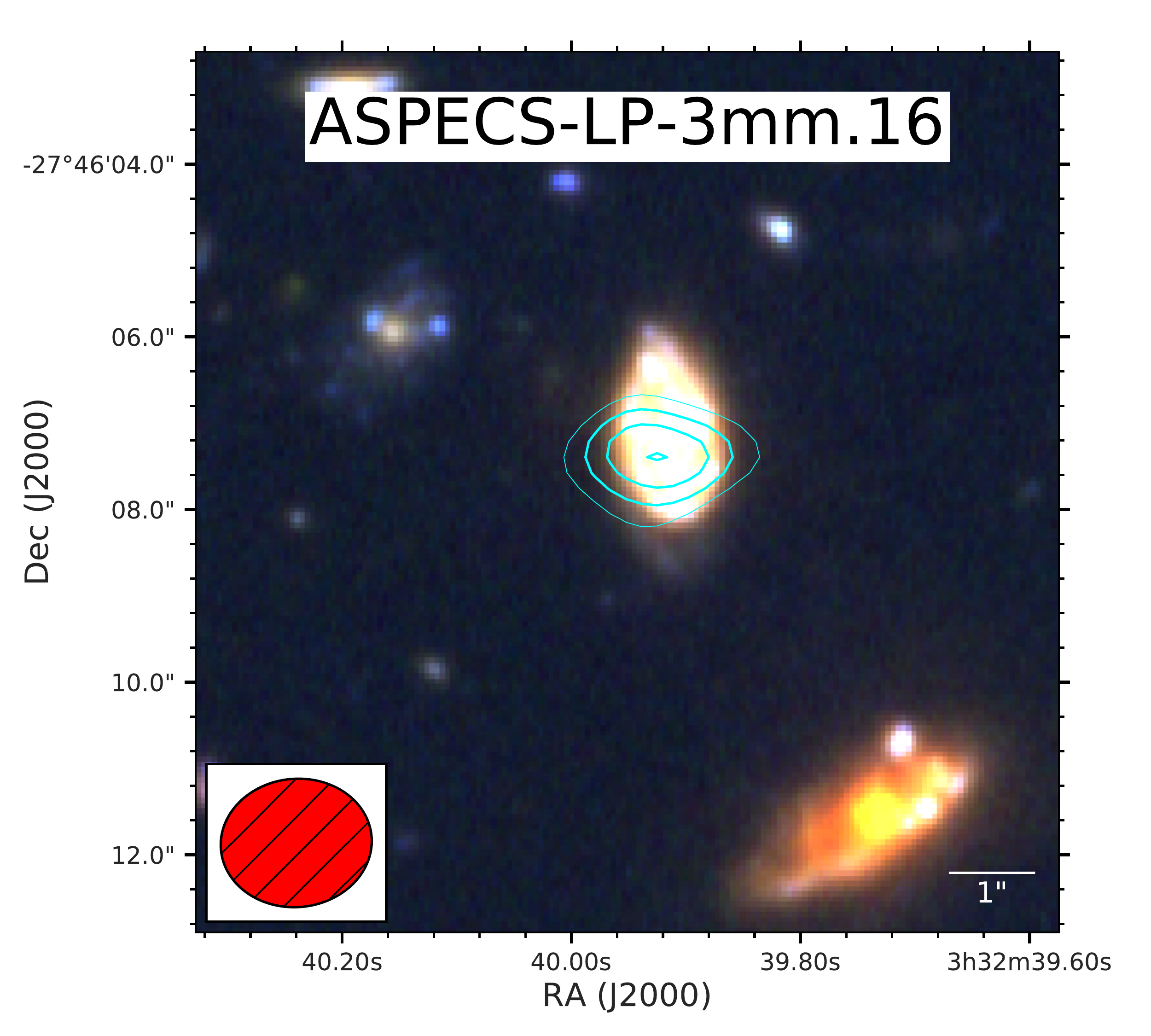}
\caption{Continuation from fig. \ref{fig:LP_spectra_postamp1}.\label{fig:LP_spectra_postamp4}}
\end{figure*}

\begin{deluxetable*}{cccccc}
\tablecaption{Emission line candidates in ASPECS-LP band 3 cube. \label{tab:LP}}
\tablehead{
\colhead{ID} & 
\colhead{R.A.} & 
\colhead{Dec} & 
\colhead{Freq.} & 
\colhead{S/N} & 
\colhead{{\rm Fidelity}} \\
\colhead{} & 
\colhead{} & 
\colhead{} & 
\colhead{[GHz]} & 
\colhead{} & 
\colhead{} 
}
\colnumbers
\startdata
ASPECS-LP-3mm.01 & 03:32:38.54 & -27:46:34.62 & 97.58 & 37.7 & $1.0_{-0.0}^{+0.0}$\\
ASPECS-LP-3mm.02 & 03:32:42.38 & -27:47:07.92 & 99.51 & 17.9 & $1.0_{-0.0}^{+0.0}$\\
ASPECS-LP-3mm.03 & 03:32:41.02 & -27:46:31.56 & 100.135 & 15.8 & $1.0_{-0.0}^{+0.0}$\\
ASPECS-LP-3mm.04 & 03:32:34.44 & -27:46:59.82 & 95.502 & 15.5 & $1.0_{-0.0}^{+0.0}$\\
ASPECS-LP-3mm.05 & 03:32:39.76 & -27:46:11.58 & 90.4 & 15.0 & $1.0_{-0.0}^{+0.0}$\\
ASPECS-LP-3mm.06 & 03:32:39.90 & -27:47:15.12 & 110.026 & 11.9 & $1.0_{-0.0}^{+0.0}$\\
ASPECS-LP-3mm.07 & 03:32:43.53 & -27:46:39.47 & 93.548 & 10.9 & $1.0_{-0.0}^{+0.0}$\\
ASPECS-LP-3mm.08 & 03:32:35.58 & -27:46:26.16 & 96.775 & 9.5 & $1.0_{-0.0}^{+0.0}$\\
ASPECS-LP-3mm.09 & 03:32:44.03 & -27:46:36.05 & 93.517 & 9.3 & $1.0_{-0.0}^{+0.0}$\\
ASPECS-LP-3mm.10 & 03:32:42.98 & -27:46:50.45 & 113.199 & 8.7 & $1.0_{-0.0}^{+0.0}$\\
ASPECS-LP-3mm.11 & 03:32:39.80 & -27:46:53.70 & 109.972 & 7.9 & $1.0_{-0.0}^{+0.0}$\\
ASPECS-LP-3mm.12 & 03:32:36.21 & -27:46:27.78 & 96.76 & 7.0 & $1.0_{-0.0}^{+0.0}$\\
ASPECS-LP-3mm.13 & 03:32:35.56 & -27:47:04.32 & 100.213 & 6.8 & $1.0_{-0.0}^{+0.0}$\\
ASPECS-LP-3mm.14 & 03:32:34.84 & -27:46:40.74 & 109.886 & 6.7 & $1.0_{-0.0}^{+0.0}$\\
ASPECS-LP-3mm.15 & 03:32:36.48 & -27:46:31.92 & 109.964 & 6.5 & $0.99_{-0.0}^{+0.0}$\\
ASPECS-LP-3mm.16 & 03:32:39.92 & -27:46:07.44 & 100.502 & 6.4 & $0.92_{-0.02}^{+0.02}$\\
\enddata
\tablecomments{
(1) Identification for emission line candidates discovered in ASPECS-LP. 
(2) Right ascension (J2000).
(3) Declination (J2000).
(4) Central frequency of the line.
(5) S/N value return by LineSeeker in ASPECS-Pilot assuming an unresolved source.
(6) Fidelity estimate using negative detection.
}
\end{deluxetable*}

\begin{deluxetable*}{ccccccc}
\tablecaption{Properties and identification for the selected emission lines candidates in the ASPECS-LP band 3 cube. \label{tab:LP_Parameters}}
\tablehead{
\colhead{ID} & 
\colhead{Central Frequency} & 
\colhead{Peak} & 
\colhead{FWHM} & 
\colhead{Integrated Flux} & 
\colhead{Size} & 
\colhead{Identification} \\
\colhead{} & 
\colhead{[GHz]} & 
\colhead{[mJy beam$^{-1}$]} & 
\colhead{[\kms]} & 
\colhead{[Jy\kms]} & 
\colhead{} & 
\colhead{}  
}
\colnumbers
\startdata
ASPECS-LP-3mm.01 & $97.584 \pm 0.003$ & $1.71 \pm 0.06$ & $517.0 \pm 21.0$ & $1.02 \pm 0.04$ & EXT & CO(3-2), $z_{\rm CO}=2.543$\\
ASPECS-LP-3mm.02 & $99.51 \pm 0.005$ & $1.38 \pm 0.11$ & $277.0 \pm 26.0$ & $0.47 \pm 0.04$ & EXT & CO(2-1), $z_{\rm CO}=1.317$\\
ASPECS-LP-3mm.03 & $100.131 \pm 0.005$ & $0.88 \pm 0.08$ & $368.0 \pm 37.0$ & $0.41 \pm 0.04$ & EXT & CO(3-2), $z_{\rm CO}=2.454$\tablenotemark{a}\\
ASPECS-LP-3mm.04 & $95.501 \pm 0.006$ & $1.44 \pm 0.13$ & $498.0 \pm 47.0$ & $0.89 \pm 0.07$ & EXT & CO(2-1), $z_{\rm CO}=1.414$\\
ASPECS-LP-3mm.05 & $90.393 \pm 0.006$ & $0.96 \pm 0.1$ & $617.0 \pm 58.0$ & $0.66 \pm 0.06$ & EXT & CO(2-1), $z_{\rm CO}=1.550$\\
ASPECS-LP-3mm.06 & $110.038 \pm 0.005$ & $1.22 \pm 0.13$ & $307.0 \pm 33.0$ & $0.48 \pm 0.06$ & EXT & CO(2-1), $z_{\rm CO}=1.095$\\
ASPECS-LP-3mm.07 & $93.558 \pm 0.008$ & $1.08 \pm 0.13$ & $609.0 \pm 73.0$ & $0.76 \pm 0.09$ & EXT & CO(3-2), $z_{\rm CO}=2.696$\tablenotemark{b}\\
ASPECS-LP-3mm.08 & $96.778 \pm 0.002$ & $2.5 \pm 0.31$ & $50.0 \pm 8.0$ & $0.16 \pm 0.03$ & EXT & CO(2-1),$z_{\rm CO}=1.382$\\
ASPECS-LP-3mm.09 & $93.517 \pm 0.003$ & $1.97 \pm 0.19$ & $174.0 \pm 17.0$ & $0.4 \pm 0.04$ & PS & CO(3-2), $z_{\rm CO}=2.698$\tablenotemark{c}\\
ASPECS-LP-3mm.10 & $113.192 \pm 0.009$ & $0.85 \pm 0.09$ & $460.0 \pm 49.0$ & $0.59 \pm 0.07$ & PS & CO(2-1), $z_{\rm CO}=1.037$\\
ASPECS-LP-3mm.11 & $109.966 \pm 0.003$ & $2.44 \pm 0.58$ & $40.0 \pm 12.0$ & $0.16 \pm 0.03$ & EXT & CO(2-1), $z_{\rm CO}=1.096$\\
ASPECS-LP-3mm.12 & $96.757 \pm 0.004$ & $0.45 \pm 0.06$ & $251.0 \pm 40.0$ & $0.14 \pm 0.02$ & PS & CO(2-1), $z_{\rm CO}=1.383$\tablenotemark{d}\\
ASPECS-LP-3mm.13 & $100.209 \pm 0.006$ & $0.29 \pm 0.04$ & $360.0 \pm 49.0$ & $0.13 \pm 0.02$ & PS & CO(4-3), $z_{\rm CO}=3.601$\tablenotemark{e}\\
ASPECS-LP-3mm.14 & $109.877 \pm 0.009$ & $0.64 \pm 0.09$ & $355.0 \pm 52.0$ & $0.35 \pm 0.05$ & PS & CO(2-1), $z_{\rm CO}=1.098$\\
ASPECS-LP-3mm.15 & $109.971 \pm 0.005$ & $0.62 \pm 0.1$ & $260.0 \pm 39.0$ & $0.21 \pm 0.03$ & PS & CO(2-1), $z_{\rm CO}=1.096$\\
ASPECS-LP-3mm.16 & $100.503 \pm 0.004$ & $0.51 \pm 0.09$ & $125.0 \pm 28.0$ & $0.08 \pm 0.01$ & PS & CO(2-1), $z_{\rm CO}=1.294$\\
\enddata
\tablenotetext{a}{Based on $z_{\rm ph}=2.553$}
\tablenotetext{b}{Based on $z_{\rm ph}=2.914$}
\tablenotetext{c}{Based on $z_{\rm ph}=2.983$}
\tablenotetext{d}{Based on $z_{\rm ph}=1.098$}
\tablenotetext{e}{Based on $z_{\rm ph}=3.400$}
\tablecomments{
(1) Identification for emission line candidates discovered in ASPECS-LP. 
(2) Central frequency based on the first moment of the line.
(3) Peak of the line of the best-fit Gaussian profile.
(4) FWHM of the best-fit Gaussian profile.
(5) Integrated line flux obtained by integrating the channels within the vertical dashed-lines in Fig. \ref{fig:LP_spectra_postamp1}.
(6) Size of the emission line candidates. EXT corresponds to resolved lines while PS to emission lines consisting with being a point source.
(7) Identification of the emission line together with the assumed redshift.
}
\end{deluxetable*}

In table \ref{tab:LP}, we present the list of reliable emission--line candidates in the band 3 cube of the ASPECS-LP. We present all the 16 emission--line candidates for which the condition of ${\rm Fidelity}\geq0.9$ is fulfilled. Based on our comparison between the ASPECS-Pilot candidates and the ASPECS-LP data, a secure sample would consist of only those candidates that show ${\rm Fidelity}=1$. With this condition we have 15 (out of 16) secure emission--line candidates in the 3 mm ASPECS-LP cube. The fidelity values tell us that out of the sixteen emission line candidates, $\approx15.9$ should be real.

In Table \ref{tab:LP_Parameters}, we present the results from fitting a Gaussian profile to the observed spectra using a Markov chain Monte Carlo (MCMC) sampling algorithm. The median values of the posterior distribution Gaussian profiles as well as the NIR postage stamps are presented in Figs. \ref{fig:LP_spectra_postamp1}, \ref{fig:LP_spectra_postamp2}, \ref{fig:LP_spectra_postamp3} and \ref{fig:LP_spectra_postamp4}. The spectra as well as the fitted properties have been corrected by the primary beam response of the mosaic. Column six from  Table \ref{tab:LP_Parameters} shows whether the line is spatially resolved or not. A line is classified as sptially resolved (EXT) when the total integrated line flux obtained by adding the flux from all the voxels with $S/N\geq2$ in the collapsed line image is at least $10\%$ higher than the ones obtained by taking only the central voxel flux value. Otherwise emission--line candidates are classified as point sources (PS). In case the emission line is classified as resolved, the spectrum and the integrated flux values listed in the table correspond to those values measured in the voxels with $S/N\geq2$ in collapsed line image and not in the central value.

The last column in Table \ref{tab:LP_Parameters} presents the identification of the emission line candidates. Eleven of the candidates have NIR counterpart galaxies with spectroscopic redshifts, allowing for a secure identification of the detected emission lines. In all cases the NIR spectroscopic redshift agrees very well with that obtain from the CO emission line. Five emission lines candidates have clear counterpart galaxies but no spectroscopic redshifts. In this case we take the photometric redshift and choose the closest redshift that would identify the detected emission line with a CO or atomic carbon (\ci) transitions presented in Table \ref{tab:LineRanges}. 

The fact that ASPECS-LP-3mm.16 has a counterpart galaxy with spectroscopic redshift that supports the observed frequency as being CO(2-1) at $z_{\rm CO}=1.294$ allows us to increase the number of secure emission lines candidates up to 16. 

We now compare our source list to that published in the pilot study by \citet{Walter2016} that covered a significantly smaller area on the sky with only one pointing in the 3mm band. Out of the 10 sources considered in the pilot program, two sources are not included in the area covered by our large program (sources 3mm.6 and 3mm.10 in \citet{Walter2016}). Out of the remaining 8 sources, we recover four at high significance in the current study (sources 3mm.1, 3mm.2, 3mm.3, and 3mm.5). We can not confirm the remaining four sources, but note that based on the improved selection discussed in the current paper (in particular treating narrow line-widths correctly when estimating the fidelity, see Sec. \ref{sec:Fidelity}), we should not have selected these sources in the pilot program (indeed, most of the unconfirmed sources have very narrow, i.e. $<$100\,km\,s$^{-1}$, line widths).

\subsection{Detected continuum sources} \label{sec:ResultsBlindContinuum}

\begin{figure*}
\epsscale{0.33}
\plotone{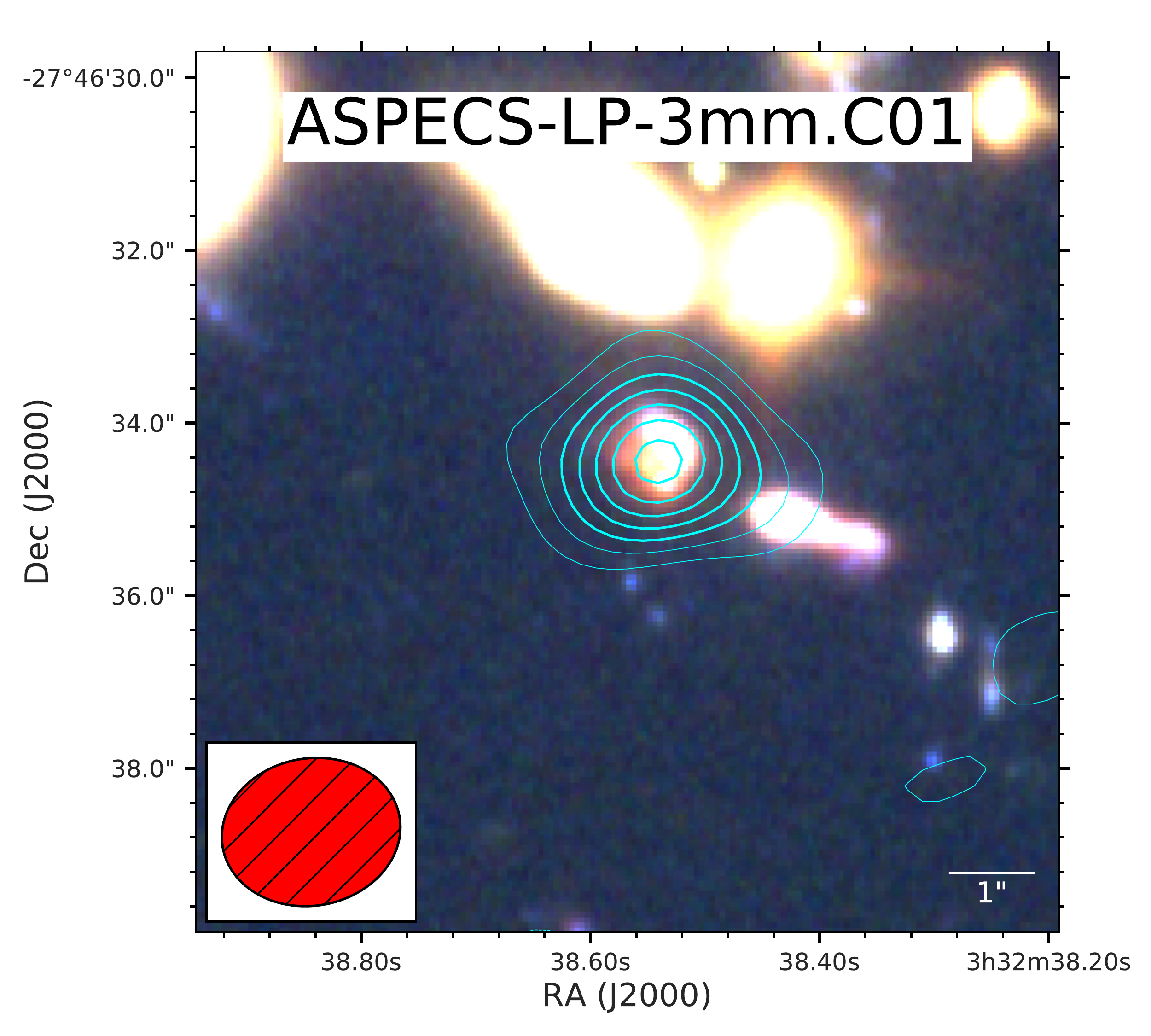}
\plotone{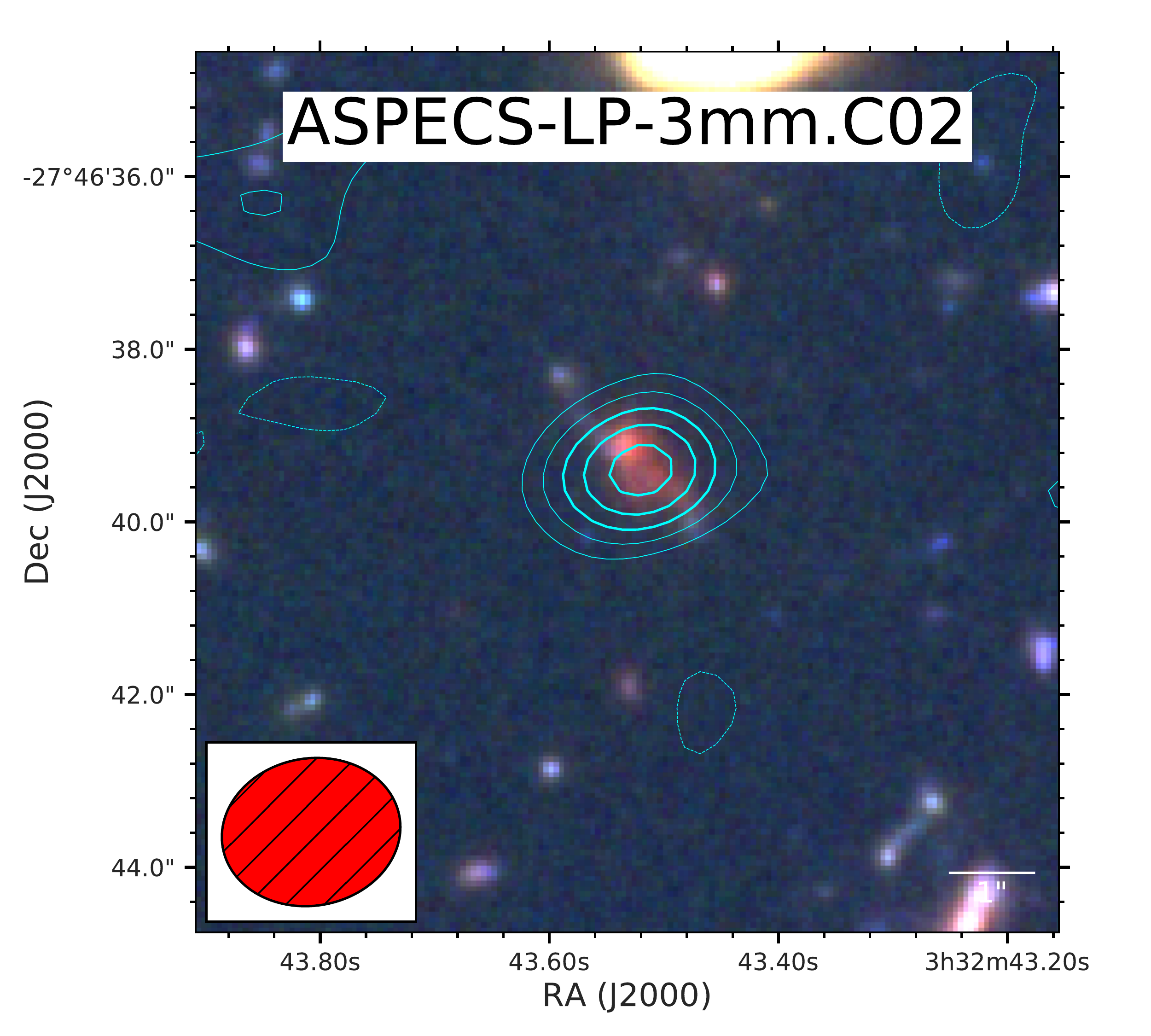}
\plotone{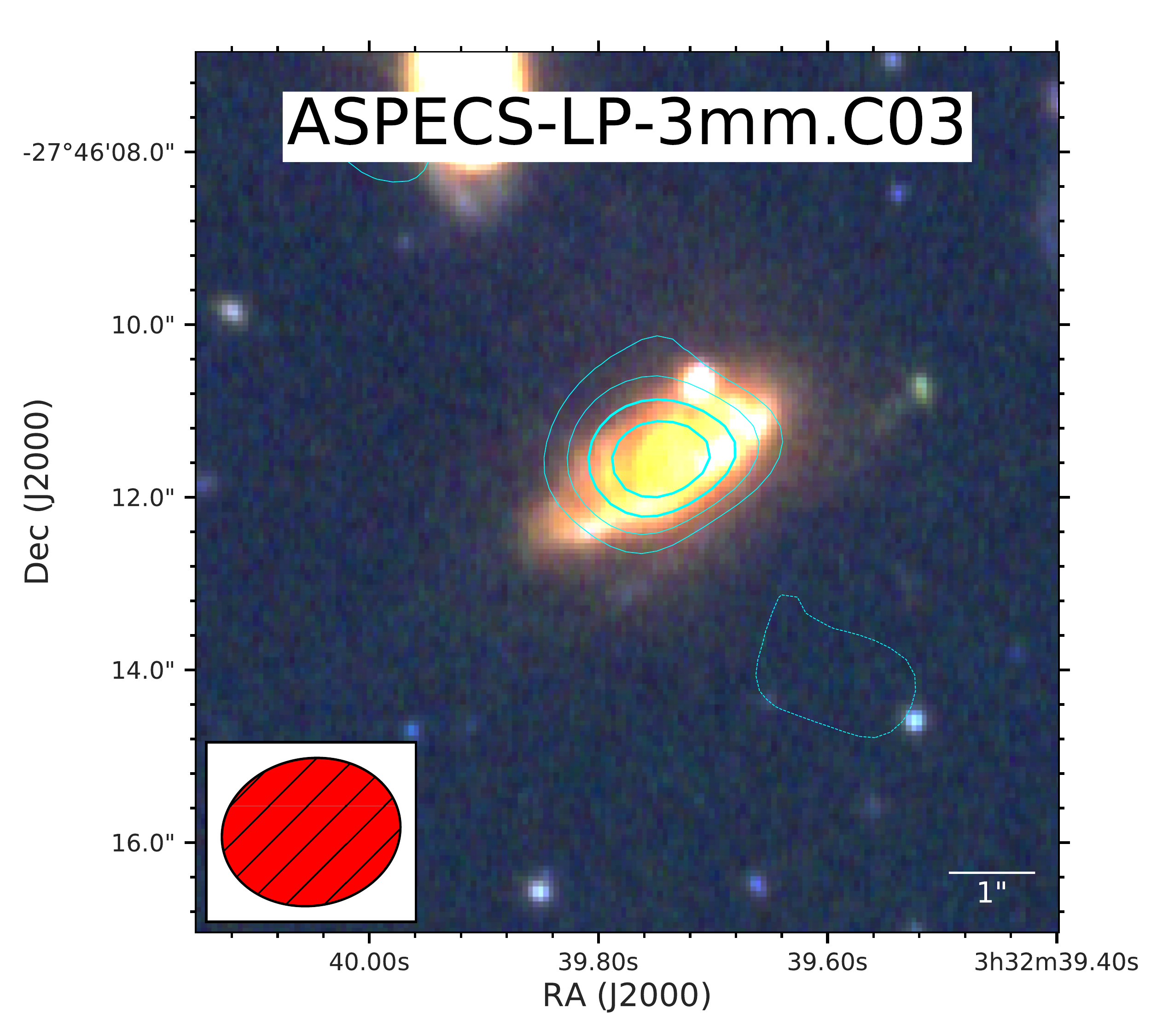}
\plotone{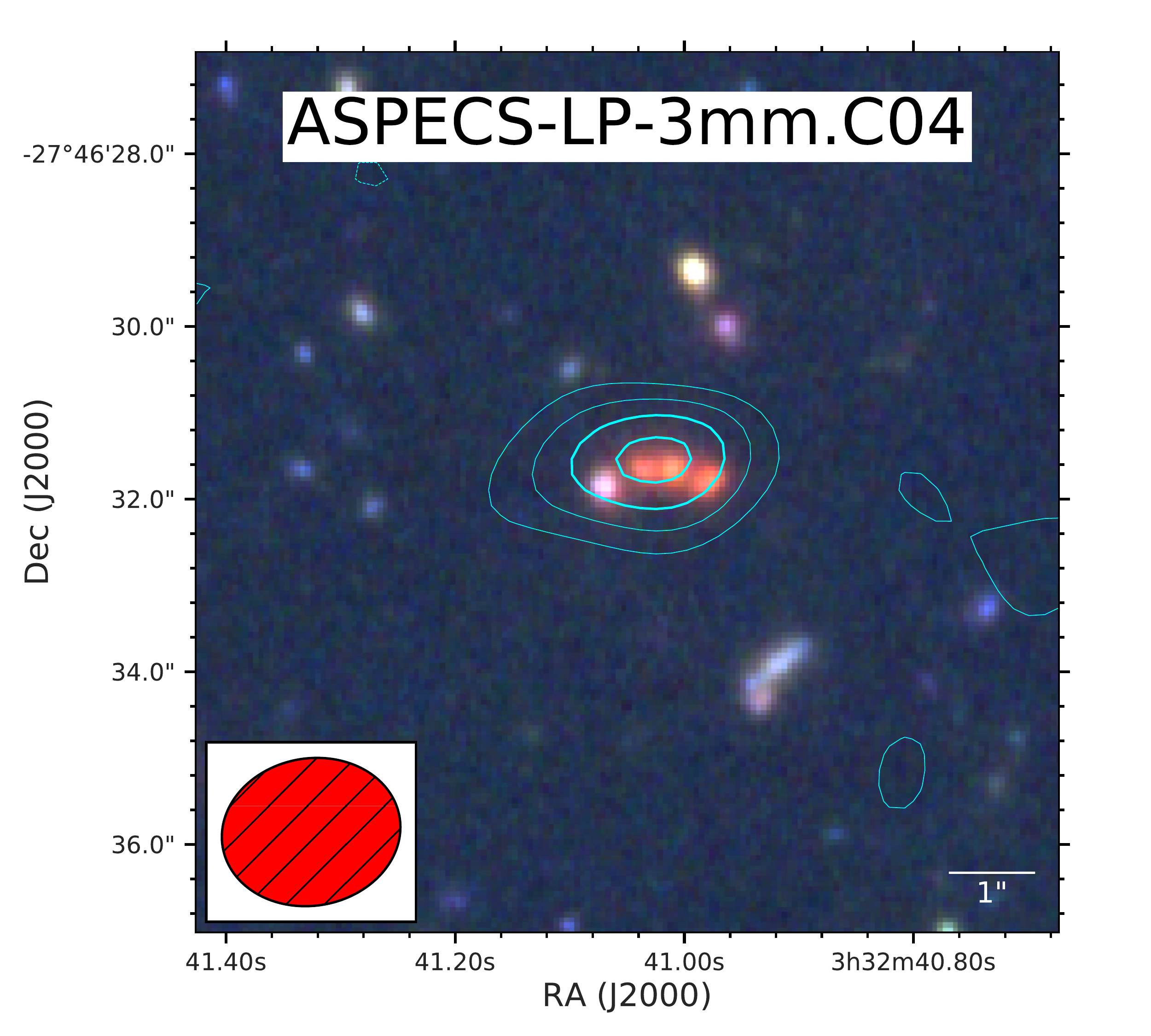}
\plotone{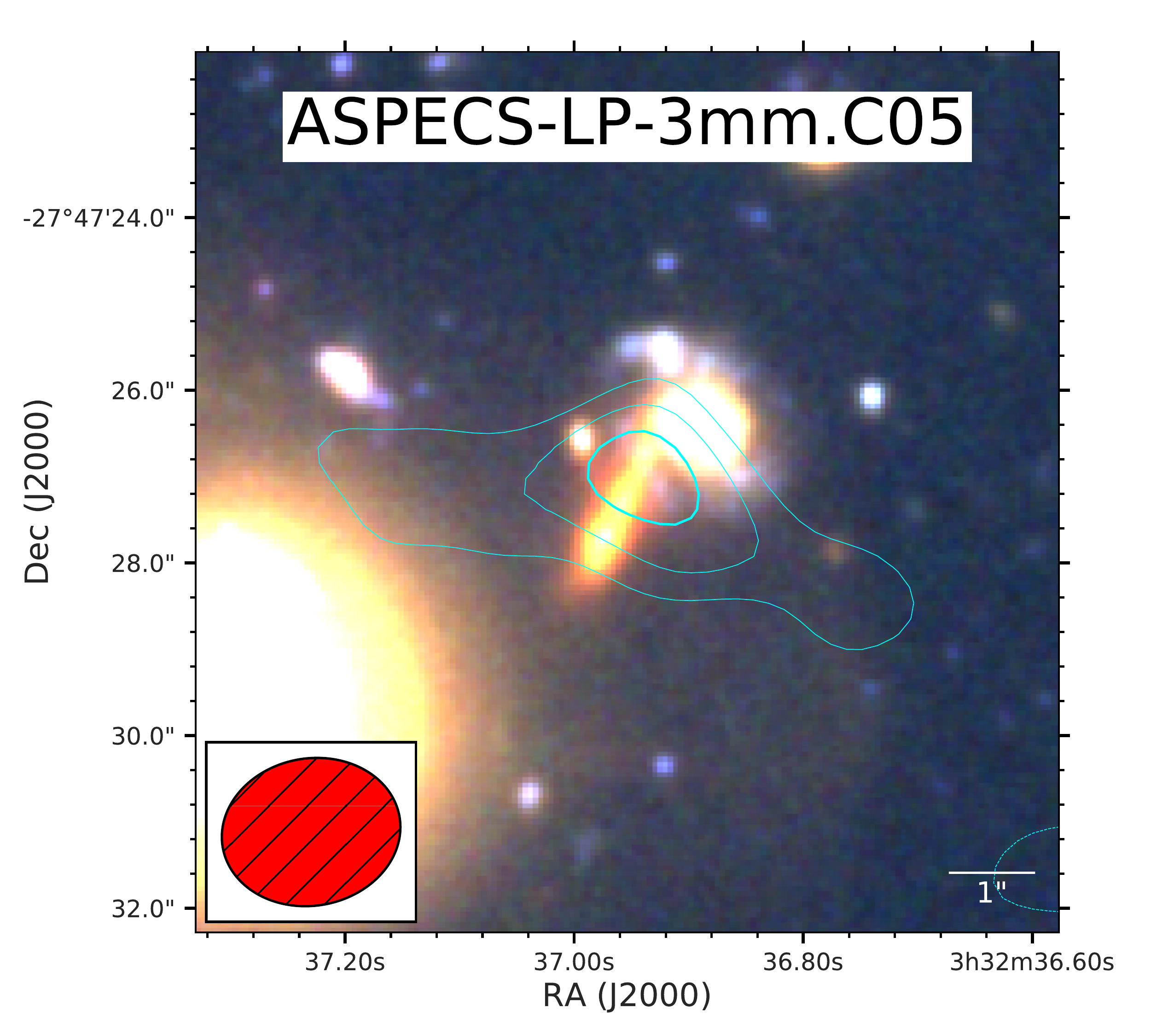}
\plotone{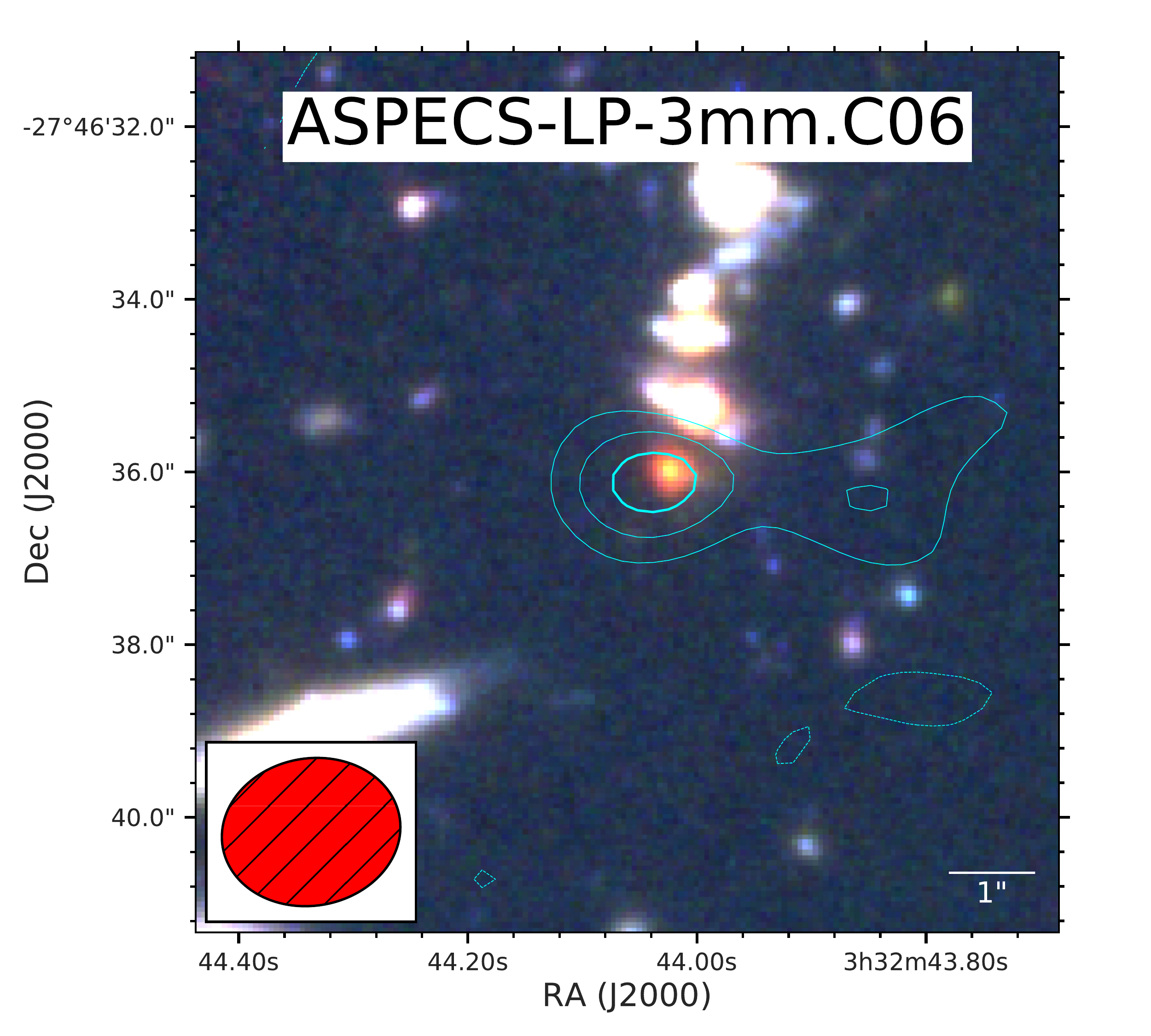}
\caption{Color postage stamp of the six continuum source candidates discovered in the ASPECS-LP continuum image. The contour levels go from $\pm3\sigma$ up to $10\sigma$ in steps of $1\sigma$. The color scale is the same as in fig. \ref{fig:footprint}. \label{fig:LP_Continuum_postamp1}}
\end{figure*}

\movetabledown=1.5in
\begin{rotatetable*}
\begin{deluxetable*}{cccccccccc}
\tablecaption{Continuum source candidates in ASPECS-LP 3mm continuum image. \label{tab:LPContinuum}}
\tablehead{
\colhead{ID} & 
\colhead{R.A.} & 
\colhead{Dec} & 
\colhead{S/N} & 
\colhead{{\rm Fidelity}} & 
\colhead{Integrated Flux 3 mm} & 
\colhead{Size} & 
\colhead{Integrated Flux 1.2 mm} & 
\colhead{$\beta$} & 
\colhead{Identification} \\
\colhead{} & 
\colhead{} & 
\colhead{} & 
\colhead{} & 
\colhead{} & 
\colhead{[$\rm \mu Jy$]} & 
\colhead{} & 
\colhead{[$\rm \mu Jy$]} & 
\colhead{} & 
\colhead{}
}
\colnumbers
\startdata
ASPECS-LP-3mm.C01 & 03:32:38.54 & -27:46:34.44 & 8.4 & $1.0_{-0.0}^{+0.0}$&  $32.5 \pm 3.8 $ &PS &$ 746 \pm 31 $&$ 1.8 \pm 0.1 $& $z=2.543$ \\
ASPECS-LP-3mm.C02 & 03:32:43.52 & -27:46:39.47 & 6.5 & $1.0_{-0.0}^{+0.0}$&  $46.5 \pm 7.1 $ &PS &$ 835 \pm 75 $&$ 1.6 \pm 0.2 $& $z=2.696$ \\
ASPECS-LP-3mm.C03 & 03:32:39.75 & -27:46:11.58 & 6.0 & $1.0_{-0.0}^{+0.0}$&  $27.4 \pm 4.6 $ &PS &$ 376 \pm 45 $&$ 1.9 \pm 0.3 $& $z=1.550$ \\
ASPECS-LP-3mm.C04 & 03:32:41.02 & -27:46:31.56 & 5.4 & $1.0_{-0.0}^{+0.0}$&  $22.7 \pm 4.2 $ &PS &$ 292 \pm 38 $&$ 1.2 \pm 0.3 $& $z=2.454$ \\
ASPECS-LP-3mm.C05 & 03:32:36.94 & -27:47:27.00 & 4.7 & $0.95_{-0.02}^{+0.02}$&  $29.6 \pm 6.3 $ &EXT &$ 481 \pm 47 $&$ 1.7 \pm 0.3 $& $z=1.759$ \\
ASPECS-LP-3mm.C06 & 03:32:44.03 & -27:46:36.05 & 4.6 & $0.97_{-0.01}^{+0.01}$&  $44.5 \pm 9.7 $ &PS &$ 798 \pm 84 $&$ 1.6 \pm 0.3 $& $z=2.698$ 
\enddata
\tablecomments{
(1) Identification for continuum source candidates discovered in ASPECS-LP 3mm continuum image.
(2) Right ascension (J2000).
(3) Declination (J2000).
(4) S/N value return by LineSeeker assuming an unresolved source.
(5) Fidelity estimate using negative detection and Poisson statistics.
(6) Integrated flux density 3 mm obtained after removing the channels with bright emission lines.
(7) Size of the continuum source candidate. EXT corresponds to resolved source while PS to source candidates consisting with being a point source.
(8) Integrated flux density 1.2 mm.
(9) Dust emissivity index ($\beta$) estimated assuming a dust temperature of 35 K.
(10) Redshift of the NIR counterpart.
}
\end{deluxetable*}
\end{rotatetable*}

In Tab. \ref{tab:LPContinuum}, we present the list of significant continuum source candidates in the ASPECS-LP 3 mm continuum image. We present all the continuum source candidates for which the condition of ${\rm Fidelity}\geq0.9$ is fulfilled. The NIR postage stamps of the continuum source candidates are presented in Fig. \ref{fig:LP_Continuum_postamp1}.

In table \ref{tab:LPContinuum}, we also present the integrated flux density (corrected by the mosaic primary beam response) and whether the continuum source candidate is spatially resolved or not (in the same way as for the emission--line candidates). We also present the integrated line flux density measured in the 1.2 mm continuum map created by combining the observations from ASPECS-Pilot at 1.2 mm and from 1.3 mm ALMA map in the UDF \citep{Dunlop2017}. The six candidates are detected in the 1.2 mm map, fully supporting the fidelity estimates and the reliability of the sample. 
The last column in Table \ref{tab:LPContinuum} presents the redshifts for the NIR counterparts to the 3 mm continuum source candidate, which show a median redshift of $z_{\rm m}\approx 2.5$. The fidelity values tells us that $\approx5.9$ out of the six source candidates should be real. 

%%%%%%%%%%%%%%%%%%%%%%%%%%%%%%%%%%%%%%%%%%%%%%%%%%%%%%%%%%%%%%%%%%%%%%%%%%%%%%%%%%%%%%%%%%%%%%
%%%%%%%%%%%%%%%%%%%%%%%%%%%%%%%%%%%%%%%%%%%%%%%%%%%%%%%%%%%%%%%%%%%%%%%%%%%%%%%%%%%%%%%%%%%%%%
%%%%%%%%%%%%%%%%%%%%%%%%%%%%%%%%%%%%%%%%%%%%%%%%%%%%%%%%%%%%%%%%%%%%%%%%%%%%%%%%%%%%%%%%%%%%%%
\section{Discussion} \label{sec:Discussion}

\subsection{The distribution of emission line widths \label{sec:FWHMDistribution}}

\begin{figure}
\epsscale{1.2}
\plotone{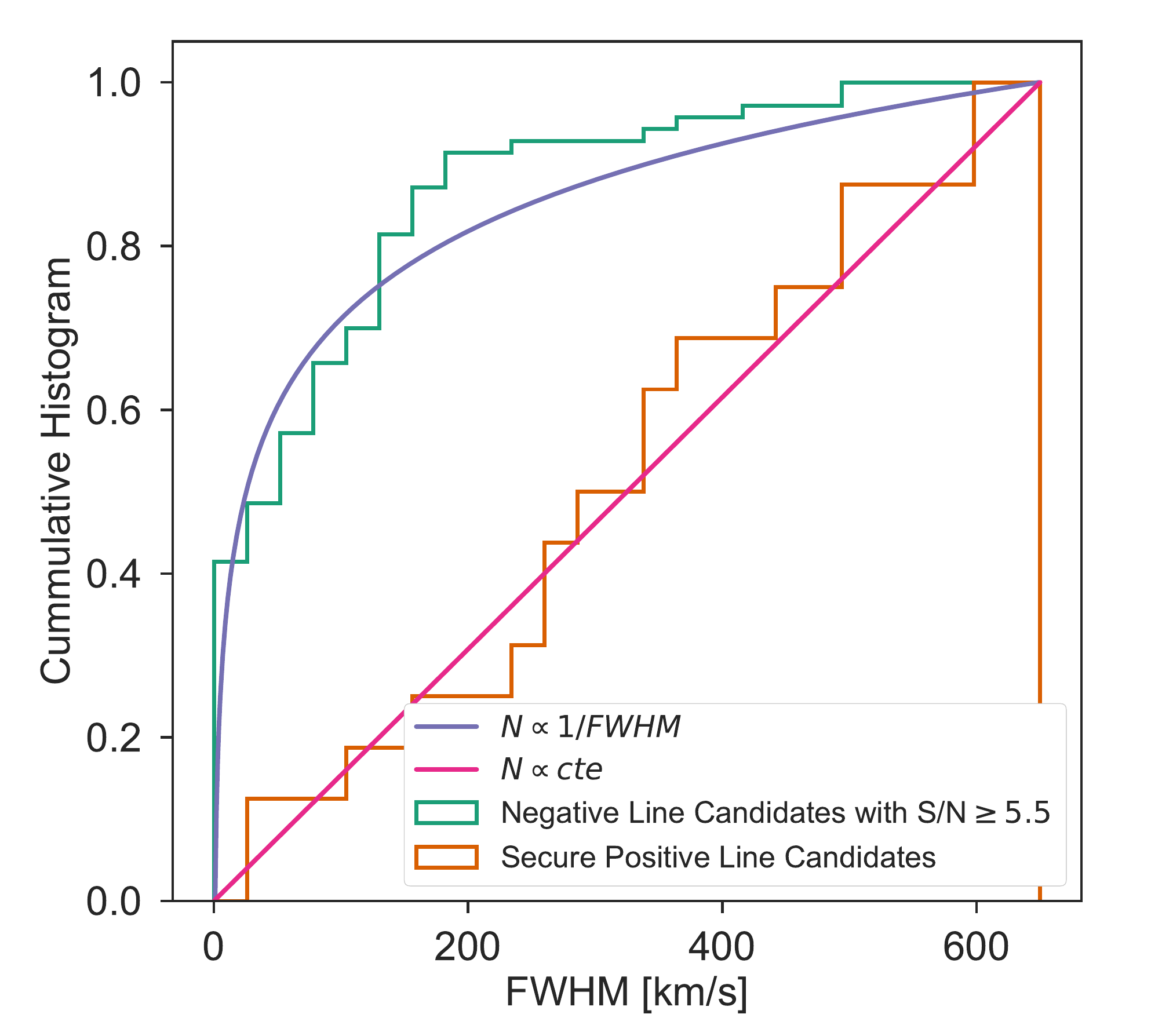}
\caption{Cumulative histogram of the FWHM distribution of negative line candidates (green) and secure positive line candidates (orange). The blue line corresponds to the distribution expected for line candidates width based on the number of independent elements as described in \$\ref{sec:Fidelity} ($\propto 1/{\rm FWHM}$). The red line corresponds to a FWHM flat distribution. \label{fig:HistogramFWHMNegatives}}
\end{figure}

In Figure \ref{fig:HistogramFWHMNegatives}, we compare the distribution of line candidates widths found in the negative data with the widths of secure emission lines. The green histogram shows the distribution of line widths for 71 negative line candidates detected with $S/N\geq5.5$ while the orange histogram shows the cumulative histogram of the 16 secure detection presented in Tables \ref{tab:LP} and \ref{tab:LP_Parameters}.  It is clear from just a visual inspection that both samples have different line widths distributions. The negative lines are strongly weighted towards narrower lines and have a median line FWHM of $\approx52\kms$ while the secure positive lines have a flat FWHM distribution and a median value of $\approx331\kms$. 

We expect the negative lines to be produced only by noise and should follow a FWHM distribution determined by the number of independent elements for each FWHM value. The number of independent elements will be inversely proportional to the FWHM of the line. Based on this, the distribution of widths for the negative lines should follow a shape similar to $\propto 1/{\rm FWHM}$ and the cumulative distribution will then have the following shape $\propto \log({\rm FWHM})$. Figure \ref{fig:HistogramFWHMNegatives} shows that the distribution of observed negative line candidates widths with $S/N\geq5.5$ follows closely a curve $\propto \log({\rm FWHM})$ (blue line). Applying higher S/N threshold cut results in distributions similar to the blue line. We only tested down to $S/N\geq5.5$ because for lower S/N values the number of line candidates increases dramatically. 

From Fig. \ref{fig:HistogramFWHMNegatives} and Table \ref{tab:LP_Parameters}, we can see that the secure emission lines follow a flat distribution in FWHM, with the cumulative histogram follows closely a curve $\propto {\rm FWHM}$. In our line search the distribution of line widths is flat within the FWHM range of $\sim40\kms$ to $\sim620\kms$. 

\subsection{The importance of looking for spatially resolved emission lines. \label{sec:SizeDistribution}}

\begin{deluxetable*}{cccccc}
\tablecaption{ASPECS-LP 3mm continuum number counts. \label{tab:3mmNumberCounts}}
\tablehead{
\colhead{${\rm S_{\nu}}$ range} & 
\colhead{$\log{\rm S_{\nu}}$} & 
\colhead{$d{\rm N}/d\log{\rm S_{\nu}}$} & 
\colhead{${\rm N}({\geq \rm S_{\nu}})$} & 
\colhead{$\delta{\rm N}_{-}$} & 
\colhead{$\delta{\rm N}^{+}$} \\
\colhead{[$\times 10^{-3}$ mJy]} & 
\colhead{[mJy]} & 
\colhead{[mJy$^{-1}$]} & 
\colhead{$[{\rm deg^{-2}}]$} & 
\colhead{$[{\rm deg^{-2}}]$} & 
\colhead{$[{\rm deg^{-2}}]$}
}
\colnumbers
\startdata
17.78--31.62 & -1.625 & 3&  5940 & 1995&  2579\\
31.62--56.23 & -1.375 & 3&  2433 & 1046&  1479\\
56.23--100.0 & -1.125 & $<$1.83 &  $<$1246& \nodata & \nodata\\
\enddata
\tablecomments{
(1) Flux density bin. 
(2) Flux density bin center .
(3) Number of sources per bin (before fidelity and completeness correction). In the case of no sources, an upper limit of $<$1.83 is used.
(4) Cumulative number count of sources per square degree. In the case of no sources, a $1\sigma$ upper limit is used. 
(5) Lower uncertainty in the number counts.
(6) Upper uncertainty in the number counts.
}
\end{deluxetable*}

In \$\ref{Sec:ComparisonMethods} we discussed how MF3D was the only method designed to look for extended emission lines while LineSeeker and FindClump focused mainly on unresolved emission lines. 
We can use the spatial distribution of the detected lines to test whether focusing the search for unresolved emission lines is a good choice. 
In Table \ref{tab:LP_Parameters} we found that at least half of the detected emission lines were consistent with being resolved with some angular extension beyond the synthesized beam size (at least in this image plane analysis). Unsurprisingly, most of the extended emission lines correspond to the brightest emission lines detected while the fainter emission lines are dominated by lines consisting with being unresolved. 
These results suggest that, at least to a first order, the extended galaxies with a high CO luminosities will be easily detected in a search for unresolved emission lines, mainly because of they intrinsic high line luminosity. At the same time, more compact galaxies with lower CO luminosities, as those in the bottom half of the lines detected in the LP, will also be detected in the search for unresolved lines. 

Emission lines with integrated flux in the order of $\sim0.1$ Jy$\kms$ are close to the limit of what we can reliable identify within our data. According to the completeness levels in Table. \ref{tab:CompletenessFluxPeakFWHM}, we should be able to detect some of the emission lines with that level of integrated flux if they appear as unresolved by the synthesized beam. The ability to identify an emission lines of $\sim0.1$ Jy$\kms$ with out methods will decrease if the total emission is distributed over an angular scale larger than 1.5-2.0\arcsec. We argue that this scenario, despite being possible, should not represent a major problem for our current analysis. Only a few of our emission line counterpart galaxies show emission beyond 2.0\arcsec, and these emission lines also belong to the bright end of our sample. Most of the fainter emission line counterpart galaxies show emission sizes smaller or within the synthesized beam. At the same time, we expect that some of the potential faint emission lines could correspond to galaxies at even higher redshifts, which should have smaller angular scales than the bulk of galaxies at $z=1-2$ already detected. Because of all of this, we should not be missing a population of faint extended emission lines.

\subsection{3 mm number counts}

\begin{figure*}
\epsscale{1.2}
\plotone{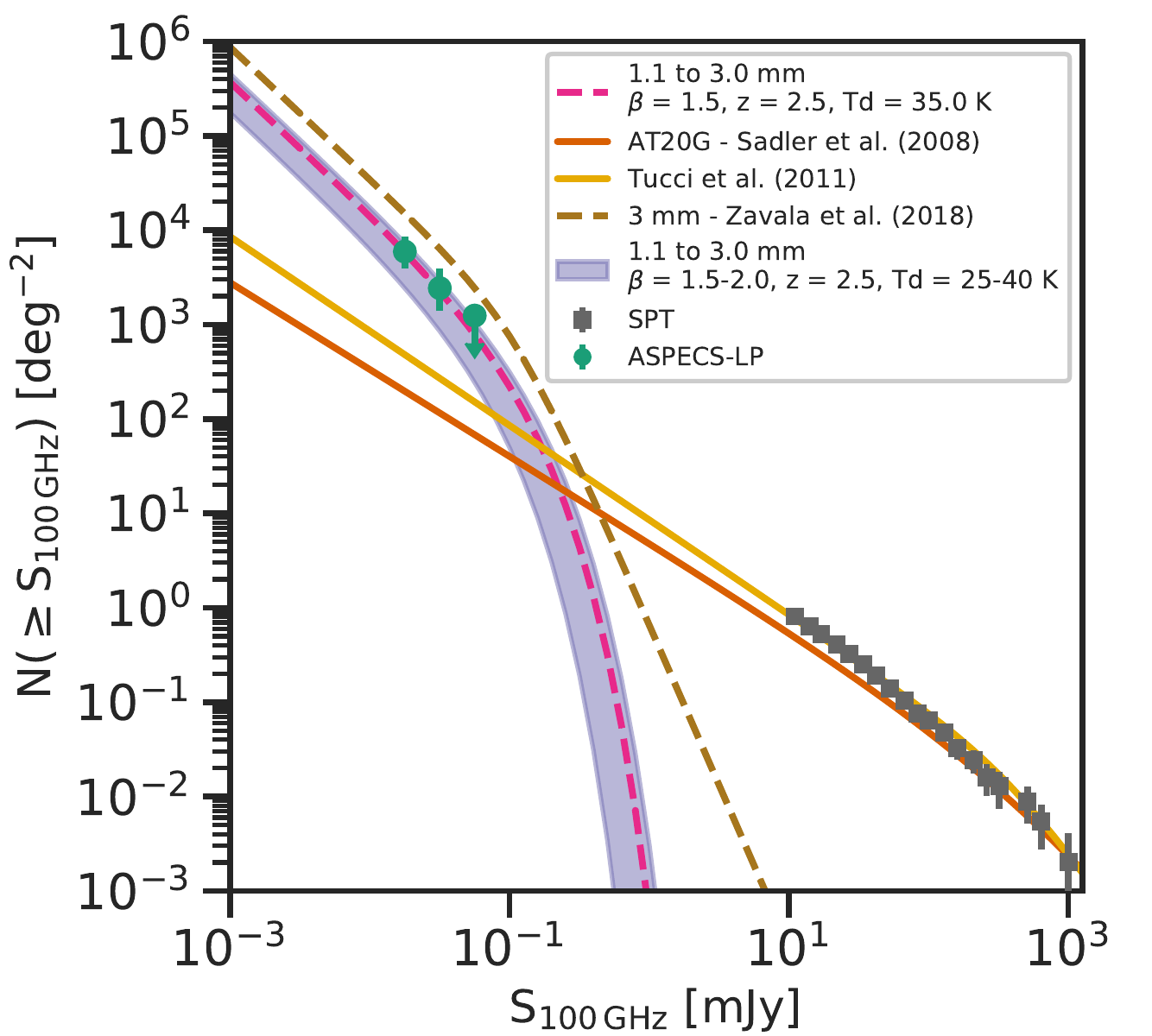}
\caption{Cumulative number counts of ASPECS-LP 3 mm continuum observations. The green points show the number counts computed as part of this work. The gray squares show the 3 mm number counts obtained by SPT \citep{Mocanu2013}. The orange and yellow lines show the number counts at 100 GHz extrapolated from 20 and 5 GHz respectively \citep{Sadler2008, Tucci2011}. The blue shaded region shows the 1.1 mm number counts extrapolated to 3 mm using a range of parameters \citep{Franco2018}. The magenta dashed-line shows the 3 mm extrapolated number counts assuming a dust temperature of 35 K, $\beta=1.5$ and $z=2.5$. The brown dashed-line corresponds to the best-fit 3 mm number counts presented by \citet{Zavala2018}. \label{fig:ContinuumNumberCounts}}
\end{figure*}

In this section we estimate the number counts of sources discovered in the ASPECS-LP 3 mm continuum observations. The number counts per bin ($N(S_{i})$) are computed as follows

\begin{equation}
    N(S_{i}) = \frac{1}{A}\sum_{j=1}^{X_{i}}\frac{P_{i}}{C_{i}},
\end{equation}

where $A$ is the total area of the observations ($1.46\times10^{-3}\ts{\rm deg^{2}}$), 
$P_{i}$ is the probability of each source of being real (Fidelity) and $C_{i}$ is the completeness correction for the corresponding intrinsic flux density. The cumulative number counts are obtained by summing up each $N(S_{i})$ over all the possible $\geq{/rm S}_{i}$. The size of the bins is $\log{S_{\nu}}=0.25$ and we use all the source candidates with ${\rm Fidelity}\geq0.9$ listed in Tab. \ref{tab:LPContinuum}. 
In Tab. \ref{tab:3mmNumberCounts} we present the differential and cumulative number counts obtained in this work. We used the flux density values presented in Tab. \ref{tab:LPContinuum}, with the ${\rm Fidelity}$ values used for $P_{i}$ and the completeness values obtained from Tab. \ref{tab:CompletenessContinuum} and corrected by the PB response. 

The cumulative number counts are also presented in Fig. \ref{fig:ContinuumNumberCounts}, where we compare the observed 3 mm number counts with predictions obtained at other wavelength. We show the number counts at 95 GHz as predicted by \citet{Sadler2008}, which used simultaneous measurements at 20 and 95 GHz using The Australia Telescope Compact Array (ATCA) to obtain the spectral index of extragalactic sources and extrapolate the number counts obtained by the Australian Telescope 20 GHz (AT30G) survey \citep{Ricci2004,Sadler2006} to 95 GHz. We also show the 100 GHz number counts extrapolated from 5 GHz observations using a spectral index of $\alpha=-0.23$ as given by the model C2Ex, which allows for different distribution in the break frequencies for the synchrotron emission of sources \citep{Tucci2011}. Both number counts prediction from lower frequencies are in agreement with the bright end of the number counts measurements at 95 GHz obtained by the South Pole Telescope (SPT). This agreement is expected since the population of 95 GHz sources is dominated by synchrotron emission \citep{Mocanu2013}. 
Despite the agreement at the bright end, we see that the extrapolation towards lower frequencies underpredicts the number counts at 100 GHz in the faint end ($\leq0.1$ mJy) when compared to our results, which indicates that the 3 mm population is not dominated by synchrotron-dominates sources. 

We also compare our number counts with the observed 3 mm number counts presented by \citet{Zavala2018}. These number counts were obtained by analyzing 3 mm ALMA archival data and correspond to the only sub-mJy number counts presented to date. We find that our number counts are $\sim3\times$ lower than the best-fit function shown as a brown dashed-line in Fig. \ref{fig:ContinuumNumberCounts}. This difference could indicate that the ALMA archival data is biased towards overdensities of galaxies because of the nature of the targeted observations. In addition to that, the difference could be enhanced by the fact that the UDF appears to be underdense in mm number counts when compared to the blank population \citep{Aravena2016a}. We notice that ASPECS-LP-3mm.C01 is also used by \citet{Zavala2018} for the estimate of their number counts; however, they used for this source a 3 mm flux density of $57\pm7$ $\mu$Jy, which is twice the value derived here (Tab. \ref{tab:LPContinuum}) or in the ASPECS-Pilot \citep[$31.1\pm5$ $\mu$Jy - ][]{Aravena2016a}. This could partially explain the difference between the ASPECS 3 mm number counts and the results presented in \citet{Zavala2018}.  
Additionally, there is the possibility that our observed counts do not represent the real population of sources at 3 mm. We could be missing a large population of sources at 3 mm if their emission is extended beyond our already coarse beam size. If we use as reference the sizes of sources detected at $\sim1$ mm, we find that in fact the bulk of measured size show effective radii $<0.6\arcsec$ \citep{Fujimoto2017,Ikarashi2017}, which should be easily detected by our 3 mm continuum observations. 
Furthermore, a positive correlation has been found between dust emission size and IR luminosity as well as between size and redshift \citep{Fujimoto2017,Ikarashi2017}. This would indicate that the 3 mm source population in the sub-mJy regime should be even more compact than the 1 mm source population, because of their fainter intrinsic luminosity  and higher median redshift, which would not support a missing extended 3 mm population. Despite the latter, larger sizes measured for some fainter gravitationally-lensed galaxies could indicate the existence of an extended fainter population of dust emitting galaxies \citep{GL2017a}.

Galaxy models that fit the number counts simultaneously at different wavelengths find that the number counts of galaxies at 3 mm should be dominated by dust emission from unlensed main sequence galaxies \citep{Bethermin2011,Cai2013}. We use the well constrained number counts at higher frequencies ($\geq$200 GHz) and extrapolate them to 100 GHz. The extrapolation to 100 GHz assumes a modified black body emission with a dust emissivity index $\beta$ parameter for the Rayleigh--Jeans tail as well as the effects of the CMB on the observations \citep{daCunha2013}. We make use of the results obtained by \citet{Franco2018}, which takes the number counts of several published studies at wavelength between $850\,\micron$ and 1.3 mm to estimate the number counts at 1.1 mm \citep{Lindner2011,Scott2012,Karim2013,Hatsukade2013,Ono2014,Simpson2015,Carniani2015,Oteo2016,Hatsukade2016,Aravena2016a,Fujimoto2016,Umehata2017,Geach2017,Dunlop2017}. 
In Fig. \ref{fig:ContinuumNumberCounts} we present the 3 mm number counts scaled from 1.1 mm using a range of parameters motivated by previous studies (blue shaded region). For the dust emissivity index we use a range of $\beta=1.5-2.0$ \citep{DunneAndEales2001,Chapin2009,Clements2010,Draine2011,Planck2011a,Planck2011b}, for the dust temperature we take a range of $25-40$ K \citep{Magdis2012,Magnelli2014,Schreiber2018} and for the redshift we take the median redshift of $z=2.5$ found for sources detected in this work (Tab. \ref{tab:LPContinuum}). We also show in Fig. \ref{fig:ContinuumNumberCounts} the 3 mm number counts scaled from 1.1 mm when assuming a dust temperature of 35 K, which is the average dust temperature expected for star-forming galaxies at $z=2.5$ as presented by \citet{Schreiber2018}. We find that a dust emissivity of $\beta=1.5$ is needed to match the observed number counts, which is in agreement with the average value $\beta=1.6\pm0.2$ obtained for our continuum sources when assuming a dust temperature of 35 K (Tab. \ref{tab:LPContinuum}).

We conclude that our observed 3 mm number counts are consistent with those observed at shorter wavelengths based on the expected spectral energy distribution of galaxies at $z\sim2.5$.
In fact, \citet{Casey2018} presented three backward evolution galaxy models which predict the median redshift of 3 mm observations to be at $z=2.3-3.2$ for a flux density cut-off similar to the one presented here.

%%%%%%%%%%%%%%%%%%%%%%%%%%%%%%%%%%%%%%%%%%%%%%%%%%%%%%%%%%%%%%%%%%%%%%%%%%%%%%%%%%%%%%%%%%%%%%
%%%%%%%%%%%%%%%%%%%%%%%%%%%%%%%%%%%%%%%%%%%%%%%%%%%%%%%%%%%%%%%%%%%%%%%%%%%%%%%%%%%%%%%%%%%%%%
%%%%%%%%%%%%%%%%%%%%%%%%%%%%%%%%%%%%%%%%%%%%%%%%%%%%%%%%%%%%%%%%%%%%%%%%%%%%%%%%%%%%%%%%%%%%%%
\section{Conclusion} \label{sec:Conclusion}

In this paper we present the results from the search for emission lines and continuum sources in the observations covering an area of 4.6 arcmin$^2$ across a frequency range of 30.75 GHz in the ALMA band 3 as part of the ASPECS-LP. We used both the ALMA band 3 observations obtained as part of ASPECS-Pilot and ASPECS-LP to compare and test different methods to search for emission lines in large data cubes. The comparison of the three search methods, LineSeeker, FindClump and MF3D has shown that three methods all return similar values when tested on simulated data cubes with injected emission lines and used in real data cubes. 

We also present new methods to obtain reliable fidelity estimates for emission lines detected in data cubes and explain the rationale to use the S/N ratio as well as the width of emission line candidates when estimating their fidelity. We show that fidelity values obtained from the negative data return reliable results for the selection of real emission lines. The sames results were applied to the search of sources in the continuum image.

Based on these methods, we identified in the data cube sixteen emission line candidates with a ${\rm Fidelity}\geq0.9$, fifteen of them having high probabilities of being real based on the fidelity limits ${\rm Fidelity}=1$. Another emission line candidate is also found to have a high probability of being real based on the fidelity values and the fact that the NIR counterpart galaxy has a matching spectroscopic redshift. 

The new algorithms and findings presented in this paper are crucial for the creation of a reliable CO luminosity function which will help us understand the distribution of molecular gas across cosmic time. In \citet{Decarli2019} we present the CO luminosity function for the detected emission lines together with the cosmic molecular gas density across time. In \citet{Boogaard2019} we derive the properties of the emission line galaxies and their optical counterparts observed by MUSE. Finally, in \citet{Aravena2019} we discuss in a global context the properties of the molecular gas content of these galaxies. 

The same algorithms used for the emission line search were used to obtain a sample of reliable continuum source candidates in the ASPECS-LP 3 mm continuum image. We identified six continuum source candidates with a ${\rm Fidelity}\geq0.9$. All six sources have a clear NIR counterpart and redshift estimates, with a median redshift of $z_{m}=2.5$. 

Finally, using the list of significant continuum sources we derived the 3 mm number counts at flux density range $<0.06$ mJy, three order of magnitudes lower than previous large area 3 mm observations. We find that the observed 3 mm number counts are inconsistent with the extrapolation obtained from lower frequencies surveys assuming synchrotron. However, they are consistent with the number counts obtained from dusty star-forming galaxies scaled from 1.1 mm results and assuming a dust emissivity index of $\beta=1.5$ a dust temperature of 35 K and a median redshift of $z=2.5$. These values are in good agreement with the  galaxy population expected to be detected within our 3 mm continuum observations.  

Our number counts represent one of the first constraints to the faint end of the 3 mm number counts and offer a unique window for revealing the different emission processes in galaxies at redshifts $z>2$.

\acknowledgments
This Paper makes use of the ALMA data \newline ADS/JAO.ALMA\#2016.1.00324.L. ALMA is a partnership of ESO (representing its member states), NSF (USA) and NINS (Japan), together with NRC (Canada), NSC and ASIAA (Taiwan), and KASI (Republic of Korea), in cooperation with the Republic of Chile. The Joint ALMA Observatory is operated by ESO, AUI/NRAO and NAOJ. The National Radio Astronomy Observatory is a facility of the National Science Foundation operated under cooperative agreement by Associated Universities, Inc.

We have made available online LineSeeker, the code used to search for emission lines in this work. The code is available as a set of Python scripts that search for emission line candidates and returns the different fidelities estimates described above. The website for the download can be found in the software section below.
The Geryon cluster at the Centro de Astro-Ingenieria UC was extensively used for the 
calculations performed in this paper. BASAL CATA PFB-06, the Anillo ACT-86, 
FONDEQUIP AIC-57, and QUIMAL 130008 provided funding for several improvements 
to the Geryon cluster. 
``Este trabajo cont\'o con el apoyo de CONICYT + Programa de Astronom\'ia+ Fondo CHINA-CONICYT''.
JGL acknowledges partial support from ALMA-CONICYT project 31160033.
D.R. and R.P. acknowledge support from the National Science Foundation 
under grant number AST-1614213.
IRS acknowledges support from the ERC Advanced Grant DUSTYGAL (321334) and STFC (ST/P000541/1).
FEB acknowledges support from CONICYT grant Basal AFB-170002, and the Ministry of Economy, Development, and Tourism's Millennium Science Initiative through grant IC120009, awarded to The Millennium Institute of Astrophysics, MAS.
T.D-S. acknowledges support from ALMA-CONICYT project 31130005 and FONDECYT project 1151239.
JH acknowledges support of the VIDI research programme with project 
number 639.042.611, which is (partly) financed by the Netherlands 
Organisation for Scientific Research (NWO).

\vspace{5mm}
\facilities{ALMA}

\software{astropy \citep{Astropy},  
          SExtractor \citep{Bertin1996}, \href{https://github.com/jigonzal/LineSeeker}{LineSeeker}
         }

\end{document}